\documentclass[11pt]{article}
\usepackage{amsmath,bm,amssymb,setspace,subcaption,natbib,adjustbox,bbm,array,multirow,tablefootnote,authblk,amsthm}
\usepackage[hidelinks]{hyperref}
\usepackage{pgf,tikz}
\usetikzlibrary{arrows,shapes.arrows,shapes.geometric,shapes.multipart,decorations.pathmorphing,positioning,shapes.swigs}
\usepackage[normalem]{ulem}
\usepackage[left=1in,top=1in,right=1in,bottom=1in]{geometry}
\theoremstyle{plain}
\newtheorem{lemma}{Lemma}
\newtheorem{theorem}{Theorem}
\newtheorem{corollary}{Corollary}
\newtheorem{remark}{Remark}
\newtheorem*{assumption*}{\assumptionnumber}
\providecommand{\assumptionnumber}{}
\makeatletter
\newenvironment{assumption4}[2]
{%
	\renewcommand{\assumptionnumber}{Assumption #1$'$}%
	\begin{assumption*}%
		\protected@edef\@currentlabel{#1$'$}%
	}
	{%
	\end{assumption*}
}
\makeatother

\newtheorem{assumption}{Assumption}

\newcommand{\myh}{h(\ttrt,\ccov)}
\def\transpose{^{\sf \scriptscriptstyle{T}}}
\newcommand{\ind}{\perp\!\!\!\perp}

\newcommand{\pathwiseparam}{\theta}
\newcommand{\Mgest}{\mathcal{M}_1}
\newcommand{\Mipw}{\mathcal{M}_2}
\newcommand{\Mor}{\mathcal{M}_3}
\newcommand{\myeta}{\xi}

\newcommand{\textdr}{\text{dr}}
\newcommand{\WA}{WA}
\newcommand{\WAZ}{WAZ}
\newcommand{\WiA}{W_iA}
\newcommand{\WiZ}{W_iZ}
\newcommand{\WiAZ}{W_iAZ}
\newcommand{\WZ}{WZ}

\newcommand{\zdiff}{Z}
\newcommand{\zjdiff}{z_j}
\newcommand{\zonediff}{z_1}
\newcommand{\ztwodiff}{z_2}
\newcommand{\zkdiff}{z_k}
\newcommand{\bigzdiff}{\mathbf{Z}}
\newcommand{\adiff}{\trt}
\newcommand{\azdiff}{AZ}
\newcommand{\pn}{\mathbb{P}_n}
\newcommand{\trt}{A}
\newcommand{\ttrt}{a}
\newcommand{\cov}{X}
\newcommand{\ccov}{x}
\newcommand{\cons}{c}
\newcommand{\consm}{m}

\newcommand{\inv}{^{-1}}
\newcommand{\dt}{\frac{\partial}{\partial \theta}\longmid_{\theta=0}}
\newcommand{\IW}{\bm{\Gamma}_W}
\newcommand{\IWi}{\bm{\Gamma}_{Wi}}
\newcommand{\IZ}{\bm{\Gamma}_Z}
\newcommand{\IZi}{\bm{\Gamma}_{Zi}}

\newcommand{\IfZalpha}{\bm{\Pi}(Z| \trt,\cov;\hat{\alpha}^{\trt,Z}_{\text{mle}})}
\newcommand{\IfZ}{\bm{\Pi}(Z| \trt,\cov)}
\newcommand{\IfZstar}{\bm{\Pi}^*(Z| \trt,\cov)}
\newcommand{\IfZj}{\bm{\Pi}(Z| \trt,\cov)_j}
\newcommand{\R}{\bm{R}}
\newcommand{\deltaWAvec}{\bm{\delta^{W}_{\adiff}}}
\newcommand{\myetaYZvec}{\bm{\myeta^{Y}_{Z}}}
\newcommand{\myetaWZvec}{\bm{\myeta^{W}_{Z}}}
\newcommand{\myetaWiZvec}{\bm{\myeta^{w_i}_{Z}}}
\newcommand{\myetaWZjvec}{\bm{\myeta^{W}_{z_j}}}
\newcommand{\etaWAZvec}{\bm{\eta^{W}_{AZ}}}
\newcommand{\etaWiAZvec}{\bm{\eta^{w_i}_{AZ}}}

\newcommand{\dY}{\delta^Y}
\newcommand{\dW}{\delta^W}
\newcommand{\longmid}{\Bigl\lvert}
\newcommand{\gform}{\text{confounded}}
\author[1]{Xu Shi}
\author[2]{Wang Miao}
\author[3]{Jennifer C. Nelson}
\author[4]{Eric J. Tchetgen Tchetgen}
\affil[1]{Department of Biostatistics, University of Michigan}
\affil[2]{Guanghua School of Management, Peking University}
\affil[3]{Kaiser Permanente Washington Health Research Institute}
\affil[4]{Department of Statistics, the Wharton School, University of Pennsylvania}
	
\begin{document}

	\title{Multiply Robust Causal Inference with Double Negative Control Adjustment for Categorical Unmeasured Confounding}
	\date{}
\maketitle
	
\begin{abstract}
	Unmeasured confounding is a threat to causal inference in observational studies. 
	In recent years, use of negative controls to mitigate unmeasured confounding has gained increasing recognition and popularity. Negative controls have a longstanding tradition in laboratory sciences and epidemiology to rule out non-causal explanations, although they have been used primarily for bias detection. Recently, \cite{miao2018identifying} have described sufficient conditions under which a pair of negative control exposure and outcome variables can be used to nonparametrically identify the average treatment effect (ATE) from observational data subject to uncontrolled confounding. In this paper, we establish nonparametric identification of the ATE under weaker conditions in the case of categorical unmeasured confounding and negative control variables. We also provide a general semiparametric framework for obtaining inferences about the ATE while leveraging information about a possibly large number of measured covariates. In particular, we derive the semiparametric efficiency bound in the nonparametric model, and we propose multiply robust and locally efficient estimators when nonparametric estimation may not be feasible. We assess the finite sample performance of our methods in extensive simulation studies. Finally, we illustrate our methods with an application to the postlicensure surveillance of vaccine safety among children.
\end{abstract}


\begin{spacing}{1.5}
	\section{Introduction}\label{intro}
	Causal inference in observational studies often relies on the assumption of no unmeasured confounding. 
	However, as often the case in practice, when this assumption is violated, uncontrolled confounding can lead to biased estimates and invalid conclusions. Various methods have been proposed to detect and control for unmeasured confounding, among which use of negative controls has recently gained increasing recognition and popularity.
	Negative controls have a longstanding tradition in laboratory sciences and epidemiology to rule out non-causal explanation of empirical findings 
	\citep{rosenbaum1989role,weiss2002can,lipsitch2010negative,glass2014experimental}.
	Specifically, a negative control outcome is an outcome known not to be causally affected by the treatment of interest. Likewise, a negative control exposure is an exposure that does not causally affect the outcome of interest. 
	To the extent possible, both negative control exposure and outcome variables should be selected such that they share a common confounding mechanism as the exposure and outcome variables of primary interest.
	For example, in a study about the effect of influenza vaccination on influenza hospitalization, injury/trauma hospitalization was considered as a negative control outcome as it is not causally affected by influenza vaccination, but may be subject to the same confounding mechanism mainly driven by health-seeking behavior \citep{jackson2005evidence}. In this case, a non-null effect of the influenza vaccination against the negative control outcome amounts to compelling evidence of potential bias due to uncontrolled confounding.
	Another prominent example is the use of paternal exposure as a negative control exposure when determining the effect of maternal exposure during pregnancy on offspring health outcomes. Paternal exposure may have a similar association with the outcome as that of maternal exposure if there is hidden genetic or household-level confounding
	\citep{smith2008assessing,smith2012negative,lipsitch2012negative}.
	
	There is a growing literature on causal inference and statistical methods leveraging negative controls to mitigate confounding bias. \cite{rosenbaum1992detecting} considered testing and sensitivity analysis for unmeasured confounding by comparing matched treatment and control groups with respect to an unaffected outcome. \cite{tchetgen2013control} developed an outcome calibration approach based on the idea that the counterfactual primary outcomes can stand as a proxy for unmeasured confounders and suffice to account for confounding of the exposure–negative control outcome association.
	\cite{schuemie2014interpreting} proposed a $p$-value calibration approach by deriving an empirical null distribution of treatment effect using a collection of negative controls.
	\cite{sofer2016negative} generalized the difference-in-difference approach to the broader context of negative control outcome by allowing different scales for primary and negative control outcomes under a monotonicity assumption. In genetic studies, \cite{gagnon2012using} and \cite{wang2017confounder} considered removing unwanted variation or batch effects using negative control genes, which are assumed to be independent of the treatment of interest. 
	In time-series studies of air pollution, \cite{flanders2011method} and \cite{flanders2017new} considered partial correction of residual confounding using a future exposure to air pollution as a negative control exposure. 
	\cite{miao2017invited} extended their method by incorporating both past and future exposures as multiple negative control exposures to further attenuate confounding bias.
	
	The aforementioned methods rely on fairly restrictive assumptions such as rank preservation \citep{tchetgen2013control}, monotonicity 	\citep{sofer2016negative}, or linear models for the outcome and the unmeasured confounder \citep{gagnon2012using,wang2017confounder,flanders2011method,flanders2017new}. In a recent paper, \cite{miao2018identifying} proposed nonparametric identification of causal effects using a pair of negative control exposure and outcome variables under certain completeness conditions. Their work focused primarily on providing sufficient identification conditions and less so on inference. 
	Ideally, one would in principle aim to obtain inferences in the nonparametric model under which causal effects are identifiable. However, in practice, because one may wish to account for a moderate to large number of observed confounders, nonparametric inference may not be feasible due to the curse of dimensionality.
	
	In this paper, we propose to resolve this difficulty by developing a general semiparametric framework for inferences about the average treatment effect (ATE) in the context of categorical unmeasured confounding adjustment using a pair of negative control exposure and outcome variables while accounting for a possibly large number of observed confounders.
	In particular, we first extend the identification result of \cite{miao2018identifying} to a allow for a weaker set of conditions, and provide an alternative representation of the identifying functional for the ATE. The representation is a difference between the standard g-formula of \cite{robins1986new} that fails to account for unmeasured confounding, 
	and an explicit bias correction term that leverages a pair of negative controls to completely account for unmeasured confounding.
	We then characterize three semiparametric estimators of the ATE that are consistent under three different semiparametric models. Each of the estimators operates on a subset of components of the likelihood for the observed data, and therefore may be severely biased if the corresponding model is misspecified.
	We carefully combine these strategies into a multiply robust estimator that produces valid inference provided one out of three models is correct, without necessarily knowing which one is indeed correct \citep{robins1994estimation,vansteelandt2008multiply,tchetgen2012semiparametric,rotnitzky2017multiply}. 
	The multiply robust estimator operates on the union of the three semiparametric models and thus offers robustness to model misspecification. 
	Furthermore, our proposed multiply robust estimator is locally efficient in the sense that when all working models are correctly specified, our estimator achieves the semiparametric efficiency bound for estimating the ATE under the union model.
	
	The paper is organized as follows. In Section~\ref{section1} we extend the nonparametric identification results of \cite{miao2018identifying}, and provide an alternative representation of their identifying functional for the ATE, 
	which opens up an opportunity for multiply robust estimation.
	For ease of exposition, we describe our results in the simple case of binary negative controls and unmeasured confounder in Section~\ref{binary_case}, where we propose a variety of semiparametric estimators including a multiply robust estimator. In Section~\ref{simu} we assess finite sample performance of our various estimators via extensive simulations.
	We illustrate our methods with an application to the postlicensure surveillance of vaccine safety among children in Section~\ref{sec:pentacel}.
	We close with a brief discussion in Section~\ref{discussion}. In addition, we extend our results to the more general setting allowing for polytomous unmeasured confounding and negative controls in Section~\ref{cate_case} of the supplementary material.

	\section{Identification and reparameterization}\label{section1}
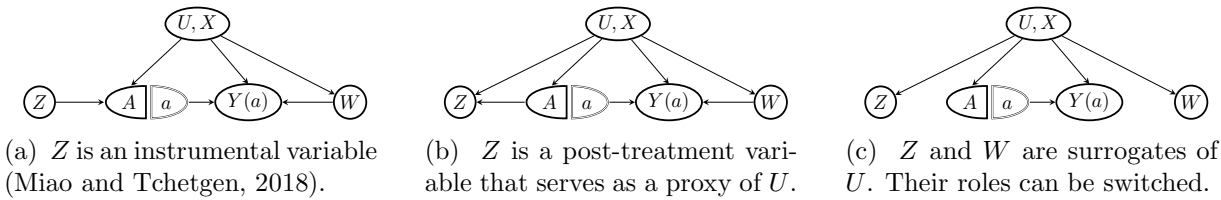
\begin{figure}[htp]
			\centering
			\begin{subfigure}[t]{0.3\textwidth}
				\hspace{0.1in}\resizebox{1.8in}{0.6in}{\begin{tikzpicture}
					\tikzset{line width=1pt,inner sep=5pt,
						swig vsplit={gap=3pt, inner line width right=0.4pt},
						ell/.style={draw, inner sep=1.5pt,line width=1pt}}
					\node[name=A,shape=swig vsplit] at (1,-1.5) {
						\nodepart{left}{$A$}
						\nodepart{right}{$a$} };
					\node[shape=ellipse,ell] (U) at (2,0) {$U,X$};
					\node[shape=circle,ell] (W) at (5,-1.5) {$W$};
					\node[shape=ellipse,ell] (Y) at (3,-1.5) {$Y(a)$};
					\node[shape=circle,ell,inner sep=2.2pt] (Z) at (-1.1,-1.5) {$Z$};
					\draw[-stealth,line width=0.5pt](A) to (Y);
					\foreach \from/\to in {U/W,Z/A,W/Y}
					\draw[-stealth,line width=0.5pt] (\from) -- (\to);
					\draw[-stealth,line width=0.5pt](U) to (0.7,-1.18);
					\draw[-stealth,line width=0.5pt](U) to (3,-1.15);
					\end{tikzpicture}}
				\caption{\label{fig:swig_exp1} $Z$ is an instrumental variable \citep{miao2018confoundingbridge}.}
			\end{subfigure}
			\hspace{0.15in}
			\begin{subfigure}[t]{0.3\textwidth}
				\hspace{0.1in}\resizebox{1.8in}{0.6in}{\begin{tikzpicture}
					\tikzset{line width=1pt,inner sep=5pt,
						swig vsplit={gap=3pt, inner line width right=0.4pt},
						ell/.style={draw, inner sep=1.5pt,line width=1pt}}
					\node[name=A,shape=swig vsplit] at (1,-1.5) {
						\nodepart{left}{$A$}
						\nodepart{right}{$a$} };
					\node[shape=ellipse,ell] (U) at (2,0) {$U,X$};
					\node[shape=circle,ell] (W) at (5,-1.5) {$W$};
					\node[shape=ellipse,ell] (Y) at (3,-1.5) {$Y(a)$};
					\node[shape=circle,ell,inner sep=2.2pt] (Z) at (-1.1,-1.5) {$Z$};
					\draw[-stealth,line width=0.5pt](A) to (Y);
					\foreach \from/\to in {U/W,U/Z,W/Y,A/Z}
					\draw[-stealth,line width=0.5pt] (\from) -- (\to);
					\draw[-stealth,line width=0.5pt](U) to (0.7,-1.18);
					\draw[-stealth,line width=0.5pt](U) to (3,-1.15);
					\end{tikzpicture}}
				\caption{\label{fig:swig_exp2} $Z$ is a post-treatment variable that serves as a proxy of $U$.}
			\end{subfigure}
			\hspace{0.15in}
			\begin{subfigure}[t]{0.3\textwidth}
				\hspace{0.1in}\resizebox{1.8in}{0.6in}{\begin{tikzpicture}
					\tikzset{line width=1pt,inner sep=5pt,
						swig vsplit={gap=3pt, inner line width right=0.4pt},
						ell/.style={draw, inner sep=1.5pt,line width=1pt}}
					\node[name=A,shape=swig vsplit] at (1,-1.5) {
						\nodepart{left}{$A$}
						\nodepart{right}{$a$} };
					\node[shape=ellipse,ell] (U) at (2,0) {$U,X$};
					\node[shape=circle,ell] (W) at (5,-1.5) {$W$};
					\node[shape=ellipse,ell] (Y) at (3,-1.5) {$Y(a)$};
					\node[shape=circle,ell,inner sep=2.2pt] (Z) at (-1.1,-1.5) {$Z$};
					\draw[-stealth,line width=0.5pt](A) to (Y);
					\foreach \from/\to in {U/W,U/Z}
					\draw[-stealth,line width=0.5pt] (\from) -- (\to);
					\draw[-stealth,line width=0.5pt](U) to (0.7,-1.18);
					\draw[-stealth,line width=0.5pt](U) to (3,-1.15);
					\end{tikzpicture}}
				\caption{\label{fig:swig_exp3} $Z$ and $W$ are surrogates of $U$. Their roles can be switched.}
			\end{subfigure}
			\caption{\label{fig:alternativeDAG} Examples of alternative single world intervention graphs. We suppressed the bi-directed arrow between $Z$ and $A$ ($Y$ and $W$) because the common causes of $Z$ and $A$ ($Y$ and $W$) do not confound the $Y$-$\trt$ relationship.}
		\end{figure}

	We consider estimating the effect of a treatment $\trt$ on an outcome $Y$ subject to confounding by both observed covariates $\cov$ and unobserved categorical variables $U$. Let $Y(\ttrt), \ttrt=0,1$ denote the counterfactual outcome that would be observed if the treatment were $\ttrt$. We are interested in the ATE defined as $E[Y(1)-Y(0)]$. Suppose that we also observe an auxiliary exposure variable $Z$ and an auxiliary outcome variable $W$, and let $Y(\ttrt,z)$ and $W(\ttrt,z)$ denote the corresponding counterfactual values that would be observed had the primary treatment and auxiliary exposure taken value $(\ttrt,z)$. 
	Then $Z$ and $W$ are negative control exposure and negative control outcome respectively if they satisfy the following assumptions.
	\begin{assumption}\label{assumption_NC}Negative control exposure: $Y(\ttrt,z)=Y(\ttrt)$, for all $z$ almost surely; Negative control outcome: $W(\ttrt,z)=W$ for all $\ttrt,z$ almost surely. 
	\end{assumption}
	\begin{figure}[htp]
		\centering
		\begin{tikzpicture}
		\tikzset{line width=1pt,inner sep=5pt,
			swig vsplit={gap=3pt, inner line width right=0.4pt},
			ell/.style={draw, inner sep=1.5pt,line width=1pt}}
		\node[name=A,shape=swig vsplit] at (1,-1.5) {
			\nodepart{left}{$A$}
			\nodepart{right}{$a$} };
		\node[shape=ellipse,ell] (U) at (2,0) {$U,X$};
		\node[shape=circle,ell] (W) at (5,-1.5) {$W$};
		\node[shape=ellipse,ell] (Y) at (3,-1.5) {$Y(a)$};
		\node[shape=circle,ell,inner sep=2.2pt] (Z) at (-1.1,-1.5) {$Z$};
		\draw[-stealth,line width=0.5pt](A) to (Y);
		\foreach \from/\to in {U/W,U/Z,W/Y,Z/A}
		\draw[-stealth,line width=0.5pt] (\from) -- (\to);
		\draw[-stealth,line width=0.5pt](U) to (0.7,-1.18);
		\draw[-stealth,line width=0.5pt](U) to (3,-1.15);
		\draw[stealth-stealth,line width=0.5pt] (-1,-1.8) to[bend right] (0.6,-1.8);
		\draw[stealth-stealth,line width=0.5pt] (3.2,-1.84) to[bend right] (5,-1.84);
		\end{tikzpicture}
		\caption{\label{fig:DAG} Single world intervention graph with unmeasured confounding $U$ and double negative control $Z$ and $W$ \citep{richardson2013single}. The bi-directed arrow between $Z$ and $A$ ($Y$ and $W$) indicates potential unmeasured common causes of $Z$ and $A$ ($Y$ and $W$).}
	\end{figure}
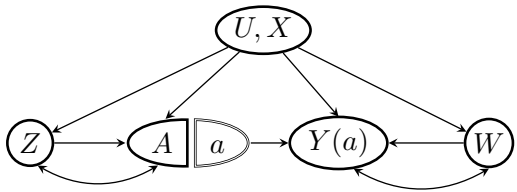 
	Figure~\ref{fig:DAG} presents a single world intervention graph (SWIG, \cite{richardson2013single}) illustrating an instance of the causal model under consideration.
	A key assumption satisfied by this graph is the conditional independence assumption stated below, which is required for identification of the causal effect.
	\begin{assumption}\label{assumption_randomization}
		Latent ignorability: $(Z,\trt)\ind(Y(\ttrt),W)\mid (U,\cov)$.
	\end{assumption}
	Assumption~\ref{assumption_randomization} states that $U$ and $\cov$ suffice to account for confounding of the relationship between $(Z,\trt)$ and $(Y(\ttrt),W)$, whereas $\cov$ alone may not. Moreover, $U$ includes all unmeasured common causes of $Z$, $\trt$, $Y$, and $W$ except for that of the $Z$-$\trt$ association and $Y$-$W$ association.
		Figure~\ref{fig:alternativeDAG} presents additional graphs all of which encode Assumption~\ref{assumption_randomization}. 
		For example, a special case is when $Z$ is an instrumental variable with the additional assumption that $Z\ind U$, as shown in Figure~\ref{fig:swig_exp1} \citep{miao2018confoundingbridge}. Alternatively $Z$ can be a post-treatment variable that serves as a proxy of $U$, as shown in Figure~\ref{fig:swig_exp2}. Furthermore, Figure~\ref{fig:swig_exp3} presents a scenario where $Z$ and $W$ can be surrogates of $U$ that satisfy the additional assumption that $(Z,W)\ind(\trt,Y)\mid (U,\cov)$, which is the nondifferential error assumption \citep{kuroki2014measurement}. In this scenario, the roles of $Z$ and $W$ can be switched.

	\begin{remark}
		In practice, specification of the unmeasured confounder is helpful for justifying the validity of negative controls. In certain scenarios, however, we do not need to know what $U$ is. For example, an underappreciated causal tenet is that the future does not affect the past. As such, with time series or longitudinal data, future exposure and past outcome may serve as $Z$ and $W$ respectively, assuming no feedback effect from past outcome to future exposure. In this case, we can control for unmeasured confounders shared over time without singling out a specific $U$ \citep{miao2017invited}.
	\end{remark}
	
	\begin{assumption}\label{assumption_cons_pos}
		Consistency: $Y(\ttrt)=Y$ almost surely when $\trt=\ttrt$; Positivity: $0<P(\trt=\ttrt,Z=z\mid \cov)<1$ for all $\ttrt,z$ almost surely.
	\end{assumption}
	The consistency assumption ensures that the exposure is defined with enough specificity such that among people with $\trt=\ttrt$, the observed outcome $Y$ is a realization of the potential outcome value $Y(\ttrt)$.
	The positivity assumption states that in all observed covariate strata there are always some individuals with treatment and negative control exposure values $(\trt=\ttrt,Z=z)$, for all $\ttrt,z$.

	\subsection{Identification with categorical negative control variables\label{sec:identification_unequal}}
	In this paper, we consider the scenario where $W$, $Z$, and $U$ are categorical. Suppose $W$, $Z$, and $U$ take on $|W|$, $|Z|$, and $|U|$ possible values denoted as $w_{i}$, $z_{j}$, and $u_{s}$, for $i=0,\dots,|W|-1$, $j=0,\dots,|Z|-1$, and $s=0,\dots,|U|-1$ respectively, where $| \cdot|$ denotes the cardinality of a categorical variable.
	Let $P(\mathbf{W}\mid  \mathbf{Z},\ttrt,\ccov)$ denote a $|W|\times |Z|$ matrix with $P(\mathbf{W}\mid  \mathbf{Z},\ttrt,\ccov)_{i,j}=P(W\!=\!w_{i-1}\mid Z\!=\!z_{j-1},\trt\!=\!\ttrt,\cov\!=\!\ccov)$,
	$P(\mathbf{W}\mid  \mathbf{U},\ccov)$ a $|W|\times|U|$ matrix with $P(\mathbf{W}\mid  \mathbf{U},\ccov)_{i,s}\!=\!P(W\!=\!w_{i-1}\mid U\!=\!u_{s-1},\cov\!=\!\ccov)$, 
	and $P(\mathbf{U}\mid  \mathbf{Z},\ttrt,\ccov)$ a $|U|\times |Z|$ matrix with $P(\mathbf{U}\mid  \mathbf{Z},\ttrt,\ccov)_{s,j}=P(U=u_{s-1}\mid Z\!=\!z_{j-1},\trt\!=\!\ttrt,\cov\!=\!\ccov)$. Similarly, let $E[Y\mid  \mathbf{Z},\ttrt,\ccov]$ denote a $1\times|Z|$ vector with $E[Y\mid  \mathbf{Z},\ttrt,\ccov]_j=E[Y\mid  Z\!=\!z_{j-1},\trt\!=\!\ttrt,\cov\!=\!\ccov]$, 
	$E[Y\mid  \mathbf{U},\ttrt,\ccov]$ a $1\times|U|$ vector with $E[Y\mid  \mathbf{U},\ttrt,\ccov]_s=E[Y\mid  U\!=\!u_{s-1},\trt\!=\!\ttrt,\cov\!=\!\ccov]$, 
	and $P(\mathbf{W}\mid \ccov)$ a $|W|\times 1$ vector with 
	$P(\mathbf{W}\mid \ccov)_{i}=P(W\!=\!w_{i-1}\mid \cov\!=\!\ccov)$.
	The following describes a sufficient condition under which the ATE is nonparametrically identified.
	\begin{assumption}\label{assumption_inverse} Both $Z$ and $W$ have at least as many categories as $U$, i.e., $|Z|\ge|U|$ and $|W|\ge|U|$. Both $P(\mathbf{W}\mid  \mathbf{U},\ccov)$ and $P(\mathbf{U}\mid  \mathbf{Z},\ttrt,\ccov)$ are full rank with rank $|U|$ at all values of $\ttrt$ and $\ccov$.
	\end{assumption}
	\begin{remark}\label{remark2} 
			Under Assumption~\ref{assumption_inverse}, $P(\mathbf{W}\mid  \mathbf{Z},\ttrt,\ccov)$ has rank $|U|$, which is proved in Section~\ref{appendix:weakercond} of the supplementary material. Thus one can infer $|U|$ from the rank of $P(\mathbf{W}\mid  \mathbf{Z},\ttrt,\ccov)$ \citep{choi2017selecting}.
	\end{remark}
	Assumption~\ref{assumption_inverse} imposes requirements on candidate negative controls for identification. Intuitively, both $Z$ and $W$ serve as proxies of $U$. Therefore, they should have at least as many possible values as $U$. They should also be strongly associated with $U$ such that variation in $U$ can be recovered from variation in $Z$ and $W$. This is reflected by the requirement that the columns of $P(\mathbf{W}\mid  \mathbf{U},\ccov)$ and the rows of $P(\mathbf{U}\mid  \mathbf{Z},\ttrt,\ccov)$ must be linearly independent vectors. 
	In practice, it is recommended to collect a negative control variable with a rich set of possible levels, or multiple negative control variables that can be combined into a composite negative control with as many categories as possible. 
	However, selection of valid negative control variable must be based on reliable subject matter knowledge because Assumptions~\ref{assumption_NC}-\ref{assumption_inverse} must be met.

	The following lemma demonstrates identification of $E[Y(\ttrt)]$, which is proved in Section~\ref{appendix:weakercond} of the supplementary material.
	\begin{lemma}\label{lemma:ate}
		Under Assumptions~\ref{assumption_NC} -- \ref{assumption_inverse},
		there exist a $1\times |W|$ vector $\myh $ such that \begin{equation}E[Y\mid \mathbf{Z},\ttrt,\ccov]=\myh P(\mathbf{W}\mid \mathbf{Z},\ttrt,\ccov),\label{eq:solve_h}\end{equation} and $E[Y(\ttrt)]$ is nonparametrically identified by $E[Y(\ttrt)]=\int_{\mathcal{\cov}}\myh P(\mathbf{W}\mid \ccov)f(\ccov)d\ccov$, where $f(\ccov)$ denotes the density function of $\cov$. Therefore, the ATE, denoted as $\Delta$, is uniquely identified by \begin{equation}\label{ate_general}
			\Delta = \int_{\mathcal{\cov}}[h(1,\ccov)-h(0,\ccov)] P(\mathbf{W}\mid \ccov)f(\ccov)d\ccov.
			\vspace{-0.02in}\end{equation}
	\end{lemma}
	As stated in Remark~\ref{remark2}, $P(\mathbf{W}\mid  \mathbf{Z},\ttrt,\ccov)$ has rank $|U|$ under Assumption~\ref{assumption_inverse}. When $|Z|\!=\!|W|\!=\!|U|$, $P(\mathbf{W}\mid  \mathbf{Z},\ttrt,\ccov)$ is full rank and the linear system (\ref{eq:solve_h}) has a unique solution \begin{equation}
		\myh =E[Y\mid  \mathbf{Z},\ttrt,\ccov]P(\mathbf{W}\mid  \mathbf{Z},\ttrt,\ccov)\inv.\label{eq:solution_h}\end{equation}
	Therefore, Lemma~\ref{lemma:ate} implies the identification result of \cite{miao2018identifying} \textcolor{black}{under a stronger assumption that} $|Z|=|W|=|U|$, which is stated in the following corollary.
	\begin{assumption4}{4}{}\label{assumption_inversenew}Completeness: 
		$P(\mathbf{W}\mid  \mathbf{Z},\ttrt,\ccov)$ is invertible with $|Z|=|W|=|U|=k+1$, $k\geq 0$.
	\end{assumption4}
	\begin{corollary}\label{coro:ate}
		Under Assumptions~\ref{assumption_NC} -- \ref{assumption_cons_pos} and~\ref{assumption_inversenew}, $E[Y(\ttrt)]$ is nonparametrically identified by
		\begin{equation*}
			E[Y(\ttrt)]=\int_{\mathcal{\cov}}E[Y\mid  \mathbf{Z},\ttrt,\ccov]P(\mathbf{W}\mid  \mathbf{Z},\ttrt,\ccov)\inv P(\mathbf{W}\mid \ccov)f(\ccov)d\ccov.
			\vspace{-0.02in}\end{equation*}
		Therefore, the ATE is given by
		\begin{equation}\begin{split}
				\Delta=&\int_{\mathcal{\cov}}E[Y\mid  \mathbf{Z},\trt=1,\cov=\ccov]P(\mathbf{W}\mid  \mathbf{Z},\trt=1,\cov=\ccov)\inv P(\mathbf{W}\mid \cov=\ccov)f(\ccov)d\ccov\\
				-&\int_{\mathcal{\cov}}E[Y\mid  \mathbf{Z},\trt=0,\cov=\ccov]P(\mathbf{W}\mid  \mathbf{Z},\trt=0,\cov=\ccov)\inv P(\mathbf{W}\mid \cov=\ccov)f(\ccov)d\ccov.
				\label{eq:functional}
			\end{split}\end{equation}
	\end{corollary}

	When $|Z|>|U|$ or $|W|>|U|$, $P(\mathbf{W}\mid  \mathbf{Z},\ttrt,\ccov)$ is rank deficient with linearly dependent rows or columns. In this case, there are infinite solutions to the linear system (\ref{eq:solve_h}). Nevertheless, $E[Y(\ttrt)]$ remains uniquely identified. 
	Note that there always exists an invertible $|U|\times |U|$ submatrix of $P(\mathbf{W}\mid  \mathbf{Z},\ttrt,\ccov)$ formed by deleting $|W|-|U|$ rows or $|Z|-|U|$ columns of $P(\mathbf{W}\mid  \mathbf{Z},\ttrt,\ccov)$ \citep{gomez2008generalized}. The $|W|-|U|$ rows or $|Z|-|U|$ columns correspond to free levels in $W$ or $Z$ that are redundant for identification but may improve efficiency. 
	
	We propose two strategies for estimation of $\Delta$ when $|Z|>|U|$ or $|W|>|U|$. 
	One is to obtain a maximum likelihood estimator of $P(\mathbf{W}\mid  \mathbf{Z},\ttrt,\ccov)$ and its Moore-Penrose inverse denoted as $P(\mathbf{W}\mid  \mathbf{Z},\ttrt,\ccov)^{+}$. A particular solution to (\ref{eq:solve_h}) is given by $\myh =E[Y\mid \mathbf{Z},\ttrt,\ccov]P(\mathbf{W}\mid  \mathbf{Z},\ttrt,\ccov)^{+}$. In fact, by Theorem 2 of \cite{james1978generalised}, the complete set of solutions to (\ref{eq:solve_h}) is given by $\myh = E[Y\mid \mathbf{Z},\ttrt,\ccov]P(\mathbf{W}\mid  \mathbf{Z},\ttrt,\ccov)^{+}+\tau(a,x)^{\transpose}[\mathbb{I}-P(\mathbf{W}\mid  \mathbf{Z},\ttrt,\ccov)P(\mathbf{W}\mid  \mathbf{Z},\ttrt,\ccov)^{+}]$, as $\tau(a,x)$, a vector function, varies over all possible values in $\{f: (a,x)\rightarrow R^{|W|}\}$.
	The second is to coarsen levels in $Z$ and $W$ until the coarsened variables
	satisfy Assumption~\ref{assumption_inversenew} \citep{kuroki2014measurement,miao2018identifying}. Suppose there are $m$ possible sets of coarsened negative control variables, then an estimator can be obtained by the generalized method of moments, i.e., ${\hat{\Delta}}=\arg\min_{\Delta}[\pn \hat{g}(\Delta)]\transpose{\hat{W}}[\pn \hat{g}(\Delta)]$, where $\hat{g}(\Delta)$ is an $m$-vector with each entry an estimating equation 
	based on an estimated influence function of $\Delta$ under a given parametric, semiparametric, or nonparametric model for a given set of coarsened negative control variables, and $\hat{W}=\pn[ \hat{g}(\Delta)\hat{g}(\Delta)\transpose]$. Such influence functions are derived in Section~\ref{binary_case}.
	
	\subsection{Reparameterization of $\Delta$ for multiply robust estimation\label{sec:estimation_equal}}
	In this section, we provide an alternative parameterization of $\Delta$ which opens up an opportunity for multiply robust estimation in the case where $|Z|=|W|=|U|=k+1$. 
	When $|Z|>|U|$ or $|W|>|U|$, in order to leverage the reparameterization, we propose to use the second strategy described in the previous section, with $g(\Delta)$ being the multiply robust estimating equation in Theorem~\ref{lemma2} of Section~\ref{EIF_bin}.
	\subsubsection{Motivation for multiply robust estimation\label{motivation_MR}}
	As discussed in Section~\ref{intro}, nonparametric estimation of $\Delta$ may not be feasible when $\cov$ is high dimensional or when $Z$ and $W$ have many levels, in which case one may need to resort to estimation under working models $E[Y\mid  \mathbf{Z},\trt,\cov;\theta_1]$, $P(\mathbf{W}\mid  \mathbf{Z},\trt,\cov;\theta_2)$, and $P(\mathbf{W}\mid \cov;\theta_3)$ where $\theta_1$, $\theta_2$, and $\theta_3$ are finite dimensional, resolving the curse of dimensionality. 
	Under such specification of a model for the conditional distribution $P(Y,W,Z,\trt \mid  \cov;\theta_1,\theta_2,\theta_3)$, one could in principle estimate $\Delta$ using the plug-in estimator, which entails estimating $\theta_1$, $\theta_2$, and $\theta_3$ by standard maximum likelihood estimation (MLE) and substituting estimated parameters in Eq.~(\ref{ate_general}) or (\ref{eq:functional}), with the cumulative distribution function of $\cov$ estimated by the empirical distribution.
	This is essentially the approach suggested by \cite{miao2018identifying}. 
	However, these working models are not in themselves of scientific interest and may be prone to model misspecification. The estimator may be severely biased if any of the three models is incorrect.
	
	To resolve this difficulty, we develop a robust inferential approach grounded in semiparametric theory \citep{bickel1993efficient,newey1990semiparametric,van1998asymptotic}, detailed in Section~\ref{binary_case}. We motivate our semiparametric approach by considering the task of estimating the functional $\Delta$ without any restriction on the observed data distribution. We characterize the efficient influence function (EIF) for $\Delta$ in the nonparametric model. We then use the EIF to construct an estimating equation to obtain an estimator of $\Delta$. Similar to the plug-in estimator, the EIF-based estimation entails estimating the distribution of the observed data under a parametric (or semiparametric) working model and then evaluating the EIF under such working model. However, unlike the plug-in estimator, we establish that our EIF based estimator of $\Delta$ remains consistent and asymptotically normal (CAN) even when the observed data likelihood is partially misspecified. In fact, we establish the multiply robust property of our proposed estimator: it remains CAN under the union of three large semiparametric models, each of which restricts a subset of components of the likelihood, allowing the remaining likelihood components to be unrestricted and hence robust to misspecification.

	\subsubsection{Reparameterization}
	An essential step towards constructing our multiply robust estimator involves a careful reparameterization of the functional $\Delta$ in terms of variation independent components of the likelihood, such that (mis)specification of one particular component does not impose any restriction on the other components.
	As such we define the following contrasts measuring the observed effects of $Z$ on $Y$ and $W$ at any value $(\ttrt,\ccov)$ as 
	\begin{equation*}\begin{split}
			&\myeta^{w_i}_{\zjdiff}(\ttrt,\ccov)=P(W=w_i\mid  \trt=\ttrt,Z=z_j,\cov=\ccov)-P(W=w_i\mid  \trt=\ttrt,Z=z_0,\cov=\ccov),i,j=1,\dots,k;\\
			&\myeta^{Y}_{\zjdiff}(\ttrt,\ccov)=E[Y\mid  \trt=\ttrt,Z=z_j,\cov=\ccov]-E[Y\mid  \trt=\ttrt,Z=z_0,\cov=\ccov],j=1,\dots,k,
		\end{split}\end{equation*}
	respectively, where $z_0$ is a user-specified reference level for $Z$. 
	Likewise, the observed effects of $\trt$ on $Y$ and $W$ at any values $(z,\ccov)$ are
	\begin{equation*}\begin{split}
			&\delta^{w_i}_{\adiff}(z,\ccov)=P(W=w_i\mid  \trt=1,Z=z,\cov=\ccov)-P(W=w_i\mid  \trt=0,Z=z,\cov=\ccov),i=1,\dots,k;\\
			&\delta^{Y}_{\adiff}(z,\ccov)=E[Y\mid  \trt=1,Z=z,\cov=\ccov]-E[Y\mid  \trt=0,Z=z,\cov=\ccov],
		\end{split}\end{equation*}
	respectively. 
	In addition, we let
	\begin{itemize}
		\item[] $\deltaWAvec(z,\ccov)=\{\delta^{w_1}_{\adiff}(z,\ccov),\delta^{w_2}_{\adiff}(z,\ccov), \dots, \delta^{w_k}_{\adiff}(z,\ccov)\}\transpose $ denote a $k\times 1$ vector;
		\item[] $\myetaYZvec(\ttrt,\ccov)=\{\myeta^{Y}_{\zonediff}(\ttrt,\ccov),\myeta^{Y}_{\ztwodiff}(\ttrt,\ccov),\dots,\myeta^{Y}_{\zkdiff}(\ttrt,\ccov)\}\transpose $ denote a $k\times 1$ vector;
		\item[] $\myetaWZvec(\ttrt,\ccov)$ denote a $k\times k$ matrix with $\myetaWZvec(\ttrt,\ccov)_{i,j}=\myeta^{w_i}_{\zjdiff}(\ttrt,\ccov)$, $i,j=1,\dots,k$.
	\end{itemize} 
	Note that $\delta^{w_0}_{\adiff}(z,\ccov)$ and $\myeta^{w_0}_{\zjdiff}(\ttrt,\ccov)$, $j=1,\dots,k$ are omitted to avoid over-parameterization, where $w_0$ is a user-specified reference level for $W$.
	The following lemma gives our alternative representation, which we prove in Section~\ref{appendix:simplify} of the supplementary material.
	\begin{lemma}\label{lemma:simplify}
		Under Assumptions~\ref{assumption_NC} -- \ref{assumption_cons_pos} and 4$'$, $\myetaWZvec(\ttrt,\ccov)$ is invertible and $\Delta$ in Eq. (\ref{eq:functional}) admits the alternative representation 
		\begin{equation}\begin{split}
				\Delta=&\Delta_{\gform}-\Delta_{\text{bias}},\label{eq:ate}\\
				\Delta_{\gform}=E[\dY_{\adiff}(Z,\cov)]&,\;\Delta_{\text{bias}}=E[\R(1\!-\!\trt,\cov)\deltaWAvec(Z,\cov)],
			\end{split}\end{equation}
				where
		$\R(\ttrt,\ccov)=\myetaYZvec(\ttrt,\ccov)\transpose \myetaWZvec(\ttrt,\ccov)\inv$ is a $1\times k$ vector. In addition, $\Delta_{\text{bias}}$=0 if there is no unmeasured confounding.
	\end{lemma}
	
	In Eq.~(\ref{eq:ate}), $\Delta_{\gform}$ is the standard g-formula which fails to adjust for unmeasured confounding, and $\Delta_{\text{bias}}$ is a bias correction term which accounts for unmeasured confounding.
	We note that $\Delta_{\text{bias}}$ is a scaled version of the observed association between $\trt$ and $W$. In fact, by Assumptions~\ref{assumption_NC} and \ref{assumption_randomization}, 
	$\deltaWAvec(Z,\cov)$ should be zero if there is no unmeasured confounding, and thus a nonzero $\deltaWAvec(Z,\cov)$ captures confounding bias. The scaling factor $\R(1-\trt,\cov)$ accounts for the fact that the effect of $U$ on $Y$ may not be on the same scale as the effect of $U$ on $W$, and therefore the bias captured by $\deltaWAvec(Z,\cov)$ needs to be carefully rescaled.
	To identify the ratio of the effects of $U$ on $Y$ and $U$ on $W$, we note that 	conditional on $\trt$ and $\cov$, any association between $Z$ and $Y$ ($Z$ and $W$) is governed by the effect of $U$ on $Y$ ($U$ on $W$). Therefore the ratio of the observed $Z$ effects, i.e., $\R(1-\trt,\cov)$, recovers the ratio of the unobserved $U$ effects. We further illustrate the intuition behind the reparameterization with an example in Section~\ref{illustrationofidentification} of the supplementary material.
	
	Decomposition of the causal effect estimand into the standard g-formula and an explicit bias correction term simplifies our inferential task, because semiparametric estimation of $\Delta_{\gform}$ has been extensively studied \citep{robins1994estimation,robins2000,scharfstein1999adjusting,van2003unified,bang2005doubly,tan2006distributional,tsiatis2007semiparametric}. Therefore we mainly study robust estimation of $\Delta_{\text{bias}}$, which together with $\Delta_{\gform}$ provides robust estimation of the ATE. 
	For ease of exposition, in the following sections we develop our semiparametric approach in the setting where $W$, $Z$, and $U$ are binary variables. We extend our results to general settings with polytomous $W$, $Z$, and $U$ in Section~\ref{cate_case} of the supplementary material.
	
	\section{Semiparametric estimation in the binary case}\label{binary_case}
	When $Z,W,U$ are binary, i.e., $k=1$, $\deltaWAvec(z,\ccov)$, $\myetaYZvec(\ttrt,\ccov)$, $\myetaWZvec(\ttrt,\ccov)$, and $\R(\ttrt,\ccov)$ simplify to the following scalar functions
	\begin{equation}\begin{split}\label{def}
			\dW_{\adiff}(z,\ccov)=&E[W\mid  \trt=1,Z=z,\cov=\ccov]-E[W\mid  \trt=0,Z=z,\cov=\ccov],\\
			\myeta^{Y}_{\zdiff}(\ttrt,\ccov)=&E[Y\mid  \trt=\ttrt,Z=1,\cov=\ccov]-E[Y\mid  \trt=\ttrt,Z=0,\cov=\ccov],\\
			\myeta^{W}_{\zdiff}(\ttrt,\ccov)=&E[W\mid  \trt=\ttrt,Z=1,\cov=\ccov]-E[W\mid  \trt=\ttrt,Z=0,\cov=\ccov],\\
			R(\ttrt,\ccov)=&\frac{\myeta^{Y}_{\zdiff}(\ttrt,\ccov)}{\myeta^{W}_{\zdiff}(\ttrt,\ccov)},
		\end{split}\end{equation}
	and representation of $\Delta$ in Eq.~(\ref{eq:ate}) is accordingly simplified.
	Note that careful specification of $R(\trt,\cov)$, $\myeta^{W}_{\zdiff}(\trt,\cov)$, and $\myeta^{Y}_{\zdiff}(\trt,\cov)$ is critical as they are in general not variation independent; that is, model specification for $R(\trt,\cov)$ and $\myeta^{W}_{\zdiff}(\trt,\cov)$ would imply a model for $\myeta^{Y}_{\zdiff}(\trt,\cov)$.

	\subsection{Parametric working models}\label{param_working_models}
	We now formally introduce variation independent components of the observed data likelihood for estimation of $\Delta$ to facilitate robust estimation. First, we note that the mean of $W$ given $\trt$, $Z$, and $\cov$ can be written as
	\begin{equation}\label{OR_W}
		E[W\mid \trt,Z,\cov]=E[W\mid \trt=0,Z=0,\cov]+\myeta^W_{\zdiff}(\trt=0,\cov)Z+\dW_{\adiff}(Z=0,\cov)\trt+\eta^W_{\azdiff}(\cov)\trt Z,
		\end{equation}
	where $\eta^W_{\azdiff}(\cdot)$ is the additive interaction of $A$ and $Z$ given $X$ with 
	\begin{equation}\label{interaction_restriction}
		\eta^W_{\azdiff}(\cov)\trt Z=[\myeta^W_{\zdiff}(\trt,\cov)-\myeta^W_{\zdiff}(\trt=0,\cov)]Z=[\dW_{\adiff}(Z,\cov)-\dW_{\adiff}(Z=0,\cov)]\trt.
		\end{equation}
	Furthermore, it is straightforward to verify that
	\begin{equation}
		E[Y\mid Z,\trt,\cov]=E[Y\mid Z=0,\trt,\cov]+R(\trt,\cov)\myeta^W_{\zdiff}(\trt,\cov)Z,\label{OR_Y}
		\end{equation}
	which implies that 
	\begin{equation}\begin{split}\label{deltayaparam}
			\delta^{Y}_{\adiff}(Z,\cov)=&\left[E[Y\mid Z=0,\trt=1,\cov]+R(\trt=1,\cov)\myeta^W_{\zdiff}(\trt=1,\cov)Z\right]-\\
			&\left[E[Y\mid Z=0,\trt=0,\cov]+R(\trt=0,\cov)\myeta^W_{\zdiff}(\trt=0,\cov)Z\right].
		\end{split}\end{equation}
	
	As we show below, multiply robust estimation requires positing working models for the following quantities: $E[Y\mid Z=0,\trt,\cov]$, $E[W\mid \trt=0,Z=0,\cov]$, $\myeta^W_{\zdiff}(\trt=0,\cov)$, $\dW_{\adiff}(Z=0,\cov)$, $\eta^W_{\azdiff}(\cov)$, $R(\trt,\cov)$, and $f(\trt,Z\mid \cov)$, where $f(\trt,Z\mid \cov)$ is the joint density of $\trt$ and $Z$ conditional on $\cov$.
	As $\cov$ may be high-dimensional and $Z$ and $W$ may have many levels, parametric working models are used to avoid the curse of dimensionality in practice. Clearly, these working models are not in themselves of scientific interest and estimators relying on a subset of these models may be biased when the corresponding models are misspecified. In order to motivate and clarify our doubly robust estimator, in Section~\ref{three_est} we will introduce three classes of semiparametric estimators of $\Delta$, which are CAN under the following working models with finite-dimensional indexing parameters:
	\begin{itemize}
		
		\item[$\Mgest$:] Working models $f(\trt,Z\mid \cov;\alpha^{\trt,Z})$ and $R(\trt,\cov;\beta^R)$ are correctly specified.
		
		\item[$\Mipw$] Working models $f(\trt,Z\mid \cov;\alpha^{\trt,Z})$, and $\myeta^W_{\zdiff}(\trt,\cov;\beta^{\WZ})$ and $\dW_{\adiff}(Z,\cov;\beta^{\WA})$ satisfying restriction (\ref{interaction_restriction}) are correctly specified. The interaction model $\eta^W_{\azdiff}(\cov;\beta^{\WAZ})$ is indexed by $\beta^{\WAZ}$, which is a sub-vector shared by $\beta^{\WZ}$ and $\beta^{\WA}$.

		\item[$\Mor$:] Working models $R(\trt,\cov;\beta^R)$, and $E[Y\mid Z=0,\trt,\cov;\beta^Y]$ and $E[W\mid \trt,Z,\cov;\beta^W]$ with $\beta^W=(\beta^{W0},\beta^{\WZ},\beta^{\WA})$ are correctly specified, where $E[W\mid \trt,Z,\cov;\beta^W]$ is parameterized by Eq.~(\ref{OR_W}) and $\beta^{W0}$ denotes the sub-vector of $\beta^W$ that indexes the baseline $E[W\mid \trt=0,Z=0,\cov]$.

	\end{itemize}
	Compared to the full list of variation independent components, we can see that in $\Mgest$, $E[Y\mid Z=0,\trt,\cov]$, $E[W\mid \trt=0,Z=0,\cov]$, $\myeta^W_{\zdiff}(\trt=0,\cov)$, $\dW_{\adiff}(Z=0,\cov)$, and $\eta^W_{\azdiff}(\cov)$  are unrestricted; in $\Mipw$, $R(\trt,\cov)$, $E[Y\mid Z=0,\trt,\cov]$, and $E[W\mid \trt=0,Z=0,\cov]$ are unrestricted; while in model $\Mor$, $f(\trt,Z\mid \cov)$ is unrestricted.
	
	\subsection{Three classes of semiparametric estimators of $\Delta$}\label{three_est}
	We describe three semiparametric estimators which are consistent under $\Mgest$, $\Mipw$, and $\Mor$, respectively.
	Let ${\gamma}_{i}, i=1,\dots,3$ denote the collection of indexing parameters in the corresponding semiparametric working model $\mathcal{M}_i$, which can be estimated under $\mathcal{M}_i$ as detailed in Appendix~\ref{appendixA} of the main manuscript. Let $\hat{\gamma}_{i}$ denote the estimated parameters, we have
	\begin{equation*}\begin{split}
			\hat{\Delta}_{1}&=\mathbb{P}_n\left\{
			\frac{(2\trt-1)Y}{f(\trt\mid Z,\cov;\hat{\gamma}_{1})}\right\} - \mathbb{P}_n\left\{
			E[R(1\!-\!\trt,\cov)\mid Z,\cov;\hat{\gamma}_{1}]\frac{(2\trt-1)W}{f(\trt\mid Z,\cov;\hat{\gamma}_{1})}
			\right\}\\
			\hat{\Delta}_{2}&=\mathbb{P}_n\left\{
			\frac{(2\trt-1)Y}{f(\trt\mid Z,\cov;\hat{\gamma}_{2})}\right\} - \mathbb{P}_n\left\{
			\frac{(2Z-1)Y}{f(Z\mid \trt,\cov;\hat{\gamma}_{2})}
			\frac{E[\delta^{W}_{\adiff}(Z,\cov)\mid 1\!-\!\trt,\cov;\hat{\gamma}_{2}]}{\myeta^{W}_{\zdiff}(\trt,\cov;\hat{\gamma}_{2})}
			\frac{f(1-\trt\mid \cov;\hat{\gamma}_{2})}{f(\trt\mid \cov;\hat{\gamma}_{2})}
			\right\}\\
			\hat{\Delta}_{3}&=\mathbb{P}_n\left\{
			E[Y\mid \trt\!=\!1,Z,\cov;\hat{\gamma}_{3}]\!-\!E[Y\mid \trt\!=\!0,Z,\cov;\hat{\gamma}_{3}]\right\} - \mathbb{P}_n\left\{
			R(1-\trt,\cov;\hat{\gamma}_{3})\dW_{\adiff}(Z,\cov;\hat{\gamma}_{3})
			\right\},
		\end{split}\end{equation*}
	where $\mathbb{P}_n$ is the empirical average operator, i.e., $\mathbb{P}_n(V) = \frac{1}{n}\sum_{i=1}^n V_i$.
	
	Each of the three estimators above may be severely biased if their corresponding model $\Mgest$, $\Mipw$, or $\Mor$ is misspecified. For example, $\hat{\Delta}_{1}$ and $\hat{\Delta}_{2}$ will generally fail to be consistent if $f(\trt\mid Z, \cov)$ is misspecified, even if the rest of the likelihood is correctly specified. Therefore, it is critical to develop a multiply robust estimator that remains CAN provided that one, but not necessarily more than one of models $\Mgest$, $\Mipw$, $\Mor$ is correctly specified, without necessarily knowing which is indeed correct. 
	
	\subsection{Efficient influence function in the nonparametric model}\label{EIF_bin}
	As discussed above, we aim to construct an estimator that is CAN under the union model $\mathcal{M}_{\text{union}}=\Mgest \cup \Mipw \cup \Mor$. 
	To this end, we first characterize the EIF for $\Delta$ in the nonparametric model $\mathcal{M}_{\text{nonpar}}$ which does not impose any restriction on the observed data distribution. We then use the EIF as an estimating equation and evaluate it under a working model to obtain an estimator of $\Delta$. We establish multiple robustness and asymptotic normality of this estimator. We also provide a consistent estimator of the asymptotic variance for the proposed estimators.
	
	It is well know that the efficient influence function of $\Delta_{\gform}$ in $\mathcal{M}_{\text{nonpar}}$ \citep{robins1994estimation} is 
	\begin{equation}\label{EIF_gformula}
		EIF_{\Delta_{\gform}}\!=\!\frac{2\trt-1}{f(\trt\mid Z,\cov)}\big(Y\!-\!E[Y\mid  \trt,Z,\cov]\big)\!+\!\big(E[Y\mid  \trt\!=\!1,Z,\cov]\!-\!E[Y\mid  \trt\!=\!0,Z,\cov]\big)\!-\!\Delta_{\gform}.
		\end{equation}
	In the theorem below, we derive the efficient influence function of $\Delta_{\text{bias}}$ in $\mathcal{M}_{\text{nonpar}}$, which is combined with $EIF_{\Delta_{\gform}}$ to obtain the efficient influence function of $\Delta$. Theorem~\ref{lemma2} is proved in Section \ref{appendix:if} of the supplementary material. 
	\begin{theorem}\label{lemma2}
		Under Assumptions~\ref{assumption_NC} -- \ref{assumption_cons_pos} and 4$'$, the efficient influence function of the bias correction term $\Delta_{\text{bias}}$ in the nonparametric model $\mathcal{M}_{\text{nonpar}}$ is
		\begin{equation*}\begin{split}
				EIF_{\Delta_{\text{bias}}}=&E[R(1\!-\!\trt,\cov)\mid Z,\cov] \frac{2\trt-1}{f(\trt\mid Z,\cov)} \Big(W-E[W\mid \trt,Z,\cov]\Big) \\
				+&\frac{2Z-1}{f(Z\mid \trt,\cov)}\Big(Y\!-\!E[Y\mid Z,\trt,\cov]\Big) \frac{
					E[\delta^{W}_{\adiff}(Z,\cov)\mid 1\!-\!\trt,\cov]}
				{\myeta^{W}_{\zdiff}(\trt,\cov)}
				\frac{f(1-\trt\mid \cov)}{f(\trt\mid \cov)}
				\\
				+&R(1-\trt,\cov)\dW_{\adiff}(Z,\cov) -\Delta_{\text{bias}}.
			\end{split}\end{equation*}
		The efficient influence function of $\Delta$ is given by
		\begin{equation*}
			EIF_{\Delta}(O)=EIF_{\Delta_{\gform}}-EIF_{\Delta_{\text{bias}}},
			\end{equation*} 
		where $O=(Y,A,Z,W,Z)$ denotes the observed data.
		The semiparametric efficiency bound for estimating the ATE in $\mathcal{M}_{\text{nonpar}}$ is $E[EIF_{\Delta}(O)^2]\inv$.
	\end{theorem}
	\begin{remark}\label{same_point_estimate}
		Theorem~\ref{lemma2} implies that if $\hat{\Delta}$ is a regular and asymptotically linear estimator of $\Delta$ in $\mathcal{M}_{\text{nonpar}}$, then $\sqrt{n}(\hat{\Delta}-\Delta) = \frac{1}{\sqrt{n}}\sum_{i=1}^n EIF_{\Delta}(O_i) + o_p(1)$ \citep{bickel1993efficient}.
	\end{remark}

	\subsection{Multiply robust estimation of $\Delta$}\label{multiple_robust_bin}
	In this section, we consider the scenario where estimation under $\mathcal{M}_{\text{nonpar}}$ is not feasible due to potentially large number of measured covariates, and proceed to estimation under $\mathcal{M}_{\text{union}}$. Specifically, we construct a multiply robust and locally efficient estimator of $\Delta$ by taking $EIF_{\Delta}(O)$ as an estimating equation and evaluating it under a working model for the observed data distribution to solve for $\Delta$. 
	Let 
	\begin{equation*}\theta=\{({\alpha}^{\trt,Z})\transpose ,({\beta}^Y)\transpose ,({\beta}^{W0})\transpose ,(\hat{\beta}^{\WA})\transpose ,(\hat{\beta}^{\WZ})\transpose ,(\hat{\beta}^{R})\transpose \}\transpose \end{equation*}
	denote the nuisance parameters of the working models in $\mathcal{M}_{\text{union}}$. We estimate $\theta$ as the solution of the following collection of estimating equations.
	
	First, we define the following score functions for maximum likelihood estimation of $f(\trt,Z\mid \cov;\alpha^{\trt,Z})$, $E[Y\mid \trt,Z=0,\cov;\beta^{Y}]$, and $E[W\mid \trt=0,Z=0,\cov;\beta^{W0}]$
	\begin{equation*}\begin{split}
			U_{\alpha^{\trt,Z}}=&\frac{\partial}{\partial\alpha^{\trt,Z}} \log f(\trt,Z\mid \cov;\alpha^{\trt,Z}); \\
			U_{\beta^{Y}}=&\frac{\partial}{\partial\beta^{Y}}\mathbbm{1}(Z=0)\log f(Y\mid \trt,Z=0,\cov;\beta^{Y});\\
			U_{\beta^{W0}}=&\frac{\partial}{\partial\beta^{W0}}\mathbbm{1}(\trt=0,Z=0) \log f(W\mid \trt=0,Z=0,\cov;\beta^{W0}),
		\end{split}\end{equation*}
	where $f(\trt,Z\mid \cov;\alpha^{\trt,Z})$ is the conditional likelihood of $(\trt,Z)$, $f(Y\mid \trt,Z=0,\cov;\beta^{Y})$ is the conditional likelihood of $Y$ restricted to the subsample with $Z=0$, and $f(W\mid \trt=0,Z=0,\cov;\beta^{W0})$ is the conditional likelihood of $W$ restricted to the subsample with $A=0,Z=0$. 
	
	Second, because $\dW_{\adiff}(Z,\cov;\beta^{\WA})$, $\myeta^W_{\zdiff}(\trt,\cov;\beta^{\WZ})$, and $R(\trt,\cov;\beta^R)$ do not by themselves give rise to a likelihood function, we estimate them by constructing the following doubly robust g-estimation equations constructed under the union model $\mathcal{M}_{\text{union}}$
	\begin{equation*}\begin{split}
			&U_{\beta^{\WA},\beta^{\WZ}}=\Big(g_0(\trt,Z,\cov)\!-\!E[g_0(\trt,Z,\cov)\mid \cov;{\alpha}^{\trt,Z}]\Big) \Big(W\!-\!E[W\mid \trt,Z,\cov; {\beta}^{W0},\beta^{\WZ},\beta^{\WA}]\Big)\\
			&U_{\beta^{R};\beta^Y,\beta^{W0},\beta^{\WA}}=\Big(g_1(\trt,Z,\cov)\!-\!E[g_1(\trt,Z,\cov)\mid \trt,\cov;{\alpha}^{\trt,Z}]\Big)\Big(Y\!-\!E[Y\mid Z,\trt,\cov;\beta^R,\!\beta^Y,\!\beta^{W0},\!\beta^{\WA}]\Big),
		\end{split}\end{equation*}
	where 
	$g_0(\trt,Z,\cov)$ and $g_1(\trt,Z,\cov)$ are user-specified vector functions; 
	$E[g_0(\trt,Z,\cov)\mid \cov;{\alpha}^{\trt,Z}]$ and $E[g_1(\trt,Z,\cov)\mid \cov;{\alpha}^{\trt,Z}]$ are evaluated under $f(\trt,Z\mid \cov;{\alpha}^{\trt,Z})$; 
	$E[W\mid \trt,Z,\cov; {\beta}^{W0},\beta^{\WZ},\beta^{\WA}]$ and $E[Y\mid Z,\trt,\cov;\beta^R,\!\beta^Y,\!\beta^{W0},\!\beta^{\WA}]$ are parameterized as in (\ref{OR_W})-(\ref{deltayaparam}).
	Let dim($v$) denote the length of a vector $v$. We require that $g_0(\trt,Z,\cov)$ is of dimension $\text{dim}(\beta^{\WA})+\text{dim}(\beta^{\WZ})-\text{dim}(\beta^{\WAZ})$, and $g_1(\trt,Z,\cov)$ is of dimension $\text{dim}(\beta^R)$ to generate adequate number of estimating equations.
	
	In summary, let \begin{equation*}
		U_{\theta}(O;\theta)=(U_{\alpha^{\trt,Z}}\transpose ,U_{\beta^{Y}}\transpose ,U_{\beta^{W0}}\transpose ,U_{\beta^{\WA},\beta^{\WZ}}\transpose ,U_{\beta^{R}}\transpose )\transpose \end{equation*}
	denote the collection of the above defined estimating equations. We estimate $\theta$ by solving $\mathbb{P}_n\Big\{U_\theta(\theta)\Big\}=0$, and we denoted the estimator as \begin{equation*}\hat{\theta}=\{(\hat{\alpha}^{\trt,Z}_{\text{mle}})\transpose ,(\hat{\beta}^Y_{\text{mle}})\transpose ,(\hat{\beta}^{W0}_{\text{mle}})\transpose ,(\hat{\beta}^{\WA}_{\textdr})\transpose ,(\hat{\beta}^{\WZ}_{\textdr})\transpose ,(\hat{\beta}^{R}_{\textdr})\transpose \}\transpose .\end{equation*}
	In particular, $\hat{\beta}_{\textdr}^{\WA}$ and $\hat{\beta}_{\textdr}^{\WZ}$ are CAN under the union model $\Mipw\cup\Mor$, and $\hat{\beta}^R_{\textdr}$ is CAN under the union model $\Mgest\cup\Mor$ \citep{robinsrotnitzky2001comment,wang2018bounded}, which is proved in Section~\ref{appendix:proof_robustness_binary} of the supplementary material.
	We obtain the estimated working models by plugging in $\hat{\theta}$ to equations (\ref{OR_W})-(\ref{deltayaparam}), which is detailed in Appendix~\ref{appendixB}.
	
	The proposed multiply robust estimator solves $\mathbb{P}_n\Big\{EIF_{\Delta}(O;\Delta,\hat{\theta})\Big\}=0$, where $EIF_{\Delta}(O;\Delta,\hat{\theta})$ is equal to $EIF_{\Delta}(O)$ evaluated at $(\Delta,\hat{\theta})$. That is, the multiply robust estimator is
	\begin{equation*}
		\hat{\Delta}_{\text{mr}} = \hat{\Delta}_{\text{confounded,mr}}-\hat{\Delta}_{\text{bias,mr}}
		\end{equation*}where
	\begin{equation*}\begin{split}
			\hat{\Delta}_{\text{confounded,mr}}=\mathbb{P}_n\Big\{&\frac{2\trt-1}{f(\trt\mid  Z,\cov;\hat{\theta})}\big(Y\!-\!E[Y\mid  \trt,Z,\cov;\hat{\theta}]\big)+\big(E[Y\mid  \trt\!=\!1,Z,\cov;\hat{\theta}]\!-\!E[Y\mid  \trt\!=\!0,Z,\cov;\hat{\theta}]\big)\Big\}\\
			\hat{\Delta}_{\text{bias,mr}}=\mathbb{P}_n\Big\{&E[R(1\!-\!\trt,\cov)\mid Z,\cov;\hat{\theta}] \frac{2\trt-1}{f(\trt\mid Z,\cov;\hat{\theta})} \Big(W-
			E[W\mid \trt,Z,\cov;\hat{\theta}]\Big) \\
			+&\frac{2Z-1}{f(Z\mid \trt,\cov;\hat{\theta})} \Big(Y-E[Y\mid \trt,Z,\cov;\hat{\theta}]\Big)
			\frac{E[\delta^{W}_{\adiff}(Z,\cov)\mid 1-\trt,\cov;\hat{\theta}]}{\myeta^{W}_{\zdiff}(\trt,\cov;\hat{\theta})} 
			\frac{f(1-\trt\mid \cov;\hat{\theta})}{f(\trt\mid \cov;\hat{\theta})}
			\\
			+&R(1-\trt,\cov;\hat{\theta})\dW_{\adiff}(Z,\cov;\hat{\theta})\Big\}.
		\end{split}\end{equation*}
	Note from Section~\ref{three_est} that each of the three semiparametric estimators $\hat{\Delta}_{1}$, $\hat{\Delta}_{2}$, and $\hat{\Delta}_{3}$ can be obtained by setting the unrestricted components in $\Mgest$, $\Mipw$, and $\Mor$ respectively in the above multiply robust estimator to zero. 
	Specifically, $\hat{\Delta}_{1}$ is obtained by setting $E[Y\mid Z=0,\trt,\cov]$, $E[W\mid \trt=0,Z=0,\cov]$, $\myeta^W_{\zdiff}(\trt=0,\cov)$, $\dW_{\adiff}(Z=0,\cov)$, and $\eta^W_{\azdiff}(\cov)$ to zero, $\hat{\Delta}_{2}$ sets $E[Y\mid Z=0,\trt,\cov]$, $E[W\mid  \trt=0,Z=0,\cov]$, and $R( \trt,\cov)$ to zero, and $\hat{\Delta}_{3}$ sets $1/f(\trt\mid Z,\cov)$ and $1/f(Z\mid \trt,\cov)$ to zero.
	Therefore, the multiply robust estimator combines three estimation strategies to produce robust inference provided one out of three models is correct, without necessarily knowing which one is indeed correct. For example, our multiply robust estimator of $\Delta_{\text{bias}}=E[R(1-\trt,\cov)\dW_{\adiff}(Z,\cov)]$ does not require correct specification of both $R(1-\trt,\cov)$ and $\dW_{\adiff}(Z,\cov)$. In fact, we improve robustness by incorporating the propensity of both exposures such that when $f(\trt,Z\mid \cov)$ is correctly specified, $\hat{\Delta}_{\text{bias}}$ is consistent if either $R(1-\trt,\cov)$ or $\dW_{\adiff}(Z,\cov)$ is correctly specified. The theorem below summarizes the multiply robust and locally efficient property of $\hat{\Delta}_{\text{mr}}$.
	
	\begin{theorem}\label{theorem_MR}
		Under Assumptions~\ref{assumption_NC} -- \ref{assumption_cons_pos} and 4$'$ and standard regularity conditions stated in Section~\ref{appendix:proof_robustness_binary} of the supplementary material, $\sqrt{n}(\hat{\Delta}_{\text{mr}}-\Delta)$ is regular and asymptotic linear under $\mathcal{M}_{\text{union}}$ with influence function
		\begin{equation*}
			IF_{\text{union}}(O;\Delta,\theta^*)=EIF_{\Delta}(O;\Delta,\theta^*)-\frac{\partial EIF_{\Delta}(O;\Delta,\theta)}{\partial \theta\transpose }\longmid_{\theta^*}
			E\Big\{\frac{\partial U_{\theta}(O;\theta)}{\partial \theta\transpose }\longmid_{\theta^*}\Big\}^{-1} U_{\theta}(O;\theta^*),
			\end{equation*} 
		and thus $\sqrt{n}(\hat{\Delta}_{\text{mr}}-\Delta)\rightarrow_d N(0,\sigma^2_{\Delta})$, where $\sigma^2_{\Delta}(\Delta,\theta^*)=E[IF_{\text{union}}(O;\Delta,\theta^*)^2]$ and $\theta^*$ denotes the probability limit of $\hat{\theta}$.
		Furthermore, $\hat{\Delta}_{\text{mr}}$ is locally semiparametric efficient in the sense that it
		achieves the semiparametric efficiency bound for $\Delta$ in $\mathcal{M}_{\text{union}}$ at the intersection submodel $\mathcal{M}_{\text{intersect}}=\Mgest\cap\Mipw\cap\Mor$ where $\Mgest$, $\Mipw$, and $\Mor$ are all correctly specified.
	\end{theorem}
	We prove Theorem~\ref{theorem_MR} in Section~\ref{appendix:proof_robustness_binary} of the supplementary material. The rationale behind multiple robustness is based on the following key observation.
		A multiply robust estimator is bound to exist if one can describe an unbiased estimating equation in each of the submodels that form the union model. It then suffices to show that the multiply robust estimating equation (i.e. the efficient influence function) reduces to each estimating equation under the corresponding submodel of the union model, by setting components which are left unrestricted by the submodel to a singleton value.
	For inference on $\Delta$, a consistent standard error estimator follows from standard M-estimation theory, which is detailed in Section~\ref{appendix:var_delta} of the supplementary material. We implemented the standard error estimator in both simulation and application studies. Alternatively, nonparametric bootstrap may be used in practice, which is justified by the asymptotic linearity of the estimator \citep{cheng2010bootstrap}.

	\section{Simulation study}\label{simu}
	We investigate the finite sample performance of the various estimators of ATE described in Section~\ref{binary_case}. We simulate 4000 samples of size $n=2000$ under the following data generating mechanism
	\begin{itemize}
		\item $X\!=\!(X_1,\dots,X_{8},X_{7}X_{8})$ where $X_j \stackrel{\mathrm{iid}}{\sim} \text{Uniform}[0,1], j\!=\!1,\dots,8$;
		\item $A$ is Bernoulli with $P(A\!=\!1\mid X)\!=\!\text{expit}(-0.01+\alpha\transpose X)$;
		\item $Z$ is Bernoulli with $P(Z\!=\!1\mid \trt,\cov)\!=\!\text{expit}(-0.01-0.2A+\alpha\transpose X)$;
		\item $U$ is Bernoulli with $E[U\mid Z,\trt,\cov]\!=\!0.4Z\!+\!0.4AZ$;
		\item $W$ is Bernoulli with $E[W\mid U\!=\!0,X]\!=\text{expit}(\!-\!1+\beta\transpose \!X)$, $E[W\mid U\!=\!1,X]-E[W\mid U\!=\!0,X]\!=\!0.5$;
		\item $Y$ is Bernoulli with $E[Y\mid \trt\!=\!0,U\!=\!0,\cov]\!=\text{expit}(-1+\beta\transpose X)$, $E[Y\mid \trt,U\!=\!1,\cov]-E[Y\mid \trt,U\!=\!0,\cov]\!=\!0.25A$, and $E[Y\mid \trt\!=\!1,U,\cov]-E[Y\mid \trt\!=\!0,U,\cov]\!=\!0.25U$,
	\end{itemize}
	where $\alpha=-10^{-2}\!\times\!(1, 1, 1, 1, 1, 1, 1, 1, -20)$ and $\beta=-10^{-1}\!\times\!(1, 1, 1, 1, 1, 1, 1, 1, 1)$. These parameters are chosen to ensure that Pr$(U=1\mid Z,A,X)$, Pr$(W=1\mid U,X)$, and Pr$(Y=1\mid U,X)$ are between 0 and 1. 
	The above models imply
	\begin{itemize}
		\item 
		$\myeta^W_{\zdiff}(A,X)\!=\!0.2+0.2A$, $\dW_{\adiff}(Z,X)\!=\!0.2Z$, $E[W\mid Z\!=\!0,A\!=\!0,X]\!=\text{expit}(-1+\beta\transpose X)$;
		\item $\myeta^Y_{\zdiff}(A,X)\!=\! 0.2A$, $\dY_{\adiff}(Z,X)\!=\!0.2Z$, $E[Y\mid Z\!=\!0,A\!=\!0,X]\!=\text{expit}(-1+\beta\transpose X)$;
		\item $R(A,X)\!=\!0.5A$.
	\end{itemize}

	We evaluate the performance of the following five estimators of the ATE: three semiparametric estimators $\hat{\Delta}_{1}$, $\hat{\Delta}_{2}$, and $\hat{\Delta}_{3}$ which operate under $\mathcal{M}_{1}$,  $\mathcal{M}_{2}$, $\mathcal{M}_{3}$, respectively, the plug-in estimator discussed in Section~\ref{motivation_MR} which we refer to as the MLE estimator hereafter, and the multiply robust (MR) estimator $\hat{\Delta}_{\text{mr}}$. The true ATE is 0.07 on the risk difference scale.
	We consider the following scenarios to investigate the impact of modeling error.
	\begin{itemize}
		\item All models are correctly specified;
		\item $\Mipw$ and $\Mor$ are wrong: $E[W\mid \trt,Z,\cov]$ is misspecified by assuming that both $\myeta^W_{\zdiff}(A,X)$ and $\dW_{\adiff}(Z,X)$ are constant;
		\item $\Mgest$ and $\Mor$ are wrong: $R(\trt,\cov)$ is misspecified by assuming that $R(A,X)$ is a constant; 
		\item $\Mgest$ and $\Mipw$ are wrong: $f(Z\mid \trt,\cov)$ is misspecified by omitting the interaction term $\cov_{7}\cov_{8}$;
		\item All models are wrong: $f(Z\mid \trt,\cov)$ and $E[Y\mid \trt,Z,\cov]$ are misspecified by omitting the interaction term $\cov_{7}\cov_{8}$.
	\end{itemize}

	\begin{table}[h]
		\caption{Operating characteristics of estimators under different model misspecification scenarios. }
		\label{tbl_simu}
		\begin{adjustbox}{center}
			\begin{tabular}{c| cccccc}
				\hline 
				\multirow{2}{*}{\textbf{Scenario}} & \multirow{2}{*}{\textbf{Method}} & \textbf{Bias} & \textbf{Var} & \textbf{Proportion} & \textbf{MSE} & \textbf{95$\%$ CI}\tabularnewline
				& & \textbf{($\times10^{3}$)} & \textbf{($\times10^{3}$)} & \textbf{Bias ($\%$ ATE)} & \textbf{($\times10^{3}$)} & \textbf{Coverage}\tabularnewline
				\hline 
				& $\Delta_{1}$ & -0.35 & 0.45 & -0.50 & 0.45 & 0.95\tabularnewline
				\textbf{All} & $\Delta_{2}$ & -0.18 & 0.49 & -0.26 & 0.49 & 0.95\tabularnewline
				\textbf{models are} & $\Delta_{3}$ & 0.04 & 0.14 & 0.06 & 0.14 & 0.95\tabularnewline
				\textbf{correct} & MLE & -0.04 & 0.15 & -0.05 & 0.15 & 0.94\tabularnewline
				& $\Delta_{\text{mr}}$ & -0.29 & 0.50 & -0.41 & 0.50 & 0.95\tabularnewline
				\hline 
				\textbf{$\mathcal{M}_{1}$ correct} & $\Delta_{2}$ & -7.30 & 0.51 & -10.44 & 0.56 & 0.94\tabularnewline
				\textbf{$\mathcal{M}_{2},\mathcal{M}_{3}$} & $\Delta_{3}$ & -6.60 & 0.16 & -9.45 & 0.20 & 0.89\tabularnewline
				\textbf{misspecified} & MLE & -18.19 & 20.73 & -26.03 & 21.05 & 0.85\tabularnewline
				& $\Delta_{\text{mr}}$ & -0.13 & 0.48 & -0.19 & 0.48 & 0.94\tabularnewline
				\hline 
				\textbf{$\mathcal{M}_{2}$ correct} & $\Delta_{1}$ & -0.41 & 0.50 & -0.59 & 0.50 & 0.94\tabularnewline
				\textbf{$\mathcal{M}_{1},\mathcal{M}_{3}$} & $\Delta_{3}$ & -6.44 & 0.61 & -9.22 & 0.65 & 0.95\tabularnewline
				\textbf{misspecified} & $\Delta_{\text{mr}}$ & 0.87 & 0.77 & 1.25 & 0.78 & 0.94\tabularnewline
				\hline 
				\textbf{$\mathcal{M}_{3}$ correct} & $\Delta_{1}$ & -0.11 & 0.45 & -0.16 & 0.45 & 0.95\tabularnewline
				\textbf{$\mathcal{M}_{1},\mathcal{M}_{2}$} & $\Delta_{2}$ & -0.95 & 0.50 & -1.36 & 0.50 & 0.95\tabularnewline
				\textbf{misspecified} & $\Delta_{\text{mr}}$ & -0.24 & 0.51 & -0.34 & 0.51 & 0.95\tabularnewline
				\hline 
				& $\Delta_{1}$ & -0.26 & 0.45 & -0.37 & 0.45 & 0.95\tabularnewline
				\textbf{All} & $\Delta_{2}$ & -1.18 & 0.50 & -1.68 & 0.50 & 0.95\tabularnewline
				\textbf{models are} & $\Delta_{3}$ & -2.62 & 0.14 & -3.75 & 0.15 & 0.94\tabularnewline
				\textbf{misspecified} & MLE & -1.66 & 0.12 & -2.37 & 0.12 & 0.93\tabularnewline
				& $\Delta_{\text{mr}}$ & 0.70 & 0.51 & 1.01 & 0.51 & 0.95\tabularnewline
				\hline 
			\end{tabular}
		\end{adjustbox}
		\centering{\small{Note: we trimmed 1\% tail of the simulations due to extreme value of the estimates.}}
	\end{table}
	Table~\ref{tbl_simu} summarizes the operating characteristics of $\hat{\Delta}_{1}$, $\hat{\Delta}_{2}$, $\hat{\Delta}_{3}$, the MLE estimator, and the MR estimator $\hat{\Delta}_{\text{mr}}$ under the above model misspecification scenarios. 
	We evaluated these estimators in terms of mean bias (scaled by $10^3$), variance (scaled by $10^3$), bias calculated as the proportion of the true ATE, mean squared error (MSE, scaled by $10^3$), and coverage of 95\% confidence intervals based on direct standard error estimates. The performance of the MLE estimator is not shown when $R(\trt,\cov)$ or $f(Z\mid \trt,\cov)$ are misspecified because it does not require specification of $R(\trt,\cov)$ or $f(Z\mid \trt,\cov)$ and thus remains unchanged under such misspecifications.
	Our proposed multiply robust estimator remained stable with relatively small bias across all scenarios, although as expected it had slightly larger variability.
	The multiply robust estimator performs better when all models are misspecified than if $\mathcal{M}_{2}$ is correctly specified, which may not be the general case in practice as the theory does not necessarily justify it.
	In contrast, the MLE estimator and the other three semiparametric estimators that rely on $\Mgest$, $\Mipw$, and $\Mor$ can be substantially biased when their corresponding model was misspecified. 
	The 95\% CI coverages were close to the nominal level with correctly specified model which indicated that our proposed standard error estimation provided valid inference. These results confirmed our theoretical results in finite sample and demonstrated the advantages of the proposed multiply robust estimator. 
	
	\section{Observational postlicensure vaccine safety surveillance\label{sec:pentacel}}
	We apply our method to an observational vaccine safety study comparing risk of medically attended fever, a common adverse event following vaccination, among children who received a combination DTaP-IPV-Hib (diphtheria and tetanus toxoids and acellular pertussis adsorbed, inactivated poliovirus, and Haemophilus influenzae type b) vaccine with children who received other DTaP-containing comparator vaccines \citep{nelson2013adapting}.
	The study population consisted of children aged 6 weeks to 2 years enrolled at Kaiser Permanente Washington from September 2008 to January 2011. Healthcare databases routinely captured information on demographics, immunizations, and diagnosis of fever within a 5-day post-vaccination risk window based on the International Classification of Diseases, Ninth Revision (ICD-9-CM) code.

	In the absence of randomization, causal inference methods can be applied to evaluate the adverse effect of DTaP-IPV-Hib vaccine.
	However, because such administrative data are not collected for research purposes, potential bias due to unmeasured confounding can undermine the validity of causal conclusion.
	In particular, parents of infants may request separate injections or the combination vaccine due to unmeasured health-seeking preference, and such health-seeking behavior may be associated with fever diagnosis.
	To explore the possibility of confounding due to health-seeking behavior, the study monitored presence of injury/trauma (ICD-9 code 800-904, 910-959) and ringworm (ICD-9 code 110) within 30 days post vaccination, which are not expected to be related to the vaccine-outcome pair of interest. In particular, injury/trauma is unlikely to be causally affected by DTaP-IPV-Hib vaccination but may be associated with parents’ health-seeking behavior on behalf of their children. Similarly, ringworm is unlikely to be a cause of fever that occurs during the 5-day risk window but may also be associated with health-seeking behavior. Therefore, we take injury/trauma as a negative control outcome and ringworm as a negative control exposure to detect and account for potential unmeasured confounding.
	During the study, 27,064 DTaP-IPV-Hib vaccinations were administered, among which 60 fevers (0.22\%) were observed within the risk window. In contrast, 19,677 comparator vaccines were administered with 46 fevers (0.23\%) observed. There were 45 ringworm cases and 46 injury/trauma cases. Sex and age group at vaccination ($<$ 5 months or 5 months$-$2 years) were also recorded. 
	
	Because $\trt$, $Z$, and $X$ are all binary, nonparametric (NP) estimation based on cell frequencies is in fact feasible. We thus considered fitting 
	a saturated model for each component of the likelihood by including main effects and all possible interactions, such that all models are nonparametrically estimated. For example, the negative control outcome model was specified as $E[W\mid A, Z, X_1, X_2]=\alpha_0\!+\!\alpha_AA\!+\!\alpha_ZZ\!+\!\alpha_{X_1}X_1\!+\!\alpha_{X_2}X_2 \!+\! \alpha_{A:Z}AZ\!+\!\alpha_{A:X_1}AX_1\!+\!\alpha_{Z:X_1}ZX_1\!+\!\alpha_{A:X_2}AX_2\!+\!\alpha_{Z:X_2}ZX_2\!+\!\alpha_{X_1:X_2}X_1X_2\!+\!\alpha_{A:Z:X_1}AZX_1\!+\!\alpha_{A:Z:X_2}AZX_2\!+\!\alpha_{A:X_1:X_2}AX_1X_2\!+\!\alpha_{Z:X_1:X_2}ZX_1X_2\!+\!\alpha_{A:Z:X_1:X_2}AZX_1X_2$, where $X_1$ denotes age group and $X_2$ denotes sex. 
	As stated in Remark~\ref{same_point_estimate}, under $\mathcal{M}_{\text{nonpar}}$ when all nuisance parameters were nonparametrically estimated, all methods should produce exactly the same point estimate and confidence interval. 
	We thus took the NP model as the true model to illustrate robustness to departure from the NP model via model restrictions in the following scenarios
	\begin{itemize}
		\item $\Mipw$ and $\Mor$ are restricted: $E[W\mid \trt,Z,\cov]$ is fitted without age-sex interaction;
		\item $\Mgest$ and $\Mor$ are restricted: $R(\trt,\cov)$ is fitted without age-sex interaction;
		\item $\Mgest$ and $\Mipw$ are restricted: $f(Z\mid \trt,\cov)$ is fitted without age-sex interaction;
		\item All are restricted: $E[W\mid \trt,Z,\cov]$ and $R(\trt,\cov)$ are fitted without age-sex interaction.
	\end{itemize}

	\begin{table}[h]
		\caption{Adverse effect of DTaP-IPV-Hib vaccine on fever among children.}\label{pentacel}
		\begin{tabular}{l| cccccc}
			\hline 
			\multirow{2}{*}{\textbf{Scenario}} & \multirow{2}{*}{\textbf{Method}} & \textbf{ $\hat{\Delta}$ } & \textbf{Prop} & \multirow{2}{*}{\textbf{$p$-val}} & \textbf{$\hat{\Delta}_{\text{confounded}}$} & \textbf{$\hat{\Delta}_{\text{bias}}$}\tabularnewline
			& & \textbf{(95\% CI)} & \textbf{Bias} & & \textbf{(95\% CI)} & \textbf{(95\% CI)}\tabularnewline
			\hline 
			& $\Delta_{1}$ & 1.7 (-1.1, 4.5) & 0.0 & 0.2 & 0.5 (-0.5, 1.4) & -1.2 (-3.8, 1.3)\tabularnewline
			\textbf{All} & $\Delta_{2}$ & 1.7 (-1.1, 4.5) & 0.0 & 0.2 & 0.5 (-0.5, 1.4) & -1.2 (-3.8, 1.3)\tabularnewline
			\textbf{models are} & $\Delta_{3}$ & 1.7 (-1.1, 4.5) & 0.0 & 0.2 & 0.5 (-0.5, 1.4) & -1.2 (-3.8, 1.3)\tabularnewline
			\textbf{NP} & MLE & 1.7 (-1.1, 4.5) & 0.0 & 0.2 & 0.5 (-0.5, 1.4) & -1.2 (-3.8, 1.3)\tabularnewline
			& MR & 1.7 (-1.1, 4.5) & 0.0 & 0.2 & 0.5 (-0.5, 1.4) & -1.2 (-3.8, 1.3)\tabularnewline
			\hline 
			\textbf{$\mathcal{M}_{1}$ is NP} & $\Delta_{2}$ & 1.7 (-1.4, 4.8) & -0.3 & 0.3 & 0.5 (-0.5, 1.4) & -1.2 (-4.1, 1.6)\tabularnewline
			\textbf{$\mathcal{M}_{2},\mathcal{M}_{3}$} & $\Delta_{3}$ & 0.5 (-0.5, 1.4) & -72.3 & 0.3 & 0.5 (-0.5, 1.4) & -0.0 (-0.3, 0.3)\tabularnewline
			\textbf{are restricted} & MLE & 1.6 (-1.5, 4.7) & -5.6 & 0.3 & 0.5 (-0.5, 1.4) & -1.1 (-4.0, 1.7)\tabularnewline
			& MR & 1.6 (-1.3, 4.5) & -5.2 & 0.3 & 0.5 (-0.5, 1.4) & -1.2 (-3.8, 1.5)\tabularnewline
			\hline 
			\textbf{$\mathcal{M}_{2}$ is NP} & $\Delta_{1}$ & 1.7 (-1.1, 4.4) & -3.1 & 0.2 & 0.5 (-0.5, 1.4) & -1.2 (-3.7, 1.3)\tabularnewline
			\textbf{$\mathcal{M}_{1},\mathcal{M}_{3}$} & $\Delta_{3}$ & 1.7 (-1.1, 4.4) & -3.1 & 0.2 & 0.5 (-0.5, 1.4) & -1.2 (-3.7, 1.3)\tabularnewline
			\textbf{are restricted} & MR & 1.7 (-1.1, 4.6) & 1.6 & 0.2 & 0.5 (-0.5, 1.4) & -1.3 (-3.9, 1.3)\tabularnewline
			\hline 
			\textbf{$\mathcal{M}_{3}$ is NP} & $\Delta_{1}$ & 1.7 (-1.1, 4.5) & 0.0 & 0.2 & 0.5 (-0.5, 1.4) & -1.2 (-3.8, 1.3)\tabularnewline
			\textbf{$\mathcal{M}_{1},\mathcal{M}_{2}$} & $\Delta_{2}$ & 1.7 (-2.4, 5.7) & -2.4 & 0.4 & 0.5 (-0.5, 1.4) & -1.2 (-5.1, 2.7)\tabularnewline
			\textbf{are restricted} & MR & 1.7 (-1.1, 4.5) & -0.0 & 0.2 & 0.5 (-0.5, 1.4) & -1.2 (-3.8, 1.3)\tabularnewline
			\hline 
			& $\Delta_{1}$ & 1.7 (-1.1, 4.4) & -3.1 & 0.2 & 0.5 (-0.5, 1.4) & -1.2 (-3.7, 1.3)\tabularnewline
			\textbf{All} & $\Delta_{2}$ & 1.7 (-1.4, 4.8) & -0.3 & 0.3 & 0.5 (-0.5, 1.4) & -1.2 (-4.1, 1.6)\tabularnewline
			\textbf{models are} & $\Delta_{3}$ & 1.4 (-0.4, 3.2) & -19.7 & 0.1 & 0.5 (-0.5, 1.4) & -0.9 (-2.4, 0.6)\tabularnewline
			\textbf{restricted} & MLE & 1.6 (-1.5, 4.7) & -5.6 & 0.3 & 0.5 (-0.5, 1.4) & -1.1 (-4.0, 1.7)\tabularnewline
			& MR & 1.7 (-1.2, 4.7) & 0.6 & 0.2 & 0.5 (-0.5, 1.4) & -1.3 (-4.0, 1.5)\tabularnewline
			\hline 
		\end{tabular}
		\hspace{0.5in}{\small Note: all point estimates and 95\% confidence intervals (CI) are scaled by $10^3$. Prop bias (\%) is the bias calculated as the proportion of the ATE under the saturated model (NP model) taken as the true value.}
	\end{table}

	Table~\ref{pentacel} lists for each method the point estimates (scaled by $10^3$) of $\Delta$, $\Delta_{\gform}$, $\Delta_{\text{bias}}$ and their $95\%$ confidence intervals (scaled by $10^3$), the bias evaluated as the proportion of the ATE under the saturated model which is taken as the true value, and the $p$-value from a Wald-test of $H_0: \Delta=0$. 
	Similar to the original study, our results indicated a slightly elevated risk of fever among children who received DTaP-IPV-Hib vaccine relative to children who received other DTap-containing comparator vaccines, although the effect was not statistically significant. 
	In addition, there was no evidence of unmeasured confounding as the confidence interval for $\Delta_{\text{bias}}$ included zero. 
	As expected, under $\mathcal{M}_{\text{nonpar}}$, all methods provided exactly the same point estimate and confidence interval. Under model misspecification, i.e., deviation from the NP model via model restrictions, all methods produced a stable estimate of $\Delta_\text{confounded}$, while $\Delta_\text{bias}$ was estimated with larger bias. The MR estimator had generally smaller bias than other methods, which indicated that multiply robust estimation provided protection against model misspecification. A caveat is that in practice if the negative control exposure is rare, the positivity assumption in Assumption~\ref{assumption_cons_pos} may be violated.

	\section{Final remarks}\label{discussion}
	In this paper, we have developed a general semiparametric framework for causal inference in the presence of unmeasured confounding leveraging a pair of negative control exposure and outcome variables. Our method provides an alternative to more conventional methods such as instrumental variable (IV) methods. Particularly, negative controls are sometimes available when a valid IV may not be, in settings such as air pollution studies \citep{miao2017invited}, genetic research \citep{gagnon2012using}, and observational studies using routinely collected healthcare databases such as electronic health records and claims data \citep{schuemie2014interpreting}. In particular, as majority of the variables in administrative healthcare data are documented by medical codes and thus are naturally categorical, we believe our application study demonstrated the promising role of double negative control for detection and control of confounding bias in observational studies using healthcare databases.
	Our paper also contributes to the literature of differential confounding misclassification since negative controls can also be viewed as mismeasured versions of the unobserved confounder \citep{kuroki2014measurement,ogburn2012nondifferential,miao2018identifying}. 
	Our findings established a theoretical basis for future research on semiparametric estimation with negative control adjustment for continuous unmeasured confounding. Another open problem is the possibility of using modern machine learning for estimation of high dimensional nuisance parameters in the context of multiply robust estimation much in the spirit of \cite{athey2017efficient,chernozhukov2016locally,van2011targeted}.
	
	\section*{Acknowledgement}\vspace{-0.2in}
	We thank the associate editor and two referees for their helpful comments.
	Research reported in this publication was supported by the National Institutes of Health under award number R01AI104459. The content is solely the responsibility of the authors and does not necessarily represent the official views of the National Institutes of Health.
	
	\vspace{-0.1in}
	\appendix
	\renewcommand\thesubsection{A.\arabic{subsection}}
	\section*{Appendix}\vspace{-0.1in}
	\subsection{Estimation under $\Mgest$-$\Mor$\label{appendixA}}\vspace{-0.1in}
	Throughout we use dim($v$) to denote the length of a vector $v$, such as $\text{dim}(\beta^R)$.
	\subsubsection{Estimation under $\Mgest$\label{deltagest}} \vspace{-0.1in}
	The first class of estimators involves models $f(\trt,Z\mid \cov;\alpha^{\trt,Z})$ and $R(\trt,\cov;\beta^R)$ under $\Mgest$, with nuisance parameter $\gamma_{1}=(\alpha^{\trt,Z},\beta^R)$. 
	Specifically, let $\hat{\alpha}^{\trt,Z}_{\text{mle}}$ denote the MLE of $\alpha^{\trt,Z}$, and define $f(\trt\mid Z,\cov;\hat{\alpha}^{\trt,Z}_{\text{mle}})=f(\trt,Z\mid \cov;\hat{\alpha}^{\trt,Z}_{\text{mle}})/\sum_a f(\trt=\ttrt,Z\mid \cov;\hat{\alpha}^{\trt,Z}_{\text{mle}})$ 
	and $f(Z\mid \trt,\cov;\hat{\alpha}^{\trt,Z}_{\text{mle}})=f(\trt,Z\mid \cov;\hat{\alpha}^{\trt,Z}_{\text{mle}})/\sum_z f(\trt,Z=z\mid \cov;\hat{\alpha}^{\trt,Z}_{\text{mle}})$. Because $R(\trt,\cov;\beta^R)$ does not by itself give rise to a likelihood function, we obtain an estimator $\hat{\beta}_{\text{gest}}^R$ of $\beta^R$ by solving the following g-estimation type equation \citep{robins1994correcting,wang2018bounded}
	\begin{equation*}
		\mathbb{P}_n\Big\{\left[h_1(\trt,Z,\cov)-E[h_1(\trt,Z,\cov)\mid \trt,\cov;\hat{\alpha}_{\text{mle}}^{\trt,Z}]\right]\left[Y- W \cdot R(\trt,\cov;\hat{\beta}_{\text{gest}}^R)\right]\Big\}=0,
		\end{equation*} 
	where $h_1(\trt,Z,\cov)$ is a vector of user-specified dim($\beta^R$) functions of $\trt$, $Z$, and $\cov$, and $E[h_1(\trt,Z,\cov)\mid \trt,\cov;\hat{\alpha}_{\text{mle}}^{\trt,Z}]$ is evaluated under $f(Z\mid \trt,\cov;\hat{\alpha}_{\text{mle}}^{\trt,Z})$. 
	We then have $\hat{\Delta}_{1}=\hat{\Delta}_{\text{confounded,ipw}}-\hat{\Delta}_{\text{bias,gest}}$, where 
	$\hat{\Delta}_{\text{confounded,ipw}}\!=\!\mathbb{P}_n \!\!\left[ \frac{(2\trt-1)Y}{f(\trt\mid Z,\cov;\hat{\alpha}_{\text{mle}}^{\trt,Z})} \right]$, $\hat{\Delta}_{\text{bias,gest}}\!=\!\mathbb{P}_n \left[ E[R(1\!-\!\trt,\cov;\hat{\beta}_{\text{gest}}^R)\!\!\mid \!\! Z,\cov;\hat{\alpha}_{\text{mle}}^{\trt,Z}]\frac{(2\trt-1)W}{f(\trt\mid Z,\cov;\hat{\alpha}_{\text{mle}}^{\trt,Z})}\right]$.

	\subsubsection{Estimation under $\Mipw$\label{deltaipw}} \vspace{-0.1in}
	The second class of estimators involves models $f(\trt,Z\mid \cov;\alpha^{\trt,Z})$, $\myeta^W_{\zdiff}(\trt,\cov;\beta^{\WZ})$, and $\dW_{\adiff}(Z,\cov;\beta^{\WA})$ under $\Mipw$, with nuisance parameter $\gamma_{2}=(\alpha^{\trt,Z},\beta^{\WZ},\beta^{\WA})$. 
	Specifically, let $\hat{\beta}_{\text{ipw}}^{\WZ}$ and $\hat{\beta}_{\text{ipw}}^{\WA}$ solve the following g-estimating equation
	\begin{equation*}\begin{split}
			\mathbb{P}_n\Big\{\Big[&h_2(\trt,Z,\cov)-E[h_2(\trt,Z,\cov)\mid \cov;\hat{\alpha}^{\trt,Z}_{\text{mle}}]\Big]\Big[W-\\
			&\myeta^W_{\zdiff}(\trt=0,\cov;\hat{\beta}_{\text{ipw}}^{\WZ})Z-\dW_{\adiff}(Z=0,\cov;\hat{\beta}_{\text{ipw}}^{\WA})\trt-\eta^W_{\azdiff}(\cov;\hat{\beta}_{\text{ipw}}^{\WAZ})\trt Z \Big]\Big\}=0
		\end{split}\end{equation*}
	where $h_2(\trt,Z,\cov)$ is a vector of user-specified functions with dimension $\text{dim}(\beta^{\WZ})+\text{dim}(\beta^{\WA})-\text{dim}(\beta^{\WAZ})$,
	and $E[h_2(\trt,Z,\cov)\mid \cov;\hat{\alpha}^{\trt,Z}_{\text{mle}}]$ is evaluated under $f(\trt,Z\mid \cov;\hat{\alpha}^{\trt,Z}_{\text{mle}})$. 
	Then $\hat{\Delta}_{2}=\hat{\Delta}_{\text{confounded,ipw}}-\hat{\Delta}_{\text{bias,ipw}}$, where 
	$\hat{\Delta}_{\text{confounded,ipw}}=\mathbb{P}_n \left[ \frac{2\trt-1}{f(\trt\mid Z,\cov;\hat{\alpha}_{\text{mle}}^{\trt,Z})}Y \right]$, 
	$\hat{\Delta}_{\text{bias,ipw}}=\mathbb{P}_n \Big[ \frac{(2Z-1)Y}{f(Z\mid \trt,\cov;\hat{\alpha}^{\trt,Z}_{\text{mle}})}\\
	\frac{f(1-\trt\mid \cov;\hat{\alpha}^{\trt,Z}_{\text{mle}})}{f(\trt\mid \cov;\hat{\alpha}^{\trt,Z}_{\text{mle}})}
	\frac{E[\delta^{W}_{\adiff}(Z,\cov;\hat{\beta}_{\text{ipw}}^{\WA})\mid 1-\trt,\cov;\hat{\alpha}^{\trt,Z}_{\text{mle}}]}{\myeta^{W}_{\zdiff}(\trt,\cov;\hat{\beta}_{\text{ipw}}^{\WZ})}
	\Big]$.
	
	\subsubsection{Estimation under $\Mor$\label{deltaor}} \vspace{-0.1in}
	The third class of estimators involves models $E[W\mid Z,\trt,\cov;\beta^W]$, $E[Y\mid Z=0,\trt,\cov;\beta^Y]$, and $R(\trt,\cov;\beta^R)$ under $\Mor$, with nuisance parameter $\gamma_{3}=(\beta^W,\beta^Y,\beta^R)$. Specifically, let $\hat{\beta}^W_{\text{mle}}=(\hat{\beta}^{W0}_{\text{mle}},\hat{\beta}^{\WZ}_{\text{mle}},\hat{\beta}^{\WA}_{\text{mle}})$ denote the MLE of $\beta^W$, and $\hat{\beta}^Y_{\text{mle}}$ denote the restricted MLE of $\beta^Y$, where the latter is obtained by maximizing the likelihood under the working model $E[Y\mid Z=0,\trt,\cov;\beta^Y]$ restricted to the subsample with $Z=0$. Let $\hat{\beta}_{\text{or}}^R$ solve the following estimating equation
	\begin{equation*}
		\mathbb{P}_n\Big[h_3(\trt,Z,\cov)\Big(Y-E[Y\mid Z=0,\trt,\cov;\hat{\beta}^Y_{\text{mle}}] - R(\trt,\cov; \hat{\beta}_{\text{or}}^R)(W-E[W\mid Z=0,\trt,\cov;\hat{\beta}^{W}_{\text{mle}}])\Big)\Big]=0,
		\end{equation*} 
	where $h_3(\trt,Z,\cov)$ is a nonzero vector function of dimension $\text{dim}(\beta^R)$. 
	We obtain $E[Y\mid Z,\trt,\cov;\hat{\beta}^Y_{\text{mle}},\\\hat{\beta}^W_{\text{mle}};\hat{\beta}_{\text{or}}^R]$ by Eq.~(\ref{OR_Y}) using $E[Y\mid Z=0,\trt,\cov;\hat{\beta}^Y_{\text{mle}}]$, $\myeta^W_{\zdiff}(\trt,\cov;\hat{\beta}^W_{\text{mle}})$, and $R(\trt,\cov;\hat{\beta}_{\text{or}}^R)$. Combining the above estimators, we have $\hat{\Delta}_{3}=\hat{\Delta}_{\text{confounded,or}}-\hat{\Delta}_{\text{bias,or}}$, where 
	$\hat{\Delta}_{\text{confounded,or}}=\mathbb{P}_n \Big[ 
	E[Y\mid \trt=1,Z,\cov;\hat{\beta}^Y_{\text{mle}},\hat{\beta}^W_{\text{mle}};\hat{\beta}_{\text{or}}^R]-E[Y\mid \trt=0,Z,\cov;\hat{\beta}^Y_{\text{mle}},\hat{\beta}^W_{\text{mle}};\hat{\beta}_{\text{or}}^R]\Big]$ and $\hat{\Delta}_{\text{bias,or}}=\mathbb{P}_n \Big[ R(1-\trt,\cov;\hat{\beta}_{\text{or}}^R)\dW_{\adiff}(Z,\cov;\\\hat{\beta}^W_{\text{mle}})\Big]$.
	
	\subsection{Estimated working models for the multiply robust estimator}\label{appendixB}\vspace{-0.1in}
	Following the variation independent parameterization detailed in (\ref{OR_W})-(\ref{deltayaparam}), we specify the estimated working models by plugging in the corresponding components in $\theta$ as follows: $f(\trt\mid Z,\cov;\hat{\theta})=f(\trt,Z\mid \cov;\hat{\alpha}^{\trt,Z}_{\text{mle}})/\sum_a f(\trt\!=\!\ttrt,Z\mid \cov;\hat{\alpha}^{\trt,Z}_{\text{mle}})$, 
	$f(\trt\mid \cov;\hat{\theta})=\sum_z f(\trt,Z\!=\!z\mid \cov;\hat{\alpha}^{\trt,Z}_{\text{mle}})$, 
	$f(Z\mid \trt,\cov;\hat{\theta})=f(\trt,Z\mid \cov;\hat{\alpha}^{\trt,Z}_{\text{mle}})/\sum_z f(\trt,Z\!=\!z\mid \cov;\hat{\alpha}^{\trt,Z}_{\text{mle}})$, 
	$E[Y\mid  \trt\!\!=\!\!0,Z,\cov;\hat{\theta}]=E[Y\mid Z\!=\!0,\trt,\cov;\hat{\beta}^Y_{\text{mle}}]+R(\trt,\cov;\hat{\beta}^{R}_{\textdr})\\\myeta^W_{\zdiff}(\trt,\cov;\hat{\beta}^{\WA}_{\textdr})Z$, 
	$E[Y\mid Z,\trt,\cov;\hat{\theta}]=E[Y\mid Z\!=\!0,\trt,\cov;\hat{\beta}^{Y}_{\text{mle}}]+R(\trt,\cov;\hat{\beta}^R_{\textdr})\myeta^W_{\zdiff}(\trt,\cov;\hat{\beta}^{\WZ}_{\textdr})$, 
	\\$E[W\mid \trt,Z,\cov; \hat{\theta}]=E[W\mid \trt\!=\!0,Z\!=\!0,\cov; {\beta}^{W0}_{\text{mle}}]+\myeta^W_{\zdiff}(\trt\!=\!0,\cov;\beta^{\WZ}_{\textdr})Z+
	\dW_{\adiff}(Z\!=\!0,\cov;\beta^{\WA}_{\textdr})\trt+\eta^W_{\azdiff}(\cov;\beta^{\WAZ}_{\textdr})\trt Z$, 
	$E[R(1-\trt,\cov)\mid Z,\cov;\hat{\theta}]=\sum_\ttrt {R}(1-\ttrt,\cov;\hat{\beta}^R_{\textdr})f(\trt=\ttrt\mid Z,\cov;\hat{\alpha}^{\trt,Z}_{\text{mle}})$, and \\
	$E[\delta^{W}_{\adiff}(Z,\cov)\mid 1-\trt,\cov;\hat{\theta}]=\sum_z \dW_{\adiff}(z,\cov;\hat{\beta}_{\textdr}^{\WA})f(Z=z\mid  1-\trt,\cov;\hat{\alpha}^{\trt,Z}_{\text{mle}})$.
	In addition, to simplify notation, we let $R(\trt,\cov;\hat{\theta})=R(\trt,\cov;\hat{\beta}^R_{\textdr})$, $\dW_{\adiff}(Z,\cov;\hat{\theta})=\dW_{\adiff}(Z,\cov;\hat{\beta}_{\textdr}^{\WA})$, and $\myeta^{W}_{\zdiff}(\trt,\cov;\hat{\theta})=\myeta^{W}_{\zdiff}(\trt,\cov;\hat{\beta}_{\textdr}^{\WZ})$.

	\clearpage
	{\begin{center}\LARGE Supplementary Materials for \\``Multiply Robust Causal Inference with Double Negative Control Adjustment for Categorical Unmeasured Confounding"\end{center}}
	\renewcommand\thesection{S.\arabic{section}}
	\renewcommand\thesubsection{\thesection.\arabic{subsection}}
	
	\section{Proof of Lemma~\ref{lemma:ate}}\label{appendix:weakercond}
	\begin{proof}
		We first show that Assumption~\ref{assumption_randomization} indicates that 
		\begin{align}
			P(\mathbf{W}\mid  \mathbf{Z},\ttrt,\ccov)&=P(\mathbf{W}\mid  \mathbf{U},\ccov)P(\mathbf{U}\mid  \mathbf{Z},\ttrt,\ccov)\label{supp:eq:identification1}\\
			E[Y\mid \mathbf{Z},\ttrt,\ccov]&=E[Y\mid \mathbf{U},\ttrt,\ccov]P(\mathbf{U}\mid \mathbf{Z},\ttrt,\ccov)\label{supp:eq:identification2},
		\end{align}
		which intuitively states that in the observed data models, the conditional effect of $Z$ on $W$ is proportional to that of $Z$ on $Y$, as they share a factor $P(\mathbf{U}\mid \mathbf{Z},\ttrt,\ccov)$ which is the confounding mechanism.
		{First, under Assumption~\ref{assumption_randomization} we have $Y(\ttrt)\ind (\trt,Z)\mid (U,\cov)$. One one hand, we have $f(Y(\ttrt)\mid U,\cov)=f(Y(\ttrt)\mid \trt,Z,U,\cov)=f(Y\mid \trt=\ttrt,Z,U,\cov)$. On the other hand, $f(Y(\ttrt)\mid U,\cov)=f(Y(\ttrt)\mid \trt,U,\cov)=f(Y\mid \trt=\ttrt,U,\cov)$. We thus have $f(Y\mid \trt=\ttrt,U,\cov)=f(Y\mid Z,\trt=\ttrt,U,\cov)$. Therefore $Y\ind Z\mid (U,\trt,\cov)$ and 
			\[
			E[Y\mid \mathbf{Z},\ttrt,\ccov]=E[Y\mid \mathbf{U},\ttrt,\ccov]P(\mathbf{U}\mid \mathbf{Z},\ttrt,\ccov).
			\]}
		Second, under Assumption~\ref{assumption_randomization} we also have $W\ind (Z,\trt)\mid U,\cov$, therefore
		\[
		P(\mathbf{W}\mid  \mathbf{Z},\ttrt,\ccov)=P(\mathbf{W}\mid  \mathbf{U},\ccov)P(\mathbf{U}\mid  \mathbf{Z},\ttrt,\ccov)
		\]
		
		Now, by Assumption~\ref{assumption_inverse}, because $P(\mathbf{W}\mid  \mathbf{U},\ccov)$ has full column rank with $|W|\geq|U|$, it is left invertible. That is, there is a $|U|\times |W|$ matrix denoted as $P(\mathbf{W}\mid  \mathbf{U},\ccov)^{+}$ such that $P(\mathbf{W}\mid  \mathbf{U},\ccov)^{+}P(\mathbf{W}\mid  \mathbf{U},\ccov)=\mathbb{I}_{|U|}$. Therefore (\ref{supp:eq:identification1}) gives
		\[
		P(\mathbf{U}\mid  \mathbf{Z},\ttrt,\ccov)=P(\mathbf{W}\mid  \mathbf{U},\ccov)^{+}P(\mathbf{W}\mid  \mathbf{Z},\ttrt,\ccov).
		\]
		Combined with (\ref{supp:eq:identification2}) we have
		\[
		E[Y\mid \mathbf{Z},\ttrt,\ccov]=E[Y\mid \mathbf{U},\ttrt,\ccov]P(\mathbf{W}\mid  \mathbf{U},\ccov)^{+}P(\mathbf{W}\mid  \mathbf{Z},\ttrt,\ccov).
		\]
		Therefore
		there exist a $1\times |W|$ vector $\myh $ such that 
		\begin{equation}\label{appendix:proof_yazx}
			E[Y\mid \mathbf{Z},\ttrt,\ccov]=\myh P(\mathbf{W}\mid \mathbf{Z},\ttrt,\ccov).
		\end{equation}
		In particular, $\myh $ does not depend on $\mathbf{U}$ because neither $E[Y\mid \mathbf{Z},\ttrt,\ccov]$ or $P(\mathbf{W}\mid \mathbf{Z},\ttrt,\ccov)$ depend on $\mathbf{U}$. Similarly, by Assumption~\ref{assumption_inverse}, $P(\mathbf{U}\mid  \mathbf{Z},\ttrt,\ccov)$ has a right inverse denoted as $P(\mathbf{U}\mid  \mathbf{Z},\ttrt,\ccov)^{+}$, which satisfies $P(\mathbf{U}\mid  \mathbf{Z},\ttrt,\ccov)P(\mathbf{U}\mid  \mathbf{Z},\ttrt,\ccov)^{+}=\mathbb{I}_{|U|}$.
		Multiplying both sides of (\ref{supp:eq:identification1}) and (\ref{supp:eq:identification2}) by $P(\mathbf{U}\mid  \mathbf{Z},\ttrt,\ccov)^{+}$, we have
		\begin{align}
			E[Y\mid \mathbf{U},\ttrt,\ccov]&=E[Y\mid \mathbf{Z},\ttrt,\ccov]P(\mathbf{U}\mid \mathbf{Z},\ttrt,\ccov)^{+}\label{appendix:proof_yuaz}\\
			P(\mathbf{W}\mid  \mathbf{U},\ccov)&=P(\mathbf{W}\mid  \mathbf{Z},\ttrt,\ccov)P(\mathbf{U}\mid  \mathbf{Z},\ttrt,\ccov)^{+}\label{appendix:proof_pwzax}
		\end{align}

		Now consider
		\begin{equation*}
			\begin{split}
				E[Y(a)\mid \ccov]=&E[Y\mid \mathbf{U},\ttrt,\ccov]P(\mathbf{U}\mid \ccov)\\
				\stackrel{(\ref{appendix:proof_yuaz})}{=}&E[Y\mid \mathbf{Z},\ttrt,\ccov]P(\mathbf{U}\mid \mathbf{Z},\ttrt,\ccov)^{+}P(\mathbf{U}\mid \ccov)\\
				\stackrel{(\ref{appendix:proof_yazx})}{=}&\myh P(\mathbf{W}\mid \mathbf{Z},\ttrt,\ccov)P(\mathbf{U}\mid \mathbf{Z},\ttrt,\ccov)^{+}P(\mathbf{U}\mid \ccov)\\
				\stackrel{(\ref{appendix:proof_pwzax})}{=}&\myh P(\mathbf{W}\mid  \mathbf{U},\ccov)P(\mathbf{U}\mid \ccov)\\
				=&\myh P(\mathbf{W}\mid \ccov)
			\end{split}
		\end{equation*}
		Therefore
		\[
		E[Y(\ttrt)]=\int_{\mathcal{\cov}}\myh P(\mathbf{W}\mid \ccov)f(\ccov)d\ccov.\]
		Thus we complete the proof of Lemma~\ref{lemma:ate}. Below we show Corollary~\ref{coro:ate}.
		
		From (\ref{supp:eq:identification1}) we know that
		\[
		\text{rank}(P(\mathbf{W}\mid  \mathbf{Z},\ttrt,\ccov))\leq\min\{\text{rank}(P(\mathbf{W}\mid  \mathbf{U},\ccov)),\text{rank}(P(\mathbf{U}\mid  \mathbf{Z},\ttrt,\ccov))\}=|U|.\] By the Sylvester’s rank inequality \citep{gantmakher2000theory}
		we also know that 
		\[
		|U|=\text{rank}(P(\mathbf{W}\mid  \mathbf{U},\ccov))+\text{rank}(P(\mathbf{U}\mid  \mathbf{Z},\ttrt,\ccov)-|U|\leq\text{rank}(P(\mathbf{W}\mid  \mathbf{Z},\ttrt,\ccov)).\]
		Therefore \[\text{rank}(P(\mathbf{W}\mid  \mathbf{Z},\ttrt,\ccov))=|U|.\]
		Thus one can learn $|U|$ from the rank of $P(\mathbf{W}\mid  \mathbf{Z},\ttrt,\ccov)$, which is observable. In particular, when $|Z|=|W|=|U|$, $P(\mathbf{W}\mid \mathbf{Z},\ttrt,\ccov)$ is invertible under Assumption~\ref{assumption_inverse}, and the above linear system (\ref{appendix:proof_yazx}) has a unique solution 
		\[
		\myh =E[Y\mid  \mathbf{Z},\ttrt,\ccov]P(\mathbf{W}\mid  \mathbf{Z},\ttrt,\ccov)\inv.
		\]
		Thus
		\[
		E[Y(\ttrt)]=\int_{\mathcal{\cov}}E[Y\mid  \mathbf{Z},\ttrt,\ccov]P(\mathbf{W}\mid  \mathbf{Z},\ttrt,\ccov)\inv P(\mathbf{W}\mid \ccov)f(\ccov)d\ccov.\]
		In contrast, when $|Z|>|U|$ or $|W|>|U|$, $\myh $ is not unique, but $E[Y(\ttrt)]$ is still uniquely identified by
		\[
		E[Y(\ttrt)]=\int_{\mathcal{\cov}}\myh P(\mathbf{W}\mid \ccov)f(\ccov)d\ccov.
		\]

	\end{proof}

	\newpage
	\section{Proof of Lemma~\ref{lemma:simplify}}\label{appendix:simplify}
	\begin{proof}
		Because $\Delta = \int_{\mathcal{\cov}}\Big\{
		E[Y(1)\mid  \cov=\ccov]-E[Y(0)\mid  \cov=\ccov]
		\Big\}f(\ccov)d\ccov$,
		it suffice to consider $E[Y(1)\mid  \cov=\ccov]-E[Y(0)\mid  \cov=\ccov]$. To simplify notation, conditioning on $X$ is implicit in the following proof. In addition, we let the levels of $W$ and $Z$ be $w_i$ and $z_j$ respectively, $i,j=0,\dots,k$.
		We note that for general polytomous negative controls $W$ and $Z$ of $k+1$ categories, we have
		\begin{equation*}\begin{split} 
				&E[Y(1)]=E[Y\mid  \mathbf{Z},\trt\!=\!1]P(\mathbf{W}\mid  \mathbf{Z},\trt\!=\!1)\inv P(\mathbf{W})\\
				=&E[Y\mid  \mathbf{Z},\trt\!=\!1]P(\mathbf{W}\mid  \mathbf{Z},\trt\!=\!1)\inv\cdot \\
				&\Big[P(\mathbf{W}\mid  \mathbf{Z},\trt\!=\!1)_{(k+1) \times (k+1)} ,P(\mathbf{W}\mid  \mathbf{Z},\trt\!=\!0)_{(k+1) \times (k+1)} \Big]_{(k+1)\times 2(k+1)} \begin{pmatrix}P(\mathbf{Z},\trt\!=\!1)_{(k+1) \times 1} \\P(\mathbf{Z},\trt\!=\!0)_{(k+1) \times 1} \end{pmatrix}_{2(k+1) \times 1}\\
				=&E[Y\mid  \mathbf{Z},\trt\!=\!1]\Big\{
				P(\mathbf{Z},\trt\!=\!1)_{(k+1) \times 1} +P(\mathbf{W}\mid  \mathbf{Z},\trt\!=\!1)\inv\cdot P(\mathbf{W}\mid  \mathbf{Z},\trt\!=\!0)P(\mathbf{Z},\trt\!=\!0) \Big\},
			\end{split}\end{equation*}
		Thus we can simplify $E[Y(1)]-E[Y(0)]$ as follows.
		\vspace{-0.05in}\begin{equation}\begin{split}
				E[Y(1)]-E[Y(0)]=&E[Y\mid  \mathbf{Z},\trt\!=\!1][P(\mathbf{Z})-P(\mathbf{Z},\trt\!=\!0)]-E[Y\mid  \mathbf{Z},\trt\!=\!0][P(\mathbf{Z})-P(\mathbf{Z},\trt\!=\!1)]\\
				& +E[Y\mid  \mathbf{Z},\trt\!=\!1]P(\mathbf{W}\mid  \mathbf{Z},\trt\!=\!1)\inv\cdot P(\mathbf{W}\mid  \mathbf{Z},\trt\!=\!0)P(\mathbf{Z},\trt\!=\!0)\\
				& -E[Y\mid  \mathbf{Z},\trt\!=\!0]P(\mathbf{W}\mid  \mathbf{Z},\trt\!=\!0)\inv\cdot P(\mathbf{W}\mid  \mathbf{Z},\trt\!=\!1)P(\mathbf{Z},\trt\!=\!1)\\
				=&E_Z[\dY_{\adiff}(Z)]-E(Y\mid \mathbf{Z},\trt\!=\!1)\Big[\mathbb{I}-P^{-1}(\mathbf{W}\mid \mathbf{Z},\trt\!=\!1)P(\mathbf{W}\mid \mathbf{Z},\trt\!=\!0)\Big]P(\mathbf{Z},\trt\!=\!0)
				\\
				-&E(Y\mid \mathbf{Z},\trt\!=\!0)\Big[P^{-1}(\mathbf{W}\mid \mathbf{Z},\trt\!=\!0)P(\mathbf{W}\mid \mathbf{Z},\trt\!=\!1)-\mathbb{I}\Big]P(\mathbf{Z},\trt\!=\!1)\\
				=&E_Z[\dY_{\adiff}(Z)]-\sum_{\ttrt \in\{0,1\}}E(Y\mid  \mathbf{Z},1\!-\!\ttrt)\cdot P^{-1}(\mathbf{W}\mid \mathbf{Z},1\!-\!\ttrt)\cdot \deltaWAvec(\mathbf{Z})\cdot P(\mathbf{Z},\ttrt)\label{simplify_original}
			\end{split}\vspace{-0.05in}\end{equation}
		where $\mathbb{I}$ is an identity matrix, and $\deltaWAvec(\mathbf{Z})=P(\mathbf{W}\mid \mathbf{Z},\trt\!=\!1)-P(\mathbf{W}\mid \mathbf{Z},\trt\!=\!0)$ is a (k+1) by (k+1) matrix.
		We note that $E_Z[\dY_{\adiff}(Z)]$ is the g-formula of treatment effect but ignoring the unmeasured confounding $U$, whereas
		$\sum_{\ttrt \in\{0,1\}}E(Y\mid  \mathbf{Z},1\!-\!\ttrt)\cdot P^{-1}(\mathbf{W}\mid \mathbf{Z},1\!-\!\ttrt)\cdot \deltaWAvec(\mathbf{Z})\cdot P(\mathbf{Z},\ttrt)$ is a bias correction term that adjusts for the bias due to unmeasured confounding using a negative control exposure $Z$ and a negative control outcome $W$. In the following, we show that when $P\inv( \mathbf{W}\mid \mathbf{Z},1-a)$ is invertible, we have the following two conclusions:
		\begin{itemize}
			\item[(1)] If there is no unmeasured confounder $U$, then $E[Y\mid  \mathbf{Z},1\!-\!\ttrt]P^{-1}(\mathbf{W}\mid \mathbf{Z},1\!-\!\ttrt) \deltaWAvec(\mathbf{Z}) p(\mathbf{Z} \mid \ttrt )=0$. In this case, the bias correction term is equal to zero, Eq. \ref{simplify_original} reduces to the common effect estimate 
			\begin{equation*}
				E[Y(1)]-E[Y(0)]=E_Z[\dY_{\adiff}(Z)].
				\end{equation*} 
			\item[(2)] If the unmeasured confounder $U$ exist (which ensures that $E[\deltaWAvec(Z)\mid  A,\cov]\neq 0$), then
			\begin{equation*}\begin{split} 
					&E[Y\mid  \mathbf{Z},1\!-\!\ttrt]_{1\times (k+1)}P^{-1}(\mathbf{W}\mid \mathbf{Z},1\!-\!\ttrt)_{(k+1)\times (k+1)} \deltaWAvec(\mathbf{Z})_{(k+1)\times (k+1)} p(\mathbf{Z} \mid \ttrt )_{(k+1)\times 1}\\
					=&\Big(\myeta^{Y}_{\zonediff}(1\!-\!\ttrt),\myeta^{Y}_{\ztwodiff}(1\!-\!\ttrt),\dots,\myeta^{Y}_{\zkdiff}(1\!-\!\ttrt)\Big)_{1\times k}\\
					&\begin{pmatrix}
						\myeta^{w_1}_{\zonediff}(1\!-\!\ttrt)& \myeta^{w_1}_{\ztwodiff}(1\!-\!\ttrt) & \cdots &\myeta^{w_1}_{\zkdiff}(1\!-\!\ttrt) \\
						\myeta^{w_2}_{\zonediff}(1\!-\!\ttrt)& \myeta^{w_2}_{\ztwodiff}(1\!-\!\ttrt) & \cdots &\myeta^{w_2}_{\zkdiff}(1\!-\!\ttrt) \\
						\vdots & \vdots & \ddots& \vdots \\
						\myeta^{w_k}_{\zonediff}(1\!-\!\ttrt)& \myeta^{w_k}_{\ztwodiff}(1\!-\!\ttrt) & \cdots &\myeta^{w_k}_{\zkdiff}(1\!-\!\ttrt) 
					\end{pmatrix}\inv_{k\times k}
					\begin{pmatrix}
						E [\delta^{w_1}_{\adiff}(Z) \mid \ttrt ]\\
						E [\delta^{w_2}_{\adiff}(Z) \mid \ttrt ]\\
						\vdots \\
						E [\delta^{w_k}_{\adiff}(Z) \mid \ttrt ]
					\end{pmatrix}_{k\times 1},
				\end{split}\end{equation*}
			where
			\begin{equation*}\begin{split} 
					\myeta^{Y}_{\zjdiff}(1\!-\!\ttrt)=&E[Y\mid  Z\!=\!z_j,1\!-\!\ttrt]-E[Y\mid  Z\!=\!z_0,1\!-\!\ttrt], t=1,\dots,k;\\
					\myeta^{w_i}_{\zjdiff}(1\!-\!\ttrt)=& P(W\!=\!w_i\mid  Z\!=\!z_j,1\!-\!\ttrt)-P(W\!=\!w_i\mid  Z\!=\!z_0,1\!-\!\ttrt), j=1,\dots,k, t=1,\dots,k;\\
					\delta^{w_i}_{\adiff}(Z)=&P(W\!=\!w_i\mid  Z,1)-P(W\!=\!w_i\mid  Z,0), j=1,\dots,k.
				\end{split}\end{equation*}
		\end{itemize}
		
		Note that column sums of $P(\mathbf{W}\mid \mathbf{Z},1\!-\!\ttrt)$ are all equal to 1. One can show that column sums of $P^{-1}(\mathbf{W}\mid \mathbf{Z},1\!-\!\ttrt)$ are also all equal to 1. 
		This is because for an invertible matrix $\trt$ with column sums all equal to 1, we have $\mathbbm{1}\transpose \trt\trt^{-1}=\mathbbm{1}\transpose \trt^{-1}=\mathbbm{1}\transpose \mathbb{I}=\mathbbm{1}\transpose $, where $\mathbbm{1}=(1,1,\dots,1)\transpose $.
		Accordingly, we denote the $(k+1)\times (k+1)$ matrix $P^{-1}(\mathbf{W}\mid \mathbf{Z},1\!-\!\ttrt)$ as
		$
		P^{-1}(\mathbf{W}\mid \mathbf{Z},1\!-\!\ttrt) = \begin{pmatrix}
		1-\sum_{i=1}^k \cons_{i0} &\cdots& 1-\sum_{i=1}^k \cons_{ik}\\
		\cons_{10} &\cdots& \cons_{1k}\\
		\cons_{20} &\cdots& \cons_{2k}\\
		\vdots &&\vdots\\
		\cons_{k0} &\cdots& \cons_{kk}
		\end{pmatrix}$,
		and as we will show later, we have
		\vspace{-0.05in}\begin{equation}
			P^{-1}(\mathbf{W}\mid \mathbf{Z},1\!-\!\ttrt) \deltaWAvec(\mathbf{Z}) p(\mathbf{Z} \mid \ttrt )=\begin{pmatrix}
				-\mathbbm{1}_{1\times k} \\ \mathbb{I}_{k\times k} 
			\end{pmatrix}
			\begin{pmatrix}
				\consm_1\\
				\consm_2\\
				\vdots \\
				\consm_k
			\end{pmatrix}
			\label{eq:dimension_reduction}
			\vspace{-0.05in}\end{equation}
		where $\consm_t=\sum_{j=1}^k (\cons_{t,j}-\cons_{t,0})E[\delta^{w_i}_{\adiff}(Z)\mid \ttrt ]$. When there is no unmeasured confounder, because $W$ is a negative control outcome, we know that $E[\delta^{w_i}_{\adiff}(Z)\mid \ttrt ]=0,\forall j$ and thus $m_t=0,\forall t$. In this case, the bias adjustment term $\sum_{\ttrt \in\{0,1\}}E(Y\mid  \mathbf{Z},1\!-\!\ttrt)\cdot P^{-1}(\mathbf{W}\mid \mathbf{Z},1\!-\!\ttrt)\cdot \deltaWAvec(\mathbf{Z})\cdot P(\mathbf{Z},\ttrt)$ is equal to zero. When $U$ actually exists and $P(\mathbf{W}\mid \mathbf{Z},\ttrt)$ is invertible, there exists $j$ such that $E[\delta^{w_i}_{\adiff}(Z)\mid \ttrt ]\neq 0$. In this case, we solve for $m_t, t=1,\dots,k$ in
		\begin{equation}\label{eq:solvethisee}
			p( \mathbf{W}\mid \mathbf{Z}, 1\!-\!\ttrt)P^{-1}(\mathbf{W}\mid \mathbf{Z},1\!-\!\ttrt) \deltaWAvec(\mathbf{Z}) p(\mathbf{Z} \mid \ttrt )=p(\mathbf{W} \mid  \mathbf{Z}, 1\!-\!\ttrt)\begin{pmatrix}
				-\mathbbm{1}_{1\times k} \\ \mathbb{I}_{k\times k} 
			\end{pmatrix}
			\begin{pmatrix}
				\consm_1\\
				\consm_2\\
				\vdots \\
				\consm_k
			\end{pmatrix}.
			\end{equation} 
		As we will show later, the left hand side of (\ref{eq:solvethisee}) can be simplified as 
		\begin{equation}\label{eq:LHS}
			\begin{split} 
				&p( \mathbf{W}\mid \mathbf{Z}, 1\!-\!\ttrt)P^{-1}(\mathbf{W}\mid \mathbf{Z},1\!-\!\ttrt) \deltaWAvec(\mathbf{Z}) p(\mathbf{Z} \mid \ttrt )=
				\begin{pmatrix}
					-\mathbbm{1}_{1\times k}\\ \mathbb{I}_{k\times k} 
				\end{pmatrix}_{(k+1) \times k}
				\begin{pmatrix}
					E [\delta^{w_1}_{\adiff}(Z) \mid \ttrt ]\\
					E [\delta^{w_2}_{\adiff}(Z) \mid \ttrt ]\\
					\vdots \\
					E [\delta^{w_k}_{\adiff}(Z) \mid \ttrt ]
				\end{pmatrix}_{k \times 1}.
			\end{split}\end{equation}
		As we will show later, the right hand side of (\ref{eq:solvethisee}) can be simplified as 
		\vspace{-0.05in}\begin{equation}\label{eq:RHS}
			p( \mathbf{W}\mid \mathbf{Z}, 1\!-\!\ttrt)_{(k+1)\times (k+1)}\begin{pmatrix}
				-\mathbbm{1}_{1\times k} \\ \mathbb{I}_{k\times k} \end{pmatrix}\begin{pmatrix}
				\consm_1\\
				\consm_2\\
				\vdots \\
				\consm_k
			\end{pmatrix}=\begin{pmatrix}
				-\mathbbm{1}_{1\times k} \\ \mathbb{I} \end{pmatrix}
			\myetaWZvec(1-\ttrt)_{k \times k}\begin{pmatrix}
				\consm_1\\
				\consm_2\\
				\vdots \\
				\consm_k
			\end{pmatrix},
			\vspace{-0.05in}\end{equation}
		where $\myetaWZvec(1-\ttrt)=\begin{pmatrix}
		\myeta^{w_1}_{\zonediff}(1\!-\!\ttrt)& \myeta^{w_1}_{\ztwodiff}(1\!-\!\ttrt) & \cdots &\myeta^{w_1}_{\zkdiff}(1\!-\!\ttrt) \\
		\myeta^{w_2}_{\zonediff}(1\!-\!\ttrt)& \myeta^{w_2}_{\ztwodiff}(1\!-\!\ttrt) & \cdots &\myeta^{w_2}_{\zkdiff}(1\!-\!\ttrt) \\
		\vdots & \vdots & \ddots& \vdots \\
		\myeta^{w_k}_{\zonediff}(1\!-\!\ttrt)& \myeta^{w_k}_{\ztwodiff}(1\!-\!\ttrt) & \cdots &\myeta^{w_k}_{\zkdiff}(1\!-\!\ttrt) 
		\end{pmatrix}$ is a $k\times k$ matrix with element $\myeta^{w_i}_{\zjdiff}(1\!-\!\ttrt) = P(W\!=\!w_i\mid  Z\!=\!z_j,1\!-\!\ttrt)-P(W\!=\!w_i\mid  Z\!=\!z_0,1\!-\!\ttrt), i,j=1,\dots,k$. 
		
		Because $\begin{pmatrix}
		-\mathbbm{1}_{1\times k} \\ \mathbb{I}_{k\times k} \end{pmatrix}$ has rank $k$ with an identity matrix $\mathbb{I}_{k\times k}$, and $p( \mathbf{W}\mid \mathbf{Z}, 1\!-\!\ttrt)$ is invertible, we know that the lefthand side of the above Eq. (\ref{eq:RHS}) has rank $k$. Since for a $(k+1)\times k$ matrix $A$ and a $k\times k$ matrix $B$, we have rank$(AB)\leq $ $\min\{$rank$(A)$, rank$(B)\}$, we know that $\myetaWZvec(1-\ttrt)$ has to have rank $k$. Therefore, $\myetaWZvec(1-\ttrt)$ is invertible.

		Combining (\ref{eq:LHS}) and (\ref{eq:RHS}) we arrive at the following linear equations
		\begin{equation*}\begin{split} 
				&\begin{pmatrix}
					-\mathbbm{1}_{1\times k} \\ \mathbb{I}_{k\times k} \end{pmatrix}
				\begin{pmatrix}
					\myeta^{w_1}_{\zonediff}\!(1\!-\!\ttrt)\!& \myeta^{w_1}_{\ztwodiff}\!(1\!-\!\ttrt)\! & \cdots &\myeta^{w_1}_{\zkdiff}\!(1\!-\!\ttrt)\! \\
					\myeta^{w_2}_{\zonediff}\!(1\!-\!\ttrt)\!& \myeta^{w_2}_{\ztwodiff}\!(1\!-\!\ttrt)\! & \cdots &\myeta^{w_2}_{\zkdiff}\!(1\!-\!\ttrt)\! \\
					\vdots & \vdots & \ddots& \vdots \\
					\myeta^{w_k}_{\zonediff}\!(1\!-\!\ttrt)\!& \myeta^{w_k}_{\ztwodiff}\!(1\!-\!\ttrt)\! & \cdots &\myeta^{w_k}_{\zkdiff}\!(1\!-\!\ttrt)\! 
				\end{pmatrix}
				\begin{pmatrix}\consm_1\\\consm_2\\\vdots \\\consm_k\end{pmatrix}=\begin{pmatrix}
					-\mathbbm{1}_{1\times k} \\ \mathbb{I}_{k\times k} 
				\end{pmatrix}
				\begin{pmatrix}
					E [\delta^{w_1}_{\adiff}(Z) \mid \ttrt ]\\
					E [\delta^{w_2}_{\adiff}(Z) \mid \ttrt ]\\
					\vdots \\
					E [\delta^{w_k}_{\adiff}(Z) \mid \ttrt ]
				\end{pmatrix},
			\end{split}\end{equation*}
		the solution to which is
		\begin{equation}
		\begin{pmatrix}\consm_1\\\consm_2\\\vdots \\\consm_k\end{pmatrix}=\begin{pmatrix}
			\myeta^{w_1}_{\zonediff}(1\!-\!\ttrt)& \myeta^{w_1}_{\ztwodiff}(1\!-\!\ttrt) & \cdots &\myeta^{w_1}_{\zkdiff}(1\!-\!\ttrt) \\
			\myeta^{w_2}_{\zonediff}(1\!-\!\ttrt)& \myeta^{w_2}_{\ztwodiff}(1\!-\!\ttrt) & \cdots &\myeta^{w_2}_{\zkdiff}(1\!-\!\ttrt) \\
			\vdots & \vdots & \ddots& \vdots \\
			\myeta^{w_k}_{\zonediff}(1\!-\!\ttrt)& \myeta^{w_k}_{\ztwodiff}(1\!-\!\ttrt) & \cdots &\myeta^{w_k}_{\zkdiff}(1\!-\!\ttrt) 
		\end{pmatrix}\inv
		\begin{pmatrix}
			E [\delta^{w_1}_{\adiff}(Z) \mid \ttrt ]\\
			E [\delta^{w_2}_{\adiff}(Z) \mid \ttrt ]\\
			\vdots \\
			E [\delta^{w_k}_{\adiff}(Z) \mid \ttrt ]
		\end{pmatrix}.\label{eq:solution_m}
		\end{equation}
		
		Finally, we have
		\begin{equation*}\begin{split} 
				&E[Y\mid  \mathbf{Z},1\!-\!\ttrt]P^{-1}(\mathbf{W}\mid \mathbf{Z},1\!-\!\ttrt) \deltaWAvec(\mathbf{Z}) p( \mathbf{Z}\mid \ttrt )\\
				=&E[Y\mid  \mathbf{Z},1\!-\!\ttrt]
				\begin{pmatrix}
					-\mathbbm{1}_{1\times k} \\ \mathbb{I}_{k\times k} 
				\end{pmatrix}
				\begin{pmatrix}
					\consm_1\\
					\consm_2\\
					\vdots \\
					\consm_k
				\end{pmatrix}\text{by Eq. (\ref{eq:dimension_reduction})}\\
				=&\Big(E[Y\mid  Z\!=\!z_0,1\!-\!\ttrt],E[Y\mid  Z\!=\!z_1,1\!-\!\ttrt],\dots,E[Y\mid  Z\!=\!z_k,1\!-\!\ttrt]\Big)
				\begin{pmatrix}
					-\mathbbm{1}_{1\times k} \\ \mathbb{I}_{k\times k} 
				\end{pmatrix}
				\begin{pmatrix}
					\consm_1\\
					\consm_2\\
					\vdots \\
					\consm_k
				\end{pmatrix}\\
				=&\Big(\myeta^{Y}_{\zonediff}(1\!-\!\ttrt),\myeta^{Y}_{\ztwodiff}(1\!-\!\ttrt),\dots,\myeta^{Y}_{\zkdiff}(1\!-\!\ttrt)\Big)
				\begin{pmatrix}
					\consm_1\\
					\consm_2\\
					\vdots \\
					\consm_k
				\end{pmatrix},
			\end{split}\end{equation*}
		where $\myeta^{Y}_{\zjdiff}(1\!-\!\ttrt)=E[Y\mid  Z\!=\!z_j,1\!-\!\ttrt]-E[Y\mid  Z\!=\!z_0,1\!-\!\ttrt], t=1,\dots,k$.
		By Eq. (\ref{eq:solution_m}) we have
		\begin{equation*}\begin{split} 
				&E[Y\mid  \mathbf{Z},1\!-\!\ttrt]P^{-1}(\mathbf{W}\mid \mathbf{Z},1\!-\!\ttrt) \deltaWAvec(\mathbf{Z}) p( \mathbf{Z}\mid \ttrt )\\
				=&\Big(\myeta^{Y}_{\zonediff}(1\!-\!\ttrt),\myeta^{Y}_{\ztwodiff}(1\!-\!\ttrt),\dots,\myeta^{Y}_{\zkdiff}(1\!-\!\ttrt)\Big)\\
				&\begin{pmatrix}
					\myeta^{w_1}_{\zonediff}(1\!-\!\ttrt)& \myeta^{w_1}_{\ztwodiff}(1\!-\!\ttrt) & \cdots &\myeta^{w_1}_{\zkdiff}(1\!-\!\ttrt) \\
					\myeta^{w_2}_{\zonediff}(1\!-\!\ttrt)& \myeta^{w_2}_{\ztwodiff}(1\!-\!\ttrt) & \cdots &\myeta^{w_2}_{\zkdiff}(1\!-\!\ttrt) \\
					\vdots & \vdots & \ddots& \vdots \\
					\myeta^{w_k}_{\zonediff}(1\!-\!\ttrt)& \myeta^{w_k}_{\ztwodiff}(1\!-\!\ttrt) & \cdots &\myeta^{w_k}_{\zkdiff}(1\!-\!\ttrt) 
				\end{pmatrix}\inv
				\begin{pmatrix}
					E [\delta^{w_1}_{\adiff}(Z) \mid \ttrt ]\\
					E [\delta^{w_2}_{\adiff}(Z) \mid \ttrt ]\\
					\vdots \\
					E [\delta^{w_k}_{\adiff}(Z) \mid \ttrt ]
				\end{pmatrix}.
			\end{split}\end{equation*}

		Therefore, 
		\begin{equation*}\begin{split} 
				&E_Z[\dY_{\adiff}(Z)]-\sum_{\ttrt \in\{0,1\}}E(Y\mid \mathbf{Z},1\!-\!\ttrt)\cdot P^{-1}(\mathbf{W}\mid \mathbf{Z},1\!-\!\ttrt)\cdot \deltaWAvec(\mathbf{Z})\cdot P(\mathbf{Z}\mid \ttrt )P(\ttrt)\\
				=&E_Z[\dY_{\adiff}(Z)]\\
				-&E_{\trt,Z}[\Big(\myeta^{Y}_{\zonediff}(1\!-\!\trt),\myeta^{Y}_{\ztwodiff}(1\!-\!\trt),\dots,\myeta^{Y}_{\zkdiff}(1\!-\!\trt)\Big)\\
				&\begin{pmatrix}
					\myeta^{w_1}_{\zonediff}(1\!-\!\trt)& \myeta^{w_1}_{\ztwodiff}(1\!-\!\trt) & \cdots &\myeta^{w_1}_{\zkdiff}(1\!-\!\trt) \\
					\myeta^{w_2}_{\zonediff}(1\!-\!\trt)& \myeta^{w_2}_{\ztwodiff}(1\!-\!\trt) & \cdots &\myeta^{w_2}_{\zkdiff}(1\!-\!\trt) \\
					\vdots & \vdots & \ddots& \vdots \\
					\myeta^{w_k}_{\zonediff}(1\!-\!\trt)& \myeta^{w_k}_{\ztwodiff}(1\!-\!\trt) & \cdots &\myeta^{w_k}_{\zkdiff}(1\!-\!\trt) 
				\end{pmatrix}\inv
				\begin{pmatrix}
					\delta^{w_1}_{\adiff}(Z)\\
					\delta^{w_2}_{\adiff}(Z)\\
					\vdots \\
					\delta^{w_k}_{\adiff}(Z)
				\end{pmatrix}]\\
				&\equiv E_Z[\dY_{\adiff}(Z)]-E_{\trt,Z}[\R(1\!-\!\trt)\deltaWAvec(Z)],
			\end{split}\end{equation*}
		where 
		$\myetaWZvec(\ttrt)_{k \times k}=\begin{pmatrix}
		\myeta^{w_1}_{\zonediff}(\ttrt)& \myeta^{w_1}_{\ztwodiff}(\ttrt) & \cdots &\myeta^{w_1}_{\zkdiff}(\ttrt) \\
		\myeta^{w_2}_{\zonediff}(\ttrt)& \myeta^{w_2}_{\ztwodiff}(\ttrt) & \cdots &\myeta^{w_2}_{\zkdiff}(\ttrt) \\
		\vdots & \vdots & \ddots& \vdots \\
		\myeta^{w_k}_{\zonediff}(\ttrt)& \myeta^{w_k}_{\ztwodiff}(\ttrt) & \cdots &\myeta^{w_k}_{\zkdiff}(\ttrt) 
		\end{pmatrix}$, $\deltaWAvec(Z)=\begin{pmatrix}\delta^{w_1}_{\adiff}(Z)\\\delta^{w_2}_{\adiff}(Z)\\\vdots \\\delta^{w_k}_{\adiff}(Z)\end{pmatrix}$,\\
		$\myetaYZvec(\ttrt)_{k\times 1}=\Big(\myeta^{Y}_{\zonediff}(\ttrt),\myeta^{Y}_{\ztwodiff}(\ttrt),\dots,\myeta^{Y}_{\zkdiff}(\ttrt)\Big)\transpose $, and $\R(\ttrt)_{1\times k}=\myetaYZvec(\ttrt)\transpose \Big(\myetaWZvec(\ttrt)\Big)\inv$. 
		
		\paragraph*{Proof of Eq. (\ref{eq:dimension_reduction}): }
		\vspace{-0.05in}\begin{equation*}\begin{split}
				&P^{-1}(\mathbf{W}\mid \mathbf{Z},1\!-\!\ttrt) \deltaWAvec(\mathbf{Z}) p(\mathbf{Z} \mid \ttrt )\\
				=&\begin{pmatrix}
					1-\sum_{i=1}^k \cons_{i0} &\cdots& 1-\sum_{i=1}^k \cons_{ik}\\
					\cons_{10} &\cdots& \cons_{1k}\\
					\cons_{20} &\cdots& \cons_{2k}\\
					\vdots &&\vdots\\
					\cons_{k0} &\cdots& \cons_{kk}
				\end{pmatrix}_{(k+1)\times (k+1)}
				\begin{pmatrix}
					-\mathbbm{1}_{1\times k} \\ \mathbb{I}_{k\times k} 
				\end{pmatrix}_{(k+1) \times k}
				\begin{pmatrix}
					E [\delta^{w_1}_{\adiff}(Z) \mid \ttrt ]\\
					E [\delta^{w_2}_{\adiff}(Z) \mid \ttrt ]\\
					\vdots \\
					E [\delta^{w_k}_{\adiff}(Z) \mid \ttrt ]
				\end{pmatrix}_{k \times 1}\\
				=& 
				\begin{pmatrix}
					-\Big( \sum_{i=1}^k \cons_{i1}-\cons_{i0}\Big) & -\Big( \sum_{i=1}^k \cons_{i2}-\cons_{i0}\Big) & \cdots & -\Big( \sum_{i=1}^k \cons_{ik}-\cons_{i0}\Big)\\
					\cons_{11}-\cons_{10} & \cons_{12}-\cons_{10} & \cdots & \cons_{1k}-\cons_{10} \\
					\cons_{21}-\cons_{20} & \cons_{22}-\cons_{20} & \cdots & \cons_{2k}-\cons_{20} \\
					\vdots & \vdots & \ddots& \vdots \\
					\cons_{k1}-\cons_{k0} & \cons_{k2}-\cons_{k0} & \cdots & \cons_{kk}-\cons_{k0} \\
				\end{pmatrix}
				\begin{pmatrix}
					E [\delta^{w_1}_{\adiff}(Z) \mid \ttrt ]\\
					E [\delta^{w_2}_{\adiff}(Z) \mid \ttrt ]\\
					\vdots \\
					E [\delta^{w_k}_{\adiff}(Z) \mid \ttrt ]
				\end{pmatrix}\\
				=&\begin{pmatrix}
					-\mathbbm{1}_{1\times k} \\ \mathbb{I}_{k\times k} \end{pmatrix}
				\begin{pmatrix}
					\cons_{11}-\cons_{10} & \cons_{12}-\cons_{10} & \cdots & \cons_{1k}-\cons_{10} \\
					\cons_{21}-\cons_{20} & \cons_{22}-\cons_{20} & \cdots & \cons_{2k}-\cons_{20} \\
					\vdots & \vdots & \ddots& \vdots \\
					\cons_{k1}-\cons_{k0} & \cons_{k2}-\cons_{k0} & \cdots & \cons_{kk}-\cons_{k0} \\
				\end{pmatrix}
				\begin{pmatrix}
					E [\delta^{w_1}_{\adiff}(Z) \mid \ttrt ]\\
					E [\delta^{w_2}_{\adiff}(Z) \mid \ttrt ]\\
					\vdots \\
					E [\delta^{w_k}_{\adiff}(Z) \mid \ttrt ]
				\end{pmatrix}\\
				=&\begin{pmatrix}
					-\mathbbm{1}_{1\times k} \\ \mathbb{I}_{k\times k} 
				\end{pmatrix}
				\begin{pmatrix}
					\consm_1\\
					\consm_2\\
					\vdots \\
					\consm_k
				\end{pmatrix},
			\end{split}\vspace{-0.05in}\end{equation*}
		where $\consm_t=\sum_{j=1}^k (\cons_{t,j}-\cons_{t,0})E[\delta^{w_i}_{\adiff}(Z)\mid \ttrt ]$. 
		\paragraph*{Proof of of Eq (\ref{eq:LHS}): }
		\begin{equation}
			\begin{split} 
				&p( \mathbf{W}\mid \mathbf{Z}, 1\!-\!\ttrt)P^{-1}(\mathbf{W}\mid \mathbf{Z},1\!-\!\ttrt) \deltaWAvec(\mathbf{Z}) p(\mathbf{Z} \mid \ttrt )=\deltaWAvec(\mathbf{Z}) p(\mathbf{Z} \mid \ttrt )\\
				=& \begin{pmatrix}
					-\sum_{i\!=\!1}^k \delta^{W\!=\!w_i}{(Z\!=\!z_0)} & -\sum_{i\!=\!1}^k \delta^{W\!=\!w_i}{(Z\!=\!z_1)} & \cdots & -\sum_{i\!=\!1}^k \delta^{W\!=\!w_i}{(Z\!=\!z_k)}\\
					\delta^{w_1}{(Z\!=\!z_0)} & \delta^{w_1}{(Z\!=\!z_1)} & \cdots & \delta^{w_1}{(Z\!=\!z_k)}\\
					\delta^{w_2}{(Z\!=\!z_0)} & \delta^{w_2}{(Z\!=\!z_1)} & \cdots & \delta^{w_2}{(Z\!=\!z_k)}\\
					\vdots & \vdots &\ddots&\vdots\\
					\delta^{w_k}{(Z\!=\!z_0)} & \delta^{w_k}{(Z\!=\!z_1)} & \cdots & \delta^{w_k}{(Z\!=\!z_k)}
				\end{pmatrix} \begin{pmatrix}
					P(Z\!=\!z_0\mid \ttrt )\\
					P(Z\!=\!z_1\mid \ttrt ) \\
					\vdots \\
					P(Z\!=\!z_k\mid \ttrt )
				\end{pmatrix}\\
				=&\begin{pmatrix}
					-\sum_{i\!=\!1}^k E [\delta^{w_i}_{\adiff}(Z) \mid \ttrt ]\\
					E [\delta^{w_1}_{\adiff}(Z) \mid \ttrt ]\\
					E [\delta^{w_2}_{\adiff}(Z) \mid \ttrt ]\\
					\vdots \\
					E [\delta^{w_k}_{\adiff}(Z) \mid \ttrt ]
				\end{pmatrix}_{(k+1) \times 1}=
				\begin{pmatrix}
					-\mathbbm{1}_{1\times k}\\ \mathbb{I}_{k\times k} 
				\end{pmatrix}_{(k+1) \times k}
				\begin{pmatrix}
					E [\delta^{w_1}_{\adiff}(Z) \mid \ttrt ]\\
					E [\delta^{w_2}_{\adiff}(Z) \mid \ttrt ]\\
					\vdots \\
					E [\delta^{w_k}_{\adiff}(Z) \mid \ttrt ]
				\end{pmatrix}_{k \times 1}.
			\end{split}\end{equation}
		
		\paragraph*{Proof of Eq (\ref{eq:RHS}):}
		Because $p( \mathbf{W}\mid \mathbf{Z}, 1\!-\!\ttrt)$ has column sums all equal to one, similar to Eq. (\ref{eq:dimension_reduction}) we have
		\begin{equation*}\begin{split} 
				&p( \mathbf{W}\mid \mathbf{Z}, 1\!-\!\ttrt)_{(k+1)\times (k+1)}\begin{pmatrix}
					-\mathbbm{1}_{1\times k} \\ \mathbb{I}_{k\times k} 
				\end{pmatrix}_{(k+1) \times k}\\
				=&\begin{pmatrix}
					1\!-\!\sum_{i=1}^k P(W\!=\!w_i\mid  Z\!=\!z_0,1\!-\!\ttrt) &\!\cdots\!& 1\!-\!\sum_{i=1}^k P(W\!=\!w_i\mid  Z\!=\!z_k,1\!-\!\ttrt)\\
					P(W\!=\!w_1\mid  Z\!=\!z_0,1\!-\!\ttrt) &\!\cdots\!& P(W\!=\!w_1\mid  Z\!=\!z_k,1\!-\!\ttrt)\\
					P(W\!=\!w_2\mid  Z\!=\!z_0,1\!-\!\ttrt) &\!\cdots\!&P(W\!=\!w_2\mid  Z\!=\!z_k,1\!-\!\ttrt)\\
					\vdots &&\vdots\\
					P(W\!=\!w_k\mid  Z\!=\!z_0,1\!-\!\ttrt) &\!\cdots\!& P(W\!=\!w_k\mid  Z\!=\!z_k,1\!-\!\ttrt)\\
				\end{pmatrix}
				\begin{pmatrix}
					-\mathbbm{1}_{1\times k} \\ \mathbb{I}_{k\times k} 
				\end{pmatrix}\\
				=& 
				\begin{pmatrix}
					\sum_{i=1}^k \myeta^{w_i}_{\zonediff}(1\!-\!\ttrt) & -\sum_{i=1}^k \myeta^{w_i}_{\ztwodiff}(1\!-\!\ttrt) & \cdots & - \sum_{i=1}^k \myeta^{w_i}_{\zkdiff}(1\!-\!\ttrt)\\
					\myeta^{w_1}_{\zonediff}(1\!-\!\ttrt)& \myeta^{w_1}_{\ztwodiff}(1\!-\!\ttrt) & \cdots &\myeta^{w_1}_{\zkdiff}(1\!-\!\ttrt) \\
					\myeta^{w_2}_{\zonediff}(1\!-\!\ttrt)& \myeta^{w_2}_{\ztwodiff}(1\!-\!\ttrt) & \cdots &\myeta^{w_2}_{\zkdiff}(1\!-\!\ttrt) \\
					\vdots & \vdots & \ddots& \vdots \\
					\myeta^{w_k}_{\zonediff}(1\!-\!\ttrt)& \myeta^{w_k}_{\ztwodiff}(1\!-\!\ttrt) & \cdots &\myeta^{w_k}_{\zkdiff}(1\!-\!\ttrt) \\ 
				\end{pmatrix}=\begin{pmatrix}
					-\mathbbm{1}_{1\times k} \\ \mathbb{I}_{k\times k} \end{pmatrix}
				\myetaWZvec(1-\ttrt)
			\end{split}\end{equation*}
		where $\myetaWZvec(1-\ttrt)=\begin{pmatrix}
		\myeta^{w_1}_{\zonediff}(1\!-\!\ttrt)& \myeta^{w_1}_{\ztwodiff}(1\!-\!\ttrt) & \cdots &\myeta^{w_1}_{\zkdiff}(1\!-\!\ttrt) \\
		\myeta^{w_2}_{\zonediff}(1\!-\!\ttrt)& \myeta^{w_2}_{\ztwodiff}(1\!-\!\ttrt) & \cdots &\myeta^{w_2}_{\zkdiff}(1\!-\!\ttrt) \\
		\vdots & \vdots & \ddots& \vdots \\
		\myeta^{w_k}_{\zonediff}(1\!-\!\ttrt)& \myeta^{w_k}_{\ztwodiff}(1\!-\!\ttrt) & \cdots &\myeta^{w_k}_{\zkdiff}(1\!-\!\ttrt) 
		\end{pmatrix}$, and $\myeta^{w_i}_{\zjdiff}(1\!-\!\ttrt) = P(W\!=\!w_i\mid  Z\!=\!z_j,1\!-\!\ttrt)-P(W\!=\!w_i\mid  Z\!=\!z_0,1\!-\!\ttrt), i,j=1,\dots,k$. 
		
		\subsection{An alternative illustration of the representation\label{illustrationofidentification}}
		In this section, we illustrate with a toy example that the scaling factor $R(A,X)$ depends on $A$ when there is an $A$-$U$ interaction in the outcome model $E[Y\mid A,U,X]$.
		We illustrate this in the following example where $Z$, $W$, and $U$ are binary. Note that the following models are not required for the identification and estimation results in our paper.
		
		By Assumption 2 we know that $E[Y\mid A,Z,U,X]=E[Y\mid A,U,X]$ and $E[W\mid A,Z,U,X]=E[W\mid U,X]$.
		Let $\alpha(X)$, $\beta(X)$, and $\gamma(X)$ denote any arbitrary function of the observed confounders $X$. Because $A$, $Z$, and $U$ are binary, we have the following nonparametric representation of the underlying true data generating models
		\begin{equation}
			\begin{split}
				&E[Y\mid A,Z,U,X]\stackrel{NCE}{=}E[Y\mid A,U,X]=\alpha_0(X)+{\color{black}{\alpha_A(X)}}A+\alpha_U(X)U+\alpha_{AU}(X)AU\\
				&E[W\mid A,Z,U,X]\stackrel{NCO}{=}E[W\mid U,X]=\beta_0(X)+\beta_U(X)U\\
				&E[U\mid A,Z,X]=\gamma_0(X)+\gamma_A(X)A+\gamma_Z(X)Z+\gamma_{AZ}(X)AZ.
			\end{split}
		\end{equation}
		From the true models, we can derive the observed model as follows
		\begin{equation}
			\begin{split}
				E[Y\mid A,Z,X]&=\alpha_0(X)+\alpha_A(X)A\\
				&+\big[\alpha_U(X)+\alpha_{AU}(X)A\big]\big[\gamma_0(X)+\gamma_A(X)A+\gamma_Z(X)Z+\gamma_{AZ}(X)AZ\big]\\
				E[W\mid A,Z,X]&=\beta_0(X)+\beta_U(X)\big[\gamma_0(X)+\gamma_A(X)A+\gamma_Z(X)Z+\gamma_{AZ}(X)AZ\big].
			\end{split}
		\end{equation}
		Therefore by definition 
		\begin{equation}
			\begin{split}
				R(A,X) =& \frac{\big[\alpha_U(X)+\alpha_{AU}(X)A\big]\big[\gamma_Z(X)+\gamma_{AZ}(X)A\big]}{\beta_U(X)\big[\gamma_Z(X)+\gamma_{AZ}(X)A\big]}\\
				=&\frac{\big[\alpha_U(X)+\alpha_{AU}(X)A\big]}{\beta_U(X)}=\frac{\alpha_U(X)}{\beta_U(X)}+\frac{\alpha_{AU}(X)}{\beta_U(X)}A
			\end{split}
		\end{equation}
		We can see that, if there is no $A$-$U$ interaction in the outcome model $E[Y\mid A,U,X]$, i.e., $\alpha_{AU}(X)=0$, then $R(A,X)=\alpha_{U}(X)/\beta_U(X)$, which only depends on $X$. In this case, $R(\cov)$ accounts for the different scales of the effects of $U$ on $Y$ and $U$ on $W$ in $\Delta_{\text{bias}}=E[R(\cov)\dW_\adiff(Z,\cov)]$. In contrast, if there is $A$-$U$ interaction in $E[Y\mid A,U,X]$, i.e., $\alpha_{AU}(X)\neq0$, then $R(A,X)$ depends on $A$, which further accounts for the effect modification by $U$ in the outcome model. In this case, the bias adjustment term should be $\Delta_{\text{bias}}=E[R(1-\trt,\cov)\dW_\adiff(Z,\cov)]$ which we illustrate as follows.
		
		To simplify notation, hereafter we ignore covariates $X$. We have $\delta_A^W(Z) =\beta_U\big[\gamma_A+\gamma_{AZ}Z\big]$ and $\delta_A^W(Z)R(A)=[\gamma_A+\gamma_{AZ}Z][\alpha_U+\alpha_{AU}A]$. Therefore
		\begin{equation}\label{roneminusa}
			\Delta_\text{bias}=E[\delta_A^W(Z)R(1-A)]=E[\gamma_A+\gamma_{AZ}Z][\alpha_U+\alpha_{AU}]-{\underline{\alpha_{AU}E[(\gamma_A+\gamma_{AZ}Z)A]}}.
		\end{equation}

		Now we compare the true ATE and the naive ATE without accounting for unmeasured confounder. When there is $A$-$U$ interaction in $E[Y\mid A,U,X]$, the true ATE is given by
		\begin{equation}
			\text{true ATE}=\alpha_A+{\underline{\alpha_{AU}E[U]}},
		\end{equation}
		whereas what we can obtain from fitting the observed data model $E[Y\mid A,Z]$ is
		\begin{equation}\label{naiveate}
			\Delta_\text{confounded}=E[\delta_A^Y(Z)]=\alpha_A+E[\gamma_A+\gamma_{AZ}Z][\alpha_U+\alpha_{AU}]+{\underline{\alpha_{AU}E[\gamma_0+\gamma_ZZ]}}
		\end{equation}
		
		Note that $E[U]=E[(\gamma_0+\gamma_ZZ)+(\gamma_A+\gamma_{AZ}Z)A]$. Therefore, from the underlined parts of (\ref{roneminusa})-(\ref{naiveate}), we can see that using $R(1-A)$ allows us to account for the effect modification by $U$, in the scenario where $\alpha_{AU}\neq 0$. 
	\end{proof}

	\newpage
	\section{Generalization to polytomous negative controls}\label{cate_case}
	In this section, we generalize our results to allow for polytomous negative controls $Z$ and $W$, and $U$, i.e., $k$ is any positive integer. Similar to the binary case in Section~\ref{binary_case}, we first characterize the EIF for $\Delta$ in the nonparametric model. We then propose to use the EIF to construct an estimating equation to obtain a multiply robust and locally efficient estimator of $\Delta$ which requires estimating the distribution of the observed data under a parametric (or semiparametric) working model and then evaluating the EIF under such working model.
	
	\subsection{Efficient influence function in the nonparametric model}
	Recall that Lemma~\ref{lemma:simplify} provides an alternative representation of $\Delta$ given by
	\begin{equation*}\begin{split}
			\Delta=&\Delta_{\text{confounded}}-\Delta_{\text{bias}},\\
			\Delta_{\text{confounded}}=E[\dY_{\adiff}(Z,\cov)]&,\;\Delta_{\text{bias}}=E[\R(1\!-\!\trt,\cov)\deltaWAvec(Z,\cov)],
		\end{split}\end{equation*}
	where $\R(\ttrt,\ccov)=\myetaYZvec(\ttrt,\ccov)\transpose \myetaWZvec(\ttrt,\ccov)\inv$ is a $1\times k$ vector with $\myetaYZvec(\ttrt,\ccov)=\{\myeta^{Y}_{\zonediff}(\ttrt,\ccov),\myeta^{Y}_{\ztwodiff}(\ttrt,\ccov),\dots,\\\myeta^{Y}_{\zkdiff}(\ttrt,\ccov)\}\transpose $ and $\myetaWZvec(\ttrt,\ccov)$ is a $k\times k$ matrix with $\myetaWZvec(\ttrt,\ccov)_{i,j}=\myeta^{w_i}_{\zjdiff}(\ttrt,\ccov)$, and $\deltaWAvec(z,\ccov)=\{\delta^{w_1}_{\adiff}(z,\ccov),\\\delta^{w_2}_{\adiff}(z,\ccov), \dots, \delta^{w_k}_{\adiff}(z,\ccov)\}\transpose $. 
	In addition, let $\IW=\{\mathbbm{1}(W=w_1),\mathbbm{1}(W=w_2),\dots,\mathbbm{1}(W=w_k)\}\transpose $ denote a $k\times 1$ vector generalizing the binary $W$, with $\IWi=\mathbbm{1}(W=w_i)$. Let $\IfZ=\{\mathbbm{1}(Z=z_1)/f(Z=z_1\mid  \trt,\cov)-\mathbbm{1}(Z=z_0)/f(Z=z_0\mid  \trt,\cov),\mathbbm{1}(Z=z_2)/f(Z=z_2\mid  \trt,\cov)-\mathbbm{1}(Z=z_0)/f(Z=z_0\mid  \trt,\cov),\dots,\mathbbm{1}(Z=z_k)/f(Z=z_k\mid  \trt,\cov)-\mathbbm{1}(Z=z_0)/f(Z=z_0\mid  \trt,\cov)\}\transpose $ denote a $k\times 1$ vector generalizing $(2Z-1)/f(Z\mid \trt,\cov)$ in the binary case, with $\IfZj=\mathbbm{1}(Z=z_j)/f(Z=z_j\mid  \trt,\cov)-\mathbbm{1}(Z=z_0)/f(Z=z_0\mid  \trt,\cov)$.
	We begin by noting that the EIF for $\Delta_{\text{confounded}}$ in the general case is still given by Eq.~(\ref{EIF_gformula}) of the main manuscript. 
	The following theorem is a natural generalization of Theorem~\ref{lemma2} to the case of polytomous $Z$, $W$, and $U$, which reduces to Theorem~\ref{lemma2} when $k=1$. It is proved in Appendix \ref{appendix:if2}.
	\begin{theorem}\label{thm_cate}
		Under Assumptions~\ref{assumption_NC} -- \ref{assumption_inverse}, the efficient influence function of the bias correction term $\Delta_{\text{bias}}$ in the nonparametric model $\mathcal{M}_{\text{nonpar}}$ is
		\begin{equation*}\begin{split}
				EIF_{\Delta_{\text{bias}}}=&E[\R(1\!-\!\trt,\cov)\mid  Z,\cov] \cdot \frac{2\trt-1}{f(\trt\mid  Z,\cov)} \Big(\IW-\deltaWAvec(Z,\cov)\trt-E[\IW\mid  \trt\!=\!0,Z,\cov]\Big)\\
				+&\IfZ\transpose \frac{f(1-\trt\mid \cov)}{f(\trt\mid \cov)} \Big\{\Big[Y-E[Y\mid  Z\!=\!0,\trt,\cov] - \R(\trt,\cov)(\IW-\\
				&E[\IW\mid  Z\!=\!0,\trt,\cov])\Big]\myetaWZvec(\trt,\cov)\inv\Big\}\cdot E[\deltaWAvec(Z,\cov)\mid  1-\trt,\cov]\\
				+& \R(1\!-\!\trt,\cov) \deltaWAvec(Z,\cov)-\Delta_{\text{bias}}.
			\end{split}\end{equation*}
		Thus, the efficient influence function of $\Delta$ is given by
		\begin{equation*}
			EIF_{\Delta}=EIF_{\Delta_{\text{confounded}}}-EIF_{\Delta_{\text{bias}}},
			\end{equation*} 
		and the semiparametric efficiency bound in $\mathcal{M}_{\text{nonpar}}$ for estimating the ATE is $E[EIF_{\Delta}(O)^2]\inv$.
	\end{theorem}
	
	\subsection{Multiply robust estimation of $\Delta$}\label{multiple_robust_cate}
	In this section, we propose a multiply robust and locally efficient estimator using the $EIF_{\Delta}$ of Theorem~\ref{thm_cate} as an estimating equation and evaluating it under a working model of the observed data distribution. Specifically, 
	let $\etaWAZvec(\cov)$ be a $k\times k$ matrix with $\etaWAZvec(\cov)_{i,j}=\eta^{w_i}_{Az_j}(\cov)$ denoting the joint effect of $\trt$ and $\mathbbm{1}(Z=z_j)$ under the restriction that for $i,j=1,\dots,k$
	\begin{equation*}
		\eta^{w_i}_{Az_j}(\cov)\trt \mathbbm{1}\!(Z\!=\!z_j)\!=\![\myeta^{w_i}_{z_j}(\trt,\cov)-\myeta^{w_i}_{z_j}(\trt\!=\!0,\cov)]\mathbbm{1}\!(Z\!=\!z_j)\!=\![\delta^{w_i}_{\adiff}(z\!=\!z_j,\cov)-\delta^{w_i}_{\adiff}(Z\!=\!z_0,\cov)]\trt \mathbbm{1}\!(Z\!=\!z_j).
		\end{equation*} 
	It is straightforward to verify that for $i=1,\dots,k$
	\vspace{-0.05in}\begin{equation}\label{outcomeW_cate}
		E[\IWi\mid \trt,Z,\cov]\!=\!E[\IWi\mid \trt\!=\!0,Z\!=\!z_0,\cov]\!+\!\delta^{w_i}_{\adiff}(Z\!=\!z_0,\cov)\trt\!+\!\myetaWiZvec(\trt=0,\cov)\IZ\!+\!\etaWiAZvec(\cov)\trt\; \IZ,
		\vspace{-0.05in}\end{equation}
	where $\IZ=\{\mathbbm{1}(Z=z_1),\mathbbm{1}(Z=z_2),\dots,\mathbbm{1}(Z=z_k)\}\transpose $, $\myetaWiZvec(\trt,\cov)$ is the $i$-th row of $\myetaWZvec(\trt,\cov)$ with $\myetaWiZvec(\trt,\cov)\IZ=\sum_{j=1}^{k} \myeta^{w_i}_{\zjdiff}(\trt=0,\cov)\mathbbm{1}(Z=z_j)$, $\etaWiAZvec(\cov)$ is the $i$-th row of $\etaWAZvec(\cov)$ with $\etaWiAZvec(\cov)\trt\; \IZ=\sum_{j=1}^{k}\eta^{w_i}_{Az_j}(\cov)\trt \mathbbm{1}(Z=z_j)$. Likewise we have
	\vspace{-0.05in}\begin{equation}\label{outcomeY_cate}
		E[Y\mid Z,\trt,\cov]=E[Y\mid Z=0,\trt,\cov]+\R(\trt,\cov)\myetaWZvec(\trt,\cov)\IZ, 
		\vspace{-0.05in}\end{equation}
	where $\R(\trt,\cov)\myetaWZvec(\trt,\cov)\IZ=\sum_{j=1}^{k}\R(\trt,\cov)\myetaWZjvec(\trt,\cov)\mathbbm{1}(Z=z_j)$ and $\myetaWZjvec(\trt,\cov)$ is the $j$-th column of $\myetaWZvec(\trt,\cov)$.

	Similar to Section~\ref{multiple_robust_bin}, we specify parametric working model $f(\trt,Z\mid \cov;\alpha^{\trt,Z})$, $E[W\mid \trt=0,Z=0,\cov;\beta^{W0}]$, $E[Y\mid Z=0,\trt,\cov;\beta^{Y}]$, $\myetaWZvec(\trt,\cov;\beta^{\WZ})$, $\deltaWAvec(Z,\cov;\beta^{\WA})$, $\etaWAZvec(\cov;\beta^{\WAZ})$, and $\R(\trt,\cov;\beta^{R})$, with $\beta^{\WAZ}$ a common subset of $\beta^{\WZ}$ and $\beta^{\WA}$.
	We estimate the indexing parameters as follows. Let $\hat{\alpha}^{\trt,Z}_{\text{mle}}$, $\hat{\beta}^{W0}_{\text{mle}}$, and $\hat{\beta}^{Y}_{\text{mle}}$ solve
	\begin{equation*}\begin{split}
			&\mathbb{P}_n\Big\{U_{\alpha^{\trt,Z}}(\hat{\alpha}^{\trt,Z}_{\text{mle}})\Big\}=\mathbb{P}_n\Big\{\frac{\partial}{\partial\alpha^{\trt,Z}}\longmid_{\alpha^{\trt,Z}=\hat{\alpha}^{\trt,Z}_{\text{mle}}} \log f(\trt,Z\mid \cov;\alpha^{\trt,Z})  \Big\}=0,\\
			&\mathbb{P}_n\Big\{U_{\beta^{W0}}(\hat{\beta}^{W0}_{\text{mle}})\Big\}=\mathbb{P}_n\Big\{\frac{\partial}{\partial\beta^{W0}}\longmid_{\beta^{W0}=\hat{\beta}^{W0}_{\text{mle}}}\mathbbm{1}(\trt=0,Z=z_0) \log f(W\mid \trt=0,Z=z_0,\cov;\beta^{W0}) \Big\}=0,\\
			&\mathbb{P}_n\Big\{U_{\beta^{Y}}(\hat{\beta}^{Y}_{\text{mle}})\Big\}=\mathbb{P}_n\Big\{\frac{\partial}{\partial\beta^{Y}}\longmid_{\beta^{Y}=\hat{\beta}^{Y}_{\text{mle}}}\mathbbm{1}(Z=z_0)\log f(Y\mid \trt,Z=z_0,\cov;\beta^{Y})  \Big\}=0\text{ respectively.}
		\end{split}\end{equation*}
	In addition we obtain $\hat{\beta}_{\textdr}^{\WA}$, $\hat{\beta}_{\textdr}^{\WZ}$, and $\hat{\beta}_{\textdr}^{R}$ by solving the following g-estimating equations generalized to polytomous case evaluated at the above estimated nuisance models
	\begin{equation*}\begin{split}
			\mathbb{P}_n\Big\{U_{\beta^{\WiA},\beta^{\WiZ}}(\hat{\beta}_{\textdr}^{\WiA},\hat{\beta}_{\textdr}^{\WiZ})\Big \}= \mathbb{P}_n\Big\{&\Big[g^{(i)}_0(\trt,Z,\cov)-E[g^{(i)}_0(\trt,Z,\cov)\mid \cov;\hat{\alpha}^{\trt,Z}_{\text{mle}}]\Big] \Big[\IWi-\\
			&E[\IWi\mid \trt,Z,\cov; \hat{\beta}^{W0}_{\text{mle}},\beta^{\WiZ},\beta^{\WiA}]\Big]\Big\}=0,\;i=1,\dots,k,\\
		\end{split}\end{equation*}
	\begin{equation*}\begin{split}
			\mathbb{P}_n\Big\{U_{\beta^{R}}(\hat{\beta}_{\textdr}^{R})\Big\}= \mathbb{P}_n\Big\{\Big[g_1(\trt,Z,\cov)-&E[g_1(\trt,Z,\cov)\mid \trt,\cov;\hat{\alpha}^{\trt,Z}_{\text{mle}}]\Big]\Big[Y-E[Y\mid Z=0,\trt,\cov;\hat{\beta}^Y_{\text{mle}}] -\\
			\R(\trt,\cov; \beta^R)(\IW&-E[\IW\mid Z=0,\trt,\cov;\hat{\beta}^{W0}_{\text{mle}},{\hat{\beta}_{\textdr}^{\WA}}])\Big]\Big\}=0,
		\end{split}\end{equation*}
	where $g^{(i)}_0(\trt,Z,\cov)$ is a vector of $\text{dim}(\beta^{\WiA})+\text{dim}(\beta^{\WiZ})-\text{dim}(\beta^{\WiAZ})$ functions, $(\beta^{\WiA},\beta^{\WiZ},\beta^{\WiAZ})$ is the subset of $(\beta^{\WA},\beta^{\WZ},\beta^{\WAZ})$ corresponding to the $i$-th level of $W$, 
	$E[\IWi\mid \trt,Z,\cov]$ is parameterized by Eq.~(\ref{outcomeW_cate}), $E[\IW\mid \trt,Z=0,\cov]$ is a vector of $E[\IWi\mid \trt,Z=0,\cov],i=1,\dots,k$, and $g_1(\trt,Z,\cov)$ is a $k\times 1$ vector of $\text{dim}(\beta^{R})$ functions.
	It can be shown that $\hat{\beta}_{\textdr}^{\WA}$ and $\hat{\beta}_{\textdr}^{\WZ}$ are CAN under the union model $\Mipw\cup\Mor$, and $\hat{\beta}_{\textdr}^R$ is CAN under the union model $\Mgest\cup\Mor$ \citep{robinsrotnitzky2001comment,wang2018bounded}.
	The outcome model $E[Y\mid Z,\trt,\cov;\hat{\beta}^Y_{\text{mle}},\hat{\beta}_{\textdr}^{\WZ},\hat{\beta}_{\textdr}^R]$ is then obtained using Eq.~(\ref{outcomeY_cate}).

	Finally the proposed multiply robust estimator solves $\mathbb{P}_n\Big\{EIF_{\Delta}(O;\hat{\Delta}_{\text{mr}},\hat{\theta}\transpose )\Big\}=0$, where $EIF_{\Delta}(O;\Delta,\hat{\theta})$ is $EIF_{\Delta}$ evaluated at $\hat{\theta}=\{(\hat{\alpha}^{\trt,Z}_{\text{mle}})\transpose ,(\hat{\beta}^Y_{\text{mle}})\transpose ,(\hat{\beta}^{W0}_{\text{mle}})\transpose ,(\hat{\beta}_{\textdr}^{\WZ})\transpose ,(\hat{\beta}_{\textdr}^{\WA})\transpose ,(\hat{\beta}_{\textdr}^{R})\transpose \}\transpose $. 
	That is
	\begin{equation*}
		\hat{\Delta}_{\text{mr}} = \hat{\Delta}_{\text{confounded,mr}}-\hat{\Delta}_{\text{bias,mr}},
		\end{equation*}
	where
	\begin{equation*}\begin{split}
			\hat{\Delta}_{\text{confounded,mr}}=\mathbb{P}_n\Big\{&\frac{2\trt-1}{f(\trt\mid  Z,\cov;\hat{\alpha}^{\trt,Z}_{\text{mle}})}\big(Y\!-\!E[Y\mid  \trt,Z,\cov;\hat{\beta}^Y_{\text{mle}},\hat{\beta}_{\textdr}^{\WZ},\hat{\beta}^R_{\textdr}]\big)+\\
			&\big(E[Y\mid  \trt\!=\!1,Z,\cov;\hat{\beta}^Y_{\text{mle}},\hat{\beta}_{\textdr}^{\WZ},\hat{\beta}^R_{\textdr}]\!-\!E[Y\mid  \trt\!=\!0,Z,\cov;\hat{\beta}^Y_{\text{mle}},\hat{\beta}_{\textdr}^{\WZ},\hat{\beta}^R_{\textdr}]\big)\Big\}\\
			\hat{\Delta}_{\text{bias,mr}}=\mathbb{P}_n\Big\{&E[\R(1\!-\!\trt,\cov)\mid Z,\cov;\hat{\beta}^R_{\textdr},\hat{\alpha}^{\trt,Z}_{\text{mle}}] \frac{2\trt-1}{f(\trt\mid Z,\cov;\hat{\alpha}^{\trt,Z}_{\text{mle}})} \Big(\IW-E[\IW\mid Z,\cov;\hat{\beta}^{W0}_{\text{mle}},\hat{\beta}_{\textdr}^{\WA},{\hat{\beta}_{\textdr}^{\WZ}}]\Big) \\
			+&\IfZalpha\transpose \frac{f(1-\trt\mid \cov;\hat{\alpha}^{\trt,Z}_{\text{mle}})}{f(\trt\mid \cov;\hat{\alpha}^{\trt,Z}_{\text{mle}})} \Big\{\Big[Y-E[Y\mid Z=0,\trt,\cov;\hat{\beta}^Y_{\text{mle}}] - \R(\trt,\cov;\hat{\beta}^R_{\textdr})(\IW-\\
			&E[\IW\mid Z=0,\trt,\cov;\hat{\beta}^{W0}_{\text{mle}},\hat{\beta}_{\textdr}^{\WA}])\Big]\myetaWZvec(\trt,\cov;\hat{\beta}_{\textdr}^{\WZ})\inv\Big\} E[\deltaWAvec(Z,\cov)\mid 1-\trt,\cov;\hat{\beta}^{WA}_{\textdr},\hat{\alpha}^{\trt,Z}_{\text{mle}}]\\
			+&\R(1-\trt,\cov;\hat{\beta}^R_{\textdr})\deltaWAvec(Z,\cov;\hat{\beta}_{\textdr}^{\WA})\Big\}.
		\end{split}\end{equation*}
	Note that $E[\R(1\!-\!\trt,\cov)\mid Z,\cov;\hat{\beta}^R_{\textdr},\hat{\alpha}^{\trt,Z}_{\text{mle}}]= \sum_\ttrt \R(1\!-\!\ttrt,\cov;\hat{\beta}^R_{\textdr})f(\ttrt\mid Z,\cov;\hat{\alpha}^{\trt,Z}_{\text{mle}})$ is evaluated under $f(\trt\mid  Z,\cov;\hat{\alpha}^{\trt,Z}_{\text{mle}})$, and $E[\delta^{W}_{\adiff}(Z,\cov)\mid 1-\trt,\cov;\hat{\beta}^{WA}_{\textdr},\hat{\alpha}^{\trt,Z}_{\text{mle}}]=\sum_z \deltaWAvec(z,\cov;\hat{\beta}_{\textdr}^{\WA})f(z\mid  1-\trt,\cov;\hat{\alpha}^{\trt,Z}_{\text{mle}})$ is evaluated under $f(Z\mid  1-\trt,\cov;\hat{\alpha}^{\trt,Z}_{\text{mle}})$.

	The following theorem generalizes Theorem~\ref{theorem_MR} to polytomous case and is proved in Appendix~\ref{appendix:proof_robustness_cate}. The submodels $\Mgest$, $\Mipw$, and $\Mor$ are defined as in Section~\ref{param_working_models} except that instead of scalars, $E[\IW\mid  \trt,Z,\cov;\beta^W]_{k\times 1}$, $\deltaWAvec(Z,\cov;\beta^{\WA})_{k\times 1}$, $\R(\trt,\cov;\beta^{R})_{1\times k}$, and $\myetaWZvec(\trt,\cov;\beta^{\WZ})_{k\times k}$ are now vectors and matrices.
	\begin{theorem}\label{theorem_MR_2}
		Suppose Assumptions \ref{assumption_NC} -- \ref{assumption_inverse} and standard regularity conditions
		stated in Appendix~\ref{appendix:proof_robustness_cate} hold, then $\sqrt{n}(\hat{\Delta}_{\text{mr}}-\Delta)$ is regular and asymptotic linear under $\mathcal{M}_{\text{union}}$ with influence function
		\begin{equation*}
			IF_{\text{union}}(O;\Delta,\theta^*)=EIF_{\Delta}(O;\Delta,\theta^*)-\frac{\partial EIF_{\Delta}(O;\Delta,\theta)}{\partial \theta\transpose }\longmid_{\theta^*}
			E\Big\{\frac{\partial U_{\theta}(O;\theta)}{\partial \theta\transpose }\longmid_{\theta^*}\Big\}^{-1} U_{\theta}(O;\theta^*),
			\end{equation*} 
		and thus $\sqrt{n}(\hat{\Delta}_{\text{mr}}-\Delta)\rightarrow_d N(0,\sigma^2_{\Delta})$ where $\sigma^2_{\Delta}(\Delta,\theta^*)=E[IF_{\text{union}}(O;\Delta,\theta^*)^2]$, $\theta^*$ denotes the probability limit of $\hat{\theta}$, and 
		$U_{\theta}(O;\theta)=(U_{\alpha^{\trt,Z}}\transpose ,U_{\beta^{Y}}\transpose ,U_{\beta^{W0}}\transpose ,U_{\beta^{\WA},\beta^{\WZ}}\transpose ,U_{\beta^{R}}\transpose )\transpose $, where $U_{\beta^{\WA},\beta^{\WZ}}$ is the collection of $U_{\beta^{\WiA},\beta^{\WiZ}},i=1,\dots,k$.
		Furthermore, $\hat{\Delta}_{\text{mr}}$ is locally semiparametric efficient in the sense that it
		achieves the semiparametric efficiency bound for $\Delta$ in $\mathcal{M}_{\text{union}}$ at the intersection submodel $\mathcal{M}_{\text{intersect}}=\Mgest\cup\Mipw\cup\Mor$ where $\Mgest$, $\Mipw$, and $\Mor$ are all correctly specified.
	\end{theorem}

	\newpage
	\section{Proof of Theorem \ref{lemma2} (efficient influence function in $\mathcal{M}_{\text{nonpar}}$ for binary case)}\label{appendix:if}
	In this section, we show that the efficient influence function in $\mathcal{M}_{\text{nonpar}}$ for 
	\begin{equation*}\begin{split} 
			&\Delta= \int_{\mathcal{\cov}}\Big\{E_Z[\dY_{\adiff}(Z,\cov)\mid \cov\!=\!\ccov]-E_{\trt,Z}[
			R(1\!-\!\trt,\cov)\dW_{\adiff}(Z,\cov)\mid \cov\!=\!\ccov]\Big\}f(\ccov)d\ccov
		\end{split}\end{equation*}
	where
	\begin{equation*}\begin{split} 
			&\dY_{\adiff}(z,\ccov)=E[Y\mid \trt\!=\!1,Z\!=\!z,\cov\!=\!\ccov]-E[Y\mid \trt\!=\!0,Z\!=\!z,\cov\!=\!\ccov];\\
			&\dW_{\adiff}(z,\ccov)=E[W\mid \trt\!=\!1,Z\!=\!z,\cov\!=\!\ccov]-E[W\mid \trt\!=\!0,Z\!=\!z,\cov\!=\!\ccov];\\
			&\myeta^{Y}_{\zdiff}(\ttrt,\ccov)=E[Y\mid \trt=\ttrt,Z=1,\cov\!=\!\ccov]-E[Y\mid Z=0,\trt=\ttrt,\cov\!=\!\ccov];\\
			&\myeta^{W}_{\zdiff}(\ttrt,\ccov)=E[W\mid \trt=\ttrt,Z=1,\cov\!=\!\ccov]-E[W\mid \trt=\ttrt,Z=0,\cov\!=\!\ccov];\\
			&R(1\!-\!\ttrt,\ccov)=\frac{\myeta^{Y}_{\zdiff}(1\!-\!\ttrt,\ccov)}{\myeta^{W}_{\zdiff}(1\!-\!\ttrt,\ccov)}
		\end{split}\end{equation*}
	is
	\begin{equation*}\begin{split} 
			&\text{IF}_{\Delta}(Y,W,A,Z,\cov)\\
			=& \frac{2\trt-1}{f(\trt\mid Z,\cov)}\Big(Y-\dY_{\adiff}(Z,\cov)\trt-E[Y\mid \trt\!=\!0,Z,\cov]\Big)+\dY_{\adiff}(Z,\cov)\\
			-&\frac{2\trt-1}{f(\trt\mid Z,\cov)}\Big(W-\dW_{\adiff}(Z,\cov)\trt-E[W\mid \trt\!=\!0,Z,\cov]\Big)\cdot E[R(1\!-\!\trt,\cov)\mid Z,\cov]\\
			-&\frac{2Z-1}{f(Z\mid \trt,\cov)}\Big[Y-E[Y\mid Z=0,\trt,\cov] - R(\trt,\cov)\Big(W-E[W\mid Z=0,\trt,\cov]\Big)\Big]\cdot\\
			& \frac{1}{\myeta^{W}_{\zdiff}(\trt,\cov)} E[\delta^{W}_{\adiff}(Z,\cov)\mid  1-\trt,\cov]\frac{f(1-\trt\mid \cov)}{f(\trt\mid \cov)}\\
			-& R(1\!-\!\trt,\cov)\dW_{\adiff}(Z,\cov) -\Delta.
		\end{split}\end{equation*}
	\begin{proof}
		Let $f(Y,W,\trt,Z,\cov;\pathwiseparam)$ denote a one-dimensional regular parametric submodel of $\mathcal{M}_{\text{nonpar}}$ indexed by $\pathwiseparam$, under which $\Delta_\pathwiseparam=E_\pathwiseparam[\dY_{\adiff,\pathwiseparam}(Z,\cov)]-E_\pathwiseparam[R_{\pathwiseparam}(1-\trt,\cov)\dW_{\adiff,\pathwiseparam}(Z,\cov)]$.
		The efficient influence function in $\mathcal{M}_{\text{nonpar}}$ is defined as the unique mean zero, finite variance random variable $D$ satisfying
		\begin{equation*}
			\dt \Delta_\pathwiseparam = E[D\cdot S(Y,W,\trt,Z,\cov)],
			\end{equation*} 
		where $S(\cdot)$ is the score function of the path $f(Y,W,\trt,Z,\cov;\pathwiseparam)$ at $\pathwiseparam=0$, and $\dt \Delta_\pathwiseparam$ is the pathwise derivative of $\Delta$.
		To find $D$, we derive the following pathwise derivatives. 
		First, for $\dY_{\adiff,\pathwiseparam}(Z,\cov)=E_{\pathwiseparam}[Y\mid \trt=1,Z,\cov]-E_{\pathwiseparam}[Y\mid \trt=0,Z,\cov]$, we have
		{\small{\vspace{-0.05in}\begin{equation}\begin{split}
						\dt \dY_{\adiff,\pathwiseparam}(Z,\cov) =& E\Big[\frac{2\trt-1}{f(\trt\mid Z,\cov)}(Y-E[Y\mid \trt,Z,\cov])S(Y,\trt,Z,\cov)\longmid Z,\cov\Big]\\
						=& E\Big[\frac{2\trt-1}{f(\trt\mid Z,\cov)}(Y-\dY_{\adiff}(Z,\cov)\trt-E[Y\mid \trt\!=\!0,Z,\cov])S(Y,\trt,Z,\cov)\longmid Z,\cov\Big]\label{eq:pathwisederivative1}
					\end{split}\vspace{-0.05in}\end{equation}}}
		Second, for $\dW_{\adiff,\pathwiseparam}(Z,\cov)=P_{\pathwiseparam}[W\mid \trt=1,Z,\cov]-P_{\pathwiseparam}[W\mid \trt=0,Z,\cov]$, we have
		{\footnotesize{\vspace{-0.05in}\begin{equation}\begin{split}
						\dt \dW_{\adiff,\pathwiseparam}(Z,\cov) =& E\Big[\frac{2\trt-1}{f(\trt\mid Z,\cov)}(W-E[W\mid \trt,Z,\cov])S(W,\trt,Z,\cov)\longmid Z,\cov\Big]\\
						=& E\Big[\frac{2\trt-1}{f(\trt\mid Z,\cov)}(W-\dW_{\adiff}(Z,\cov)\trt-E[W\mid \trt\!=\!0,Z,\cov])S(W,\trt,Z,\cov)\longmid Z,\cov\Big]\label{eq:pathwisederivative2}
					\end{split}\vspace{-0.05in}\end{equation}}}
		Third, for $\myeta^{Y}_{\zdiff;\pathwiseparam}(1\!-\!\trt,\cov)=E_{\pathwiseparam}[Y\mid 1\!-\!\trt,Z=1,\cov]-E_{\pathwiseparam}[Y\mid 1\!-\!\trt,Z=0,\cov]$, we have
		\vspace{-0.05in}\begin{equation}\begin{split}
				\dt \myeta^{Y}_{\zdiff;\pathwiseparam}(1\!-\!\trt,\cov) =& E\Big[\frac{2Z-1}{f(Z\mid 1\!-\!\trt,\cov)}(Y-E[Y\mid Z,1\!-\!\trt,\cov])S(Y,Z,1\!-\!\trt,\cov)\longmid 1\!-\!\trt,\cov\Big]\label{dY_z}
			\end{split}\vspace{-0.05in}\end{equation}
		Forth, for $\myeta^{W}_{\zdiff;\pathwiseparam}(1\!-\!\trt,\cov)=P_{\pathwiseparam}[W\mid 1\!-\!\trt,Z=1,\cov]-P_{\pathwiseparam}[W\mid 1\!-\!\trt,Z=0,\cov]$, we have
		\vspace{-0.05in}\begin{equation}\begin{split}
				\dt \myeta^{W}_{\zdiff;\pathwiseparam}(1\!-\!\trt,\cov) =& E\Big[\frac{2Z-1}{f(Z\mid 1\!-\!\trt,\cov)}(W-E[W\mid Z,1\!-\!\trt,\cov])S(W,Z,1\!-\!\trt,\cov)\longmid 1\!-\!\trt,\cov\Big]\label{dW_z}
			\end{split}\vspace{-0.05in}\end{equation}
		Lastly, using Eq. (\ref{dY_z}) and (\ref{dW_z}), we have
		{\small{\vspace{-0.05in}\begin{equation}\begin{split}
						\dt R_\pathwiseparam(1\!-\!\trt,\cov) =& \dt \frac{\myeta^{Y}_{\zdiff;\pathwiseparam}(1\!-\!\trt,\cov)}{\myeta^{W}_{\zdiff;\pathwiseparam}(1\!-\!\trt,\cov)} \\
						=& \frac{\dt \myeta^{Y}_{\zdiff;\pathwiseparam}(1\!-\!\trt,\cov)\myeta^{W}_{\zdiff}(1\!-\!\trt,\cov) - \myeta^{Y}_{\zdiff}(1\!-\!\trt,\cov)\dt \myeta^{W}_{\zdiff;\pathwiseparam}(1\!-\!\trt,\cov)}{[\myeta^{W}_{\zdiff}(1\!-\!\trt,\cov)^2]}\\
						=&\frac{1}{\myeta^{W}_{\zdiff}(1\!-\!\trt,\cov)}\cdot E\Big[\frac{2Z-1}{f(Z\mid 1\!-\!\trt,\cov)}(Y-E[Y\mid Z,1\!-\!\trt,\cov])S(Y,Z,1\!-\!\trt,\cov)\longmid 1\!-\!\trt,\cov\Big]\\
						-&\frac{R(1\!-\!\trt,\cov)}{\myeta^{W}_{\zdiff}(1\!-\!\trt,\cov)}\cdot E\Big[\frac{2Z-1}{f(Z\mid 1\!-\!\trt,\cov)}(W-E[W\mid Z,1\!-\!\trt,\cov])S(W,Z,1\!-\!\trt,\cov)\longmid 1\!-\!\trt,\cov\Big]\\
						=& E\Big[\frac{1}{\myeta^{W}_{\zdiff}(1\!-\!\trt,\cov)}\cdot\frac{2Z-1}{f(Z\mid 1\!-\!\trt,\cov)}(Y-E[Y\mid Z,1\!-\!\trt,\cov])S(Y,Z,1\!-\!\trt,\cov)\longmid 1\!-\!\trt,\cov\Big]\\
						-& E\Big[\frac{R(1\!-\!\trt,\cov)}{\myeta^{W}_{\zdiff}(1\!-\!\trt,\cov)}\cdot\frac{2Z-1}{f(Z\mid 1\!-\!\trt,\cov)}(W-E[W\mid Z,1\!-\!\trt,\cov])S(W,Z,1\!-\!\trt,\cov)\longmid 1\!-\!\trt,\cov\Big]\\
						=& E\Big[\frac{2Z-1}{f(Z\mid 1\!-\!\trt,\cov)}[Y-E[Y\mid Z,1\!-\!\trt,\cov]-R(1\!-\!\trt,\cov)(W-E[W\mid Z,1\!-\!\trt,\cov])]\cdot\\
						&\frac{1}{\myeta^{W}_{\zdiff}(1\!-\!\trt,\cov)}S(Y,W,Z,1\!-\!\trt,\cov)\longmid 1\!-\!\trt,\cov\Big],\label{eq:pathwisederivative3}
					\end{split}\vspace{-0.05in}\end{equation}}}
		where the last equation holds because for any function $f$, $E[f(Y,Z,\trt,\cov)S(W\mid Y,Z,\trt,\cov)]=0$ and $E[f(W,Z,\trt,\cov)S(Y\mid W,Z,\trt,\cov)]=0$.

		In the following, we consider the pathwise derivative of $E[\dY_{\adiff}(Z,\cov)]$ and $E[\dW_{\adiff}(Z,\cov)\cdot R(1\!-\!\trt,\cov) ]$ respectively. By Eq. (\ref{eq:pathwisederivative1}), the pathwise derivative of $E[\dY_{\adiff}(Z,\cov)]$ is given by
		{\small{\begin{equation*}\begin{split} 
						\dt E_\pathwiseparam[\dY_{\adiff,\pathwiseparam}(Z,\cov)] =& E[\dt \dY_{\adiff}(Z,\cov) ] + E[\dY_{\adiff}(Z,\cov) S(Z,\cov)]\\
						=&E\{[\frac{2\trt-1}{f(\trt\mid Z,\cov)}(Y-\dY_{\adiff}(Z,\cov)\trt-E[Y\mid \trt\!=\!0,Z,\cov])+\\
						&\dY_{\adiff}(Z,\cov)]\cdot S(Y,\trt,Z,\cov) \}.
					\end{split}\end{equation*}}}
		Because $E[f(Y,\trt,Z,\cov)S(W\mid Y,\trt,Z,\cov)]=0$ for any function $f$, we have
		{\small{\begin{equation*}
					\dt \!E_\pathwiseparam[\dY_{\pathwiseparam}\!(Z,\cov)] =E\Big\{[\frac{2\trt-1}{f(\trt\mid Z,\cov)}(Y-\dY_{\adiff}\!(Z,\cov)\trt-E[Y\mid \trt\!=\!0,Z,\cov])+\dY_{\adiff}(Z,\cov)] S(Y,W,\trt,Z,\cov)\Big\}.
					\end{equation*} }}
		
		Now we consider the pathwise derivative of $E[\dW_{\adiff}(Z,\cov)\cdot R(1\!-\!\trt,\cov) ]$. Note that
		{\small{\vspace{-0.05in}\begin{equation}\begin{split}
						&\dt E_\pathwiseparam[\dW_{\adiff,\pathwiseparam}(Z,\cov)\cdot R_\pathwiseparam(1\!-\!\trt,\cov) ]\\
						=& E[R(1\!-\!\trt,\cov) \dt \dW_{\adiff,\pathwiseparam}(Z,\cov) ] + E[\dt R_{\pathwiseparam}(1\!-\!\trt,\cov) \dW_{\adiff}(Z,\cov)] + E[\dW_{\adiff}(Z,\cov) R(1\!-\!\trt,\cov)S(\trt,Z,\cov)].\label{eq:part0}
					\end{split}\vspace{-0.05in}\end{equation}}}
		Thus we consider the $E[R(1\!-\!\trt,\cov) \dt \dW_{\adiff,\pathwiseparam}(Z,\cov) ]$ and $E[\dt R_{\pathwiseparam}(1\!-\!\trt,\cov) \dW_{\adiff}(Z,\cov)]$ respectively. First, we consider $E[R(1\!-\!\trt,\cov) \dt \dW_{\adiff,\pathwiseparam}(Z,\cov) ]$ as follows. By Eq. (\ref{eq:pathwisederivative2})
		{\small{\vspace{-0.05in}\begin{equation}\begin{split}
						&E[R(1\!-\!\trt,\cov) \dt \dW_{\adiff,\pathwiseparam}(Z,\cov) ]=\int\!\dt \dW_{\adiff,\pathwiseparam}(z,\ccov) R(1\!-\!\ttrt,\ccov) f(\ttrt,z,\ccov) d\ttrt dzd\ccov\\
						\!=\!&\int\!E\!\Big[\!\frac{2\trt-1}{f(\trt\mid Z,\cov)}(W\!-\!E[W\mid \trt,Z,\cov]) S(W,\trt,Z,\cov)\longmid Z\!=\!z,\cov\!=\!\ccov\!\Big]\!R(1\!-\!\ttrt,\ccov) f(\ttrt,z,\ccov) d\ttrt dzd\ccov\\
						\!=\!&\int\!E\!\Big[\!\frac{2\trt-1}{f(\trt\mid Z,\cov)}(W\!-\!E[W\mid \trt,Z,\cov])E[R(1\!-\!\trt,\cov)\mid Z\!=\!z,\cov\!=\!\ccov] S(W,\trt,Z,\cov)\longmid Z\!=\!z,\cov\!=\!\ccov\!\Big]\!f(z,\ccov) dzd\ccov\\
						\!=\!&E\!\Big[\!\frac{2\trt-1}{f(\trt\mid Z,\cov)}(W\!-\!E[W\mid \trt,Z,\cov])e_R(Z,\cov) S(W,\trt,Z,\cov)\Big],
					\end{split}\vspace{-0.05in}\end{equation}}}
		where $e_R(z,\ccov)=E[R(1\!-\!\trt,\cov)\mid Z\!=\!z,\cov\!=\!\ccov]$. Because $E[f(W,\trt,Z,\cov)S(Y\mid W,\trt,Z,\cov)]=0$ for any function $f$, we have
		{\small{\vspace{-0.05in}\begin{equation}
					E[R(1\!-\!\trt,\cov) \dt \dW_{\adiff,\pathwiseparam}(Z,\cov) ]=E\!\Big[\!\frac{2\trt-1}{f(\trt\mid Z,\cov)}(W\!-\!E[W\mid \trt,Z,\cov])e_R(Z,\cov) S(Y,W,\trt,Z,\cov)\Big].\label{eq:part1}
					\vspace{-0.05in}\end{equation}}}
		
		Second, we consider $E[\dt R_{\pathwiseparam}(1\!-\!\trt,\cov) \dW_{\adiff}(Z,\cov)]$ as follows. By Eq. (\ref{eq:pathwisederivative3})
		{\small{\begin{equation*}\begin{split} 
						&E[\dt R_{\pathwiseparam}(1\!-\!\trt,\cov) \dW_{\adiff}(Z,\cov)]\\
						=&\int \dt R_{\pathwiseparam}(1\!-\!\ttrt,\ccov) \dW_{\adiff}(z,\ccov) f(\ttrt,z,\ccov) d\ttrt dzd\ccov\\
						=&\int  E\Big[ \frac{2Z-1}{f(Z\mid 1\!-\!\trt,\cov)}\Big[Y-E[Y\mid Z,1\!-\!\trt,\cov]-R(1\!-\!\trt,\cov)(W-E[W\mid Z,1\!-\!\trt,\cov])\Big]\\
						&\frac{1}{\myeta^{W}_{\zdiff}(1\!-\!\trt,\cov)}S(Y,W,Z,1\!-\!\trt,\cov)\longmid 1\!-\!\trt\!=\!1\!-\!\ttrt,\cov\!=\!\ccov\Big] \cdot
						\dW_{\adiff}(z,\ccov)f(\ttrt,z,\ccov) d\ttrt dzd\ccov
					\end{split}\end{equation*}}}
		Note that the above can not be combined with Eq. (\ref{eq:part0}) and (\ref{eq:part1}) since the score is evaluated at $S(Y,W,Z,1\!-\!\trt,\cov)$ rather than $S(Y,W,Z,\trt,\cov)$, which we solve as follows.
		Denote $E\Big[ \frac{2Z-1}{f(Z\mid 1\!-\!\trt,\cov)}\Big[Y-E[Y\mid Z,1\!-\!\trt,\cov]-R(1\!-\!\trt,\cov)(W-E[W\mid Z,1\!-\!\trt,\cov])\Big]\frac{1}{\myeta^{W}_{\zdiff}(1\!-\!\trt,\cov)}S(Y,W,Z,1\!-\!\trt,\cov)\longmid 1\!-\!\trt\!=\!1\!-\!\ttrt,\cov\!=\!\ccov\Big] $ as $h(1\!-\!\ttrt,\ccov)$, then we have
		\begin{equation*}\begin{split} 
				&\int h(1\!-\!\ttrt,\ccov) \dW_{\adiff}(z,\ccov)f(\ttrt,z,\ccov) d\ttrt dzd\ccov\\
				=&\int h(1\!-\!\ttrt,\ccov) \dW_{\adiff}(z,\ccov)[f(z,\ccov)-f(1\!-\!\ttrt, z,\ccov)] d\ttrt dzd\ccov\\
				=&\int h(1\!-\!\ttrt,\ccov) \dW_{\adiff}(z,\ccov)f(z\mid \ccov)f(\ccov) d\ttrt dz d\ccov-\int h(1\!-\!\ttrt,\ccov) \dW_{\adiff}(z,\ccov)f(z\mid 1\!-\!\ttrt,\ccov)dzf(1\!-\!\ttrt,\ccov)d\ttrt d\ccov \\
				=&\underbrace{\int h(1\!-\!\ttrt,\ccov) E[\dW_{\adiff}(z,\ccov)\mid \ccov]d\ttrt f(\ccov)d\ccov}_{\mathcal{A}}-\underbrace{\int h(1\!-\!\ttrt,\ccov) E[\dW_{\adiff}(z,\ccov)\mid 1\!-\!\ttrt,\ccov]f(1\!-\!\ttrt,\ccov)d\ttrt d\ccov}_{\mathcal{B}}.
			\end{split}\end{equation*}
		We consider simplifying $\mathcal{A}$ as follows. Because $E[\dW_{\adiff}(Z,\cov)\mid \ccov]$ is a function of $\ccov$, we have
		{\small{\vspace{-0.05in}\begin{equation}\begin{split} 
						&\int h(1\!-\!\ttrt,\ccov) E[\dW_{\adiff}(Z,\cov)\mid \ccov]d\ttrt f(\ccov)d\ccov \\
						=&\!\int\!E\Big[\frac{2Z-1}{f(Z\mid 1\!-\!\ttrt,\ccov)}\Big[Y-E[Y\mid Z,1\!-\!\trt,\cov]-R(1\!-\!\trt,\cov)(W-E[W\mid Z,1\!-\!\trt,\cov])\Big]\\
						&\frac{1}{\myeta^{W}_{\zdiff}(1\!-\!\ttrt,\ccov)} S(Y,W,Z,1\!-\!\ttrt,\ccov)\longmid 1\!-\!\ttrt,\ccov\Big] E[\dW_{\adiff}(Z,\cov)\mid \ccov]d\ttrt f(\ccov)d\ccov \\
						=&\!\int\!E\Big[\frac{2Z-1}{f(Z\mid 1\!-\!\ttrt,\ccov)}\Big[Y-E[Y\mid Z,1\!-\!\trt,\cov]-R(1\!-\!\trt,\cov)(W-E[W\mid Z,1\!-\!\trt,\cov])\Big]\\
						&\frac{1}{\myeta^{W}_{\zdiff}(1\!-\!\ttrt,\ccov)} E[\dW_{\adiff}(Z,\cov)\mid \ccov] S(Y,W,Z,1\!-\!\ttrt,\ccov) \mid 1\!-\!\ttrt,\ccov\Big] d\ttrt f(\ccov)d\ccov\\
						=&\!\int\!E\Big[\frac{2Z-1}{f(Z\mid 1\!-\!\ttrt,\ccov)}\Big[Y-E[Y\mid Z,1\!-\!\trt,\cov]-R(1\!-\!\trt,\cov)(W-E[W\mid Z,1\!-\!\trt,\cov])\Big]\\
						&\frac{1}{\myeta^{W}_{\zdiff}(1\!-\!\ttrt,\ccov)} \frac{E[\dW_{\adiff}(Z,\cov)\mid \ccov]}{f(1\!-\!\ttrt\mid \ccov)} S(Y,W,Z,1\!-\!\ttrt,\ccov) \mid 1\!-\!\ttrt,\ccov\Big] f(1\!-\!\ttrt,\ccov)d\ttrt d\ccov. \label{because1}
					\end{split}\vspace{-0.05in}\end{equation}}}
		We consider simplifying $\mathcal{B}$ as follows. Because $E[\dW_{\adiff}(Z,\cov)\mid 1\!-\!\ttrt,\ccov]$ is a function of $1\!-\!\ttrt$ and $\ccov$, we have
		{\small{\vspace{-0.05in}\begin{equation}\begin{split}
						&\!\int\!h(1\!-\!\ttrt,\ccov) E[\dW_{\adiff}(Z,\cov)\mid 1\!-\!\ttrt,\ccov]f(1\!-\!\ttrt,\ccov)d\ttrt d\ccov \\
						=&\!\int\!E\Big[\frac{2Z-1}{f(Z\mid 1\!-\!\ttrt,\ccov)}\Big[Y-E[Y\mid Z,1\!-\!\trt,\cov]-R(1\!-\!\trt,\cov)(W-E[W\mid Z,1\!-\!\trt,\cov])\Big]\\
						&\frac{1}{\myeta^{W}_{\zdiff}(1\!-\!\ttrt,\ccov)} S(Y,W,Z,1\!-\!\ttrt,\ccov)\longmid 1\!-\!\ttrt,\ccov\Big] E[\dW_{\adiff}(Z,\cov)\mid 1\!-\!\ttrt,\ccov]f(1\!-\!\ttrt,\ccov)d\ttrt d\ccov \\
						=&\!\int\!E\Big[\frac{2Z-1}{f(Z\mid 1\!-\!\ttrt,\ccov)}\Big[Y-E[Y\mid Z,1\!-\!\trt,\cov]-R(1\!-\!\trt,\cov)(W-E[W\mid Z,1\!-\!\trt,\cov])\Big]\\
						&\frac{1}{\myeta^{W}_{\zdiff}(1\!-\!\ttrt,\ccov)} E[\dW_{\adiff}(Z,\cov)\mid 1\!-\!\ttrt,\ccov] S(Y,W,Z,1\!-\!\ttrt,\ccov)\longmid 1\!-\!\ttrt,\ccov\Big] f(1\!-\!\ttrt,\ccov)d\ttrt d\ccov. \label{because2}
					\end{split}\vspace{-0.05in}\end{equation}}}
		Combining Eq. (\ref{because1}) and (\ref{because2}) we have
		{\small{\vspace{-0.05in}\begin{equation}\begin{split}
						&E[\dt R_{\pathwiseparam}(1\!-\!\trt,\cov) \dW_{\adiff}(Z,\cov)]\\
						=&\int  E\Big[\frac{2Z-1}{f(Z\mid 1\!-\!\ttrt,\ccov)}\Big[Y-E[Y\mid Z,1\!-\!\trt,\cov]-R(1\!-\!\trt,\cov)(W-E[W\mid Z,1\!-\!\trt,\cov])\Big]\\
						&\frac{1}{\myeta^{W}_{\zdiff}(1\!-\!\ttrt,\ccov)} S(Y,W,Z,1\!-\!\ttrt,\ccov)\longmid 1\!-\!\ttrt,\ccov\Big] \dW_{\adiff}(Z,\cov)f(\ttrt,z,\ccov) d\ttrt dzd\ccov\\
						=&\int E\Big[ \frac{2Z-1}{f(Z\mid 1\!-\!\ttrt,\ccov)}\Big[Y-E[Y\mid Z,1\!-\!\trt,\cov]-R(1\!-\!\trt,\cov)(W-E[W\mid Z,1\!-\!\trt,\cov])\Big]\\
						&\frac{1}{\myeta^{W}_{\zdiff}(1\!-\!\ttrt,\ccov)}\left( \frac{E[\dW_{\adiff}(Z,\cov)\mid \ccov]}{f(1\!-\!\ttrt\mid \ccov)}-E[\dW_{\adiff}(Z,\cov)\longmid 1\!-\!\ttrt,\ccov]\right)\cdot S(Y,W,Z,1\!-\!\ttrt,\ccov)\longmid 1\!-\!\ttrt,\ccov\Big] f(1\!-\!\ttrt,\ccov)d\ttrt d\ccov.\label{the_above}
					\end{split}\vspace{-0.05in}\end{equation}}}
		Let $g(1\!-\!\ttrt,\ccov)$ denote
		{\small{\begin{equation*}\begin{split} 
						&E\Big[ \frac{2Z-1}{f(Z\mid 1\!-\!\ttrt,\ccov)}\Big[Y-E[Y\mid Z,1\!-\!\trt,\cov]-R(1\!-\!\trt,\cov)(W-E[W\mid Z,1\!-\!\trt,\cov])\Big]\\
						&\frac{1}{\myeta^{W}_{\zdiff}(1\!-\!\ttrt,\ccov)}\left( \frac{E[\dW_{\adiff}(Z,\cov)\mid \ccov]}{f(1\!-\!\ttrt\mid \ccov)}-E[\dW_{\adiff}(Z,\cov)\longmid 1\!-\!\ttrt,\ccov]\right)\cdot S(Y,W,Z,1\!-\!\ttrt,\ccov)\longmid 1\!-\!\ttrt,\ccov\Big] 
					\end{split}\end{equation*}}}
		then Eq. (\ref{the_above}) is
		\begin{equation*}
			E[\dt R_{\pathwiseparam}(1\!-\!\trt,\cov) \dW_{\adiff}(Z,\cov)]=\int g(1\!-\!\ttrt,\ccov) f(1\!-\!\ttrt,\ccov)d\ttrt d\ccov .
			\end{equation*} 
		Because $\trt$ is a binary variable taking on values $0$ and $1$, we have
		\begin{equation*}\begin{split}
				\int g(1\!-\!\ttrt,\ccov) f(1\!-\!\ttrt,\ccov)d\ttrt d\ccov =& \int g(1-0,\ccov)P(\trt\!=\!1-0\mid \ccov)+g(1-1,\ccov)P(\trt\!=\!1-1\mid \ccov)f(\ccov)d\ccov\\
				=&\int g(\ttrt,\ccov) f(\ttrt\mid \ccov)d\ttrt f(\ccov)d\ccov=\int g(\ttrt,\ccov) f(\ttrt,\ccov)d\ttrt d\ccov.
			\end{split}\end{equation*}
		Therefore Eq. (\ref{the_above}) becomes
		{\small{\begin{equation*}\begin{split} 
						&E[\dt R_{\pathwiseparam}(1\!-\!\trt,\cov) \dW_{\adiff}(Z,\cov)] = \int g(\ttrt,\ccov) f(\ttrt,\ccov)d\ttrt d\ccov\\
						=& E\Big[ \frac{2Z-1}{f(Z\mid \trt,\cov)}\Big[Y-E[Y\mid Z,\trt,\cov]-R(\trt,\cov)(W-E[W\mid Z,\trt,\cov])\Big]\\
						&\frac{1}{\myeta^{W}_{\zdiff}(\trt,\cov)}\left( \frac{E[\dW_{\adiff}(Z,\cov)\mid \cov]}{f(\trt\mid \cov)}-E[\dW_{\adiff}(Z,\cov)\mid \trt,\cov] \right)S(Y,W,\trt,Z,\cov)].
					\end{split}\end{equation*}
		}}
		Because $E[f(Y,\trt,Z,\cov)S(W\mid Y,\trt,Z,\cov)]=0$, and similarly $E[f(W,\trt,Z,\cov)S(Y\mid W,\trt,Z,\cov)]=0$ for any function $f$, we have
		{\small{
				\begin{equation*}\begin{split}
						&E[\dt R_{\pathwiseparam}(1\!-\!\trt,\cov) \dW_{\adiff}(Z,\cov)]\\
						=& E\Big[ \frac{2Z-1}{f(Z\mid \trt,\cov)}(Y-E[Y\mid \trt,Z,\cov])\frac{1}{\myeta^{W}_{\zdiff}(\trt,\cov)}\Big( \frac{E[\dW_{\adiff}(Z,\cov)\mid \cov]}{f(\trt\mid \cov)}-E[\dW_{\adiff}(Z,\cov)\mid \trt,\cov] \Big)\\
						&[S(Y,\trt,Z,\cov)+S(W\mid Y,\trt,Z,\cov)]\Big]\\
						-&E\Big[\frac{2Z-1}{f(Z\mid \trt,\cov)}(W-E[W\mid \trt,Z,\cov])\frac{R(\trt,\cov)}{\myeta^{W}_{\zdiff}(\trt,\cov)} \Big( \frac{E[\dW_{\adiff}(Z,\cov)\mid \cov]}{f(\trt\mid \cov)}-E[\dW_{\adiff}(Z,\cov)\mid \trt,\cov] \Big)\\
						&[S(W,\trt,Z,\cov)+S(Y\mid W,\trt,Z,\cov)]\Big]\\
						=& E[\frac{2Z-1}{f(Z\mid \trt,\cov)}\Big[Y-E[Y\mid \trt,Z,\cov]-R(\trt,\cov)(W-E[W\mid \trt,Z,\cov])\Big]\\
						&\frac{1}{\myeta^{W}_{\zdiff}(\trt,\cov)}\Big( \frac{E[\dW_{\adiff}(Z,\cov)\mid \cov]}{f(\trt\mid \cov)}-E[\dW_{\adiff}(Z,\cov)\mid \trt,\cov] \Big) S(Y,W,\trt,Z,\cov)].
					\end{split}\end{equation*}
		}}
		In addition, because $Z$ is binary, we have $E[Y\mid \trt,Z,\cov]=\myeta^{Y}_{\zdiff}(\trt,\cov)Z+E[Y\mid Z=0,\trt,\cov]=\\R(\trt,\cov)\myeta^{W}_{\zdiff}(\trt,\cov)Z+E[Y\mid Z=0,\trt,\cov]$, and $E[W\mid \trt,Z,\cov]=\myeta^{W}_{\zdiff}(\trt,\cov)Z-E[W\mid Z=0,\trt,\cov]$.
		Therefore
		{\small{
				\begin{equation*}\begin{split}
						&E[\dt R_{\pathwiseparam}(1\!-\!\trt,\cov) \dW_{\adiff}(Z,\cov)]\\
						=& E[\frac{2Z-1}{f(Z\mid \trt,\cov)}\Big[Y-R(\trt,\cov)\myeta^{W}_{\zdiff}(\trt,\cov)Z-E[Y\mid Z=0,\trt,\cov]-R(\trt,\cov)(W-\myeta^{W}_{\zdiff}(\trt,\cov)Z-\\
						&E[W\mid Z=0,\trt,\cov])\Big]\frac{1}{\myeta^{W}_{\zdiff}(\trt,\cov)}\Big( \frac{E[\dW_{\adiff}(Z,\cov)\mid \cov]}{f(\trt\mid \cov)}-E[\dW_{\adiff}(Z,\cov)\mid \trt,\cov] \Big) S(Y,W,\trt,Z,\cov)]\\
						=& E[\frac{2Z-1}{f(Z\mid \trt,\cov)}\Big[Y-E[Y\mid Z=0,\trt,\cov]-R(\trt,\cov)(W-E[W\mid Z=0,\trt,\cov])\Big]\\
						&\frac{1}{\myeta^{W}_{\zdiff}(\trt,\cov)}\Big( \frac{E[\dW_{\adiff}(Z,\cov)\mid \cov]}{f(\trt\mid \cov)}-E[\dW_{\adiff}(Z,\cov)\mid \trt,\cov] \Big) S(Y,W,\trt,Z,\cov)].
					\end{split}\end{equation*}
		}}
		Note that
		\begin{equation*}\begin{split} 
				\frac{E[\delta^{W}_{\adiff}(Z,\cov)\mid \cov ]}{f(\trt\mid \cov )}-E[\delta^{W}_{\adiff}(Z,\cov)\mid \trt,\cov ]
				=&\int_{\mathcal{Z}}\frac{\delta^{W}_{\adiff}(z,\ccov)}{f(\ttrt\mid \ccov)}[f(z\mid \ccov)-f(z,\ttrt\mid \ccov)]dz\\
				=&\int_{\mathcal{Z}}\delta^{W}_{\adiff}(z,\ccov)\frac{f(z,1\!-\!\ttrt\mid \ccov)}{f(\ttrt\mid \ccov)}dz\\
				=&\int_{\mathcal{Z}}\delta^{W}_{\adiff}(z,\ccov)\frac{f(z\mid 1\!-\!\ttrt,\ccov)f(1\!-\!\ttrt\mid \ccov)}{f(\ttrt\mid \ccov)}dz\\
				=&E[\delta^{W}_{\adiff}(Z,\cov) \mid 1\!-\!\trt,\cov ]\frac{f(1\!-\!\trt\mid \cov )}{f(\trt\mid \cov )}.
			\end{split}\end{equation*}
		Therefore, we finally arrive at 
		{\small{
				\vspace{-0.05in}\begin{equation}\begin{split}
						&E[\dt R_{\pathwiseparam}(1\!-\!\trt,\cov) \dW_{\adiff}(Z,\cov)]\\
						=& E[\frac{2Z-1}{f(Z\mid \trt,\cov)}\Big[Y-E[Y\mid Z=0,\trt,\cov]-R(\trt,\cov)(W-E[W\mid Z=0,\trt,\cov])\Big]\\
						&\frac{1}{\myeta^{W}_{\zdiff}(\trt,\cov)}\Big(E[\delta^{W}_{\adiff}(Z,\cov)\mid  1-\trt,\cov]\frac{f(1-\trt\mid \cov)}{f(\trt\mid \cov)}\Big) S(Y,W,\trt,Z,\cov)].\label{eq:part2}
					\end{split}\vspace{-0.05in}\end{equation}
		}}
		Combining Eq. (\ref{eq:part0}), (\ref{eq:part1}), and (\ref{eq:part2}) we have
		{\small{\begin{equation*}\begin{split} 
						&\dt E_\pathwiseparam[\dW_{\adiff,\pathwiseparam}(Z,\cov)\cdot R_\pathwiseparam(1\!-\!\trt,\cov) ]\\
						=& E[\Big\{\dW_{\adiff}(Z,\cov) R(1\!-\!\trt,\cov) +\frac{2\trt-1}{f(\trt\mid Z,\cov)}(W-\dW_{\adiff}(Z,\cov)\trt-E[W\mid \trt\!=\!0,Z,\cov])E[R(1\!-\!\trt,\cov)\mid Z,\cov] \\
						+& \frac{2Z-1}{f(Z\mid \trt,\cov)}\Big[Y-E[Y\mid Z=0,\trt,\cov]-R(\trt,\cov)(W-E[W\mid Z=0,\trt,\cov])\Big]\cdot\\
						&\frac{1}{\myeta^{W}_{\zdiff}(\trt,\cov)}\Big(E[\delta^{W}_{\adiff}(Z,\cov)\mid  1-\trt,\cov]\frac{f(1-\trt\mid \cov)}{f(\trt\mid \cov)}\Big)\Big\} S(Y,W,\trt,Z,\cov)].
					\end{split}\end{equation*}}}
		
		Therefore, the efficient influence function in $\mathcal{M}_{\text{nonpar}}$ for $\Delta= E[\dY_{\adiff}(Z,\cov)] - E[\dW_{\adiff}(Z,\cov)\cdot R(1\!-\!\trt,\cov) ]$ is given by
		\begin{equation*}\begin{split} 
				\text{IF}_{\Delta}(Y,W,A,Z,\cov)
				=& \frac{2\trt-1}{f(\trt\mid Z,\cov)}\Big(Y-\dY_{\adiff}(Z,\cov)\trt-E[Y\mid \trt\!=\!0,Z,\cov]\Big)\\
				-&\frac{2\trt-1}{f(\trt\mid Z,\cov)}\Big(W-\dW_{\adiff}(Z,\cov)\trt-E[W\mid \trt\!=\!0,Z,\cov]\Big)E[R(1\!-\!\trt,\cov)\mid Z,\cov]\\
				-& \frac{2Z-1}{f(Z\mid \trt,\cov)} \Big[Y-E[Y\mid Z=0,\trt,\cov] - R(\trt,\cov)\Big(W-E[W\mid Z=0,\trt,\cov]\Big)\Big]\cdot\\
				&\frac{1}{\myeta^{W}_{\zdiff}(\trt,\cov)} \Big(E[\delta^{W}_{\adiff}(Z,\cov)\mid  1-\trt,\cov]\frac{f(1-\trt\mid \cov)}{f(\trt\mid \cov)}\Big)\\
				+&\dY_{\adiff}(Z,\cov)-R(1\!-\!\trt,\cov)\dW_{\adiff}(Z,\cov)-\Delta.
			\end{split}\end{equation*}
		
	\end{proof}
	
	\newpage
	\section{Proof of Theorem~\ref{theorem_MR}}\label{appendix:proof_robustness_binary}
	
	\begin{proof}
		Under the regularity conditions given in Theorem 3.2 of \cite{newey1994large}, the estimated nuisance parameters
		\[
		\hat{\theta}=\{(\hat{\alpha}^{\trt,Z}_{\text{mle}})\transpose ,(\hat{\beta}^Y_{\text{mle}})\transpose ,(\hat{\beta}^{W0}_{\text{mle}})\transpose ,(\hat{\beta}_{\textdr}^{\WZ})\transpose ,(\hat{\beta}_{\textdr}^{\WA})\transpose ,(\hat{\beta}_{\textdr}^{R})\transpose \}\transpose \]
		from solving the moment function vector $\mathbb{P}_n\{U_\theta(O;\theta)\}=0$ are asymptotically normal and converge at $o(n^{-1/2})$ rate to its probability limit \[{\theta}^*=\{({\alpha}^{\trt,Z}_{*})\transpose ,({\beta}^Y_{*})\transpose ,({\beta}^{W0}_{*})\transpose ,({\beta}_{*}^{\WZ})\transpose ,({\beta}_{*}^{\WA})\transpose ,({\beta}_{*}^{R})\transpose \}\transpose \] 
		regardless of whether the corresponding nuisance models are correctly specified. 
		
		The main step of the proof is to show that $EIF_{\Delta}(O)$ is an unbiased estimating equation for $\Delta$ under $\mathcal{M}_{\text{union}}$. This is completed by first showing that that $\beta_{*}^{\WA}=\beta^{\WA}$, $\beta_{*}^{\WZ}=\beta^{\WZ}$ under $\Mipw\cup\Mor$, and $\beta^R_*=\beta^R$ under $\Mgest\cup\Mor$ in Section~\ref{supp:sec_E1}; then showing that $\Delta_{\text{mr}}^*$, the probability limit of $\hat{\Delta}_{\text{mr}}$, satisfies $E[\Delta_{\text{mr}}^*]=\Delta$ in Section~\ref{subsec:can}.
		
		Now we derive the asymptotic distribution of $\hat{\Delta}_{\text{mr}}$. Assuming that the regularity conditions given in Corollary 1, Chapter 8 of \cite{manski1988analog} hold for $EIF_{\Delta}(O;\Delta,\theta)$ and $U_{\theta}(O;\theta)$, it follows from standard Taylor expansion of $\sqrt{n}\mathbb{P}_n\left\{EIF_{\Delta}(O;\Delta,\theta^*)\right\}=0$ that
		\begin{equation*}
			\begin{split}
				0=&\sqrt{n}\mathbb{P}_n\left\{EIF_{\Delta}(O;\Delta,\theta^*)\right\}+ \frac{\partial EIF_{\Delta}(O;\Delta,\theta)}{\partial \Delta\transpose }\longmid_{\Delta}\sqrt{n}(\hat{\Delta}_{\text{mr}}-\Delta) \\
				&+\sqrt{n}\mathbb{P}_n\left\{\frac{\partial EIF_{\Delta}(O;\Delta,\theta)}{\partial \theta\transpose }\longmid_{\theta^*}
				E\Big\{-\frac{\partial U_{\theta}(O;\theta)}{\partial \theta\transpose }\longmid_{\theta^*}\Big\}^{-1} U_{\theta}(O;\theta^*)\right\}+o_p(1),
			\end{split}
		\end{equation*}
		where
		$\frac{\partial EIF_{\Delta}(O;\Delta,\theta)}{\partial \Delta\transpose }\longmid_{\Delta}=-1$.
		Therefore
		\begin{equation*}
			\sqrt{n}(\hat{\Delta}_{\text{mr}}-\Delta)=\frac{1}{\sqrt{n}}\sum_{i=1}^n IF_{\text{union}}(O_i;\Delta,\theta^*)+o_p(1),
			\end{equation*} 
		where
		\begin{equation*}
			IF_{\text{union}}(O;\Delta,\theta^*)=EIF_{\Delta}(O;\Delta,\theta^*)+\frac{\partial EIF_{\Delta}(O;\Delta,\theta)}{\partial \theta\transpose }\longmid_{\theta^*}
			E\Big\{-\frac{\partial U_{\theta}(O;\theta)}{\partial \theta\transpose }\longmid_{\theta^*}\Big\}^{-1} U_{\theta}(O;\theta^*),
			\end{equation*} 
		$O_i$ stands for the $i$-th observation and $\theta^*$ is the probability limit of $\hat{\theta}$.
		By Slutsky's Theorem and the Central Limit Theorem we have $\sqrt{n}(\hat{\Delta}_{\text{mr}}-\Delta)\rightarrow_d N(0,\sigma^2_{\Delta})$, 
		where $\sigma^2_{\Delta}(\Delta,\theta^*)=E[IF_{\text{union}}(O;\Delta,\theta^*)^2]$.
		
		At the intersection submodel $\mathcal{M}_{\text{intersect}}$ where all models $\Mgest$, $\Mipw$, and $\Mor$ are correctly specified, $\theta^*=\theta$ and we have that 
		\begin{equation*}
			\frac{\partial EIF_{\Delta}(O;\Delta,\theta)}{\partial \theta\transpose }\longmid_{\theta^*}=0,
			\end{equation*} 
		and thus 
		\begin{equation*}
			IF_{\text{union}}(O;\Delta,\theta^*=\theta)=EIF_{\Delta}(O;\Delta,\theta^*=\theta).
			\end{equation*} 
		Therefore if all models $\Mgest$, $\Mipw$, and $\Mor$ are correctly specified, $\hat{\Delta}_{\text{mr}}$ achieves the semiparametric efficient bound under model $\mathcal{M}_{\text{union}}$.
		
		\subsection{Proof that  $\beta_{*}^{\WA}=\beta^{\WA}$, $\beta_{*}^{\WZ}=\beta^{\WZ}$ under $\Mipw\cup\Mor$, and $\beta^R_*=\beta^R$ under $\Mgest\cup\Mor$\label{supp:sec_E1}}
		To simplify notation, we let $\text{D}^*_{\dW_{\adiff}(Z,\cov)}$, $\text{D}^*_{R(1\!-\!\trt,\cov)}$, $R^*(1\!-\!\trt,\cov)$,
		$E^*[Y\mid Z=0,\trt,\cov]$, $E^*[W\mid \trt\!=\!0,Z=0,\cov]$, $\delta^{W^*}_{\adiff}(Z,\cov)$, $\myeta^{W^*}_{\zdiff}(\trt,\cov)$, $E^*[W\mid Z=0,\trt,\cov]$, $f^*(\trt\mid Z,\cov)$, $f^*(Z\mid \trt,\cov)$, and $f^*(\trt\mid \cov)$ denote the probability limit of the estimated nuisance models. Similarly, we let $\Delta^*_{\text{mr}}$, $\Delta^*_{\text{confounded,dr}}$, and $\Delta^*_{\text{bias,mr}}$ denote the probability limit of the estimated parameters of interest.
		
		We start with showing that $\delta^{W^*}_{\adiff}(Z,\cov)=\delta^{W}_{\adiff}(Z,\cov)$ and $\myeta^{W^*}_{\zdiff}(\trt,\cov)=\myeta^{W}_{\zdiff}(\trt,\cov)$ under $\Mipw\cup\Mor$, and $R^*(\trt,\cov)=R(\trt,\cov)$ under $\Mgest\cup\Mor$. Note that $\delta^{W}_{\adiff}(Z,\cov)$, $\myeta^{W}_{\zdiff}(\trt,\cov)$, and $R(\trt,\cov)$ do not by themselves give rise to a likelihood, and estimation of these components relies on construction of estimating equations that depends on other components of the full data likelihood such as $f(\trt,Z\mid \cov)$ and $E[W\mid \trt=0,Z=0,\cov]$ which can be estimated by the MLE. Therefore, we show such doubly robust property by showing that the constructed estimating equations are unbiased with mean zero under the union models $\Mipw\cup\Mor$ (or $\Mgest\cup\Mor$). 
		
		First, we show that $\delta^{W^*}_{\adiff}(Z,\cov)=\delta^{W}_{\adiff}(Z,\cov)$ and $\myeta^{W^*}_{\zdiff}(\trt,\cov)=\myeta^{W}_{\zdiff}(\trt,\cov)$ under $\Mipw\cup\Mor$. Under $\Mipw$ where $f(\trt,Z\mid \cov;\alpha^{\trt,Z})$, $\myeta^W_{\zdiff}(\trt,\cov;\beta^{\WZ})$ and $\dW_{\adiff}(Z,\cov;\beta^{\WA})$ are correctly specified, we have ${\alpha}^{\trt,Z}_{*}={\alpha}^{\trt,Z}$, $f^*(\trt,Z\mid \cov)=f(\trt,Z\mid \cov)$, and thus $E^*[g_0(\trt,Z,\cov)\mid \cov]=E[g_0(\trt,Z,\cov)\mid \cov]$ for any function $g_0(\trt,Z,\cov)$. Recall that $\hat{\beta}_{\textdr}^{\WA}$ and $\hat{\beta}_{\textdr}^{\WZ}$ solves $\mathbb{P}_n\Big\{U_{\beta^{\WA},\beta^{\WZ}}(\hat{\beta}_{\textdr}^{\WA},\hat{\beta}_{\textdr}^{\WZ})\Big \}=0$ with\\ $\lim_{n\rightarrow\infty}\mathbb{P}_n\Big\{U_{\beta^{\WA},\beta^{\WZ}}(\hat{\beta}_{\textdr}^{\WA},\hat{\beta}_{\textdr}^{\WZ})\Big \}=E[U_{\beta^{\WA},\beta^{\WZ}}(\beta_{*}^{\WA},\beta_{*}^{\WZ})]$. 
		Therefore we consider\\ $E[U_{\beta^{\WA},\beta^{\WZ}}(\beta_{*}^{\WA},\beta_{*}^{\WZ})]\longmid_{\beta_{*}^{\WA}=\beta^{\WA},\beta_{*}^{\WZ}=\beta^{\WZ}}$ under $\Mipw$ where $\myeta^W_{\zdiff}(\trt,\cov;\beta^{\WZ})$ and \\$\dW_{\adiff}(Z,\cov;\beta^{\WA})$ are correctly specified, i.e. $\myeta^W_{\zdiff}(\trt,\cov)=\myeta^W_{\zdiff}(\trt,\cov;\beta^{\WZ})$ and $\dW_{\adiff}(Z,\cov)=\dW_{\adiff}(Z,\cov;\beta^{\WA})$, we have
		\begin{equation*}\begin{split}
				&E[U_{\beta^{\WA},\beta^{\WZ}}(\beta^{\WA},\beta^{\WZ})]\\
				=&E\{\Big[g_0(\trt,Z,\cov)-E^*[g_0(\trt,Z,\cov)\mid \cov]\Big] \Big[W-E[W\mid \trt,Z,\cov; \beta^{W0}_{*},\beta^{\WZ},\beta^{\WA}]\Big]\}\\
				=&E\{\Big[g_0(\trt,Z,\cov)-E[g_0(\trt,Z,\cov)\mid \cov]\Big] \Big[E[W\mid \trt=0,Z=0,\cov]-E[W\mid \trt=0,Z=0,\cov; \beta_*^{W0}]+\\
				&[\myeta^W_{\zdiff}(\trt=0,\cov)-\myeta^W_{\zdiff}(\trt=0,\cov;\beta^{\WZ})]Z+[\dW_{\adiff}(Z=0,\cov)-\dW_{\adiff}(Z=0,\cov;\beta^{\WA})]\trt+\\
				&[\eta^W_{\azdiff}(\cov)-\eta^W_{\azdiff}(\cov;\beta^{\WAZ})]\trt Z\Big]\}\\
				=&E\{\Big[g_0(\trt,Z,\cov)-E[g_0(\trt,Z,\cov)\mid \cov]\Big] \Big[E[W\mid \trt=0,Z=0,\cov]-E[W\mid \trt=0,Z=0,\cov; \beta_*^{W0}]\Big]\}\\
				=&0
			\end{split}\end{equation*} 
		because $E\Big[\{g_0(\trt,Z,\cov)-E[g_0(\trt,Z,\cov)\mid \cov]\}h(X)\Big]=0$ for any function $h$. Thus, under $\Mipw$ where $\myeta^W_{\zdiff}(\trt,\cov;\beta^{\WZ})$ and $\dW_{\adiff}(Z,\cov;\beta^{\WA})$ are correctly specified, $\mathbb{P}_n\Big\{U_{\beta^{\WA},\beta^{\WZ}}(\hat{\beta}_{\textdr}^{\WA},\hat{\beta}_{\textdr}^{\WZ})\longmid_{\hat{\beta}_{\textdr}^{\WA}=\beta^{\WA},\hat{\beta}_{\textdr}^{\WZ}=\beta^{\WZ}}\Big\}$ converges to zero, i.e. $(\beta^{\WA},\beta^{\WZ})$ is a solution to the probability limit of $\mathbb{P}_n\Big\{U_{\beta^{\WA},\beta^{\WZ}}(\hat{\beta}_{\textdr}^{\WA},\hat{\beta}_{\textdr}^{\WZ})\Big\}=0$. Thus $\beta_{*}^{\WA}=\beta^{\WA}$, and $\beta_{*}^{\WZ}=\beta^{\WZ}$, and thus $\delta^{W^*}_{\adiff}(Z,\cov)=\delta^{W}_{\adiff}(Z,\cov)$ and $\myeta^{W^*}_{\zdiff}(\trt,\cov)=\myeta^{W}_{\zdiff}(\trt,\cov)$ under $\Mipw$.
		
		Similar arguments apply to the scenario under $\Mor$. Under $\Mor$ where working models $R(\trt,\cov;\beta^R)$, $E[Y\mid Z=0,\trt,\cov;\beta^Y]$, $\myeta^W_{\zdiff}(\trt,\cov;\beta^{\WZ})$, $\dW_{\adiff}(Z,\cov;\beta^{\WA})$, and $E[W\mid \trt=0,Z=0,\cov;\beta^W]$ are correctly specified, we have ${\beta}^{W0}_{*}={\beta}^{W0}$ and thus $E^*[W\mid \trt=0,Z=0,\cov]=E[W\mid \trt=0,Z=0,\cov]$. In addition, we again have $\myeta^W_{\zdiff}(\trt,\cov)=\myeta^W_{\zdiff}(\trt,\cov;\beta^{\WZ})$ and $\dW_{\adiff}(Z,\cov)=\dW_{\adiff}(Z,\cov;\beta^{\WA})$. Now consider
		\begin{equation*}\begin{split}
				&E[U_{\beta^{\WA},\beta^{\WZ}}(\beta^{\WA},\beta^{\WZ})]\\
				=&E\{\Big[g_0(\trt,Z,\cov)-E^*[g_0(\trt,Z,\cov)\mid \cov]\Big] \Big[W-E[W\mid \trt,Z,\cov; \beta^{W0}_{*},\beta^{\WZ},\beta^{\WA}]\Big]\}\\
				=&E\{\Big[g_0(\trt,Z,\cov)-E^*[g_0(\trt,Z,\cov)\mid \cov]\Big] \Big[E[W\mid \trt=0,Z=0,\cov]-E[W\mid \trt=0,Z=0,\cov; \beta_*^{W0}]+\\
				&[\myeta^W_{\zdiff}(\trt=0,\cov)-\myeta^W_{\zdiff}(\trt=0,\cov;\beta^{\WZ})]Z+[\dW_{\adiff}(Z=0,\cov)-\dW_{\adiff}(Z=0,\cov;\beta^{\WA})]\trt+\\
				&[\eta^W_{\azdiff}(\cov)-\eta^W_{\azdiff}(\cov;\beta^{\WAZ})]\trt Z\Big]\}\\
				=&E\{\Big[g_0(\trt,Z,\cov)-E^*[g_0(\trt,Z,\cov)\mid \cov]\Big] \Big[E[W\mid \trt=0,Z=0,\cov]-E[W\mid \trt=0,Z=0,\cov; \beta_*^{W0}]\Big]\}\\
				=&0
			\end{split}\end{equation*} 
		because $E[W\mid \trt=0,Z=0,\cov; \beta_*^{W0}]=E[W\mid \trt=0,Z=0,\cov]$. Therefore $\delta^{W^*}_{\adiff}(Z,\cov)=\delta^{W}_{\adiff}(Z,\cov)$ and $\myeta^{W^*}_{\zdiff}(\trt,\cov)=\myeta^{W}_{\zdiff}(\trt,\cov)$ under $\Mor$. In addition, we have that 
		\begin{equation}\begin{split}\label{underM3}
				E^*[W\mid A,Z,X]=&E^*[W\mid \trt=0,Z=0,\cov]+\myeta^{W^*}_{\zdiff}(\trt=0,\cov)Z+\delta^{W^*}_{\adiff}(Z=0,\cov)A+\eta^{W^*}_{\azdiff}(\cov)AZ\\
				=&E[W\mid \trt=0,Z=0,\cov]+\myeta^{W}_{\zdiff}(\trt=0,\cov)Z+\delta^{W}_{\adiff}(Z=0,\cov)A+\eta^{W}_{\azdiff}(\cov)AZ\\
				=&E[W\mid A,Z,X].
			\end{split}\end{equation} 
		
		Second, we show that $R^*(\trt,\cov)=R(\trt,\cov)$ under $\Mgest\cup\Mor$.
		Under $\Mgest$ where working models $f(\trt,Z\mid \cov;\alpha^{\trt,Z})$ and $R(\trt,\cov;\beta^R)$ are correctly specified, we have ${\alpha}^{\trt,Z}_{*}={\alpha}^{\trt,Z}$, $f^*(\trt,Z\mid \cov)=f(\trt,Z\mid \cov)$, and thus $E^*[g_1(\trt,Z,\cov)\mid \trt,\cov]=E[g_1(\trt,Z,\cov)\mid \trt,\cov]$ for any function $g_1(\trt,Z,\cov)$. Recall that $\hat{\beta}_{\textdr}^R$ solves $\mathbb{P}_n\Big\{U_{\beta^{R}}(\hat{\beta}_{\textdr}^{R})\Big\}=0$ with $\lim_{n\rightarrow\infty}\mathbb{P}_n\Big\{U_{\beta^{R}}(\hat{\beta}_{\textdr}^{R})\Big\}=E[U_{\beta^{R}}(\beta_{*}^{R})]$. Now consider $E[U_{\beta^{R}}(\beta^{R}_*)]\longmid_{\beta^{R}_*=\beta^{R}}$ under $\Mgest$ where $R(\trt,\cov;\beta^R)$ is correctly specified, i.e. $R(\trt,\cov)=R(\trt,\cov;\beta^R)$, we have
		\begin{equation*}\begin{split}
				&E[U_{\beta^{R}}(\beta^{R})]=E\{\Big[g_1(\trt,Z,\cov)-E^*[g_1(\trt,Z,\cov)\mid \trt,\cov]\Big]\Big[Y-E^*[Y\mid Z=0,\trt,\cov] -\\
				&R(\trt,\cov;\beta^{R})(W-E^*[W\mid Z=0,\trt,\cov])\Big]\}\\
				=&E\{\Big[g_1(\trt,Z,\cov)-E[g_1(\trt,Z,\cov)\mid \trt,\cov]\Big]\Big[\{R(\trt,\cov)-R(\trt,\cov;\beta^{R})\}\myeta^{W}_{\zdiff}(\trt,\cov)Z + \\
				&\{E[Y\mid Z=0,\trt,\cov]-E^*[Y\mid Z=0,\trt,\cov]\}+ \{E[W\mid Z=0,\trt,\cov]-E^*[W\mid Z=0,\trt,\cov]\}R(\trt,\cov;\beta^{R})\Big]\}\\
				=&E\{\Big[g_1(\trt,Z,\cov)-E[g_1(\trt,Z,\cov)\mid \trt,\cov]\Big]\Big[\{E[Y\mid Z=0,\trt,\cov]-E^*[Y\mid Z=0,\trt,\cov]\}+ \\
				&\{E[W\mid Z=0,\trt,\cov]-E^*[W\mid Z=0,\trt,\cov]\}R(\trt,\cov;\beta^{R})\Big]\}\\
				=&0
			\end{split}\end{equation*} 
		because
		$E\Big[\{g_1(\trt,Z,\cov)-E[g_1(\trt,Z,\cov)\mid \trt,\cov]\}h(A,X)\Big]=0$ for any function $h$. Thus, under $\Mgest$ where $R(\trt,\cov;\beta^R)$ is correctly specified, $\mathbb{P}_n\Big\{U_{\beta^{R}}(\hat{\beta}_{\textdr}^{R})\longmid_{\hat{\beta}_{\textdr}^{R}=\beta^R}\Big\}$ converges to zero, i.e. $\beta^R$ is a solution to the probability limit of $\mathbb{P}_n\Big\{U_{\beta^{R}}(\hat{\beta}_{\textdr}^{R})\Big\}=0$. Thus $\beta^R_*=\beta^R$ and $R^*(\trt,\cov)=R(\trt,\cov)$ under $\Mgest$.
		
		Similar arguments apply to the scenario under $\Mor$. Under $\Mor$ where working models $R(\trt,\cov;\beta^R)$, $E[Y\mid Z=0,\trt,\cov;\beta^Y]$ and $E[W\mid \trt,Z,\cov;\beta^W]$ are correctly specified, we have $E^*[Y\mid Z=0,\trt,\cov]=E[Y\mid Z=0,\trt,\cov]$ and by (\ref{underM3}) we have $E^*[W\mid \trt,Z,\cov]=E[W\mid \trt,Z,\cov]$. We again consider 
		\begin{equation*}\begin{split}
				&E[U_{\beta^{R}}(\beta^{R})]=E\{\Big[g_1(\trt,Z,\cov)-E^*[g_1(\trt,Z,\cov)\mid \trt,\cov]\Big]\Big[Y-E^*[Y\mid Z=0,\trt,\cov] -\\
				&R(\trt,\cov;\beta^{R})(W-E^*[W\mid Z=0,\trt,\cov])\Big]\}\\
				=&E\{\Big[g_1(\trt,Z,\cov)-E^*[g_1(\trt,Z,\cov)\mid \trt,\cov]\Big]\Big[\{R(\trt,\cov)-R(\trt,\cov;\beta^{R})\}\myeta^{W}_{\zdiff}(\trt,\cov)Z + \\
				&\{E[Y\mid Z=0,\trt,\cov]-E^*[Y\mid Z=0,\trt,\cov]\}+ \{E[W\mid Z=0,\trt,\cov]-E^*[W\mid Z=0,\trt,\cov]\}R(\trt,\cov;\beta^{R})\Big]\}\\
				=&E\{\Big[g_1(\trt,Z,\cov)-E^*[g_1(\trt,Z,\cov)\mid \trt,\cov]\Big]\Big[\{E[Y\mid Z=0,\trt,\cov]-E^*[Y\mid Z=0,\trt,\cov]\}+ \\
				&\{E[W\mid Z=0,\trt,\cov]-E^*[W\mid Z=0,\trt,\cov]\}R(\trt,\cov;\beta^{R})\Big]\}\\
				=&0
			\end{split}\end{equation*} 
		because $E^*[Y\mid Z=0,\trt,\cov]=E[Y\mid Z=0,\trt,\cov]$ and $E^*[W\mid Z=0,\trt,\cov]=E[W\mid Z=0,\trt,\cov]$. Therefore $\beta^R_*=\beta^R$ and $R^*(\trt,\cov)=R(\trt,\cov)$ under $\Mor$.
		
		\subsection{Proof that $E[\Delta^*_{\text{mr}}]=\Delta$ under $\mathcal{M}_{\text{union}}$}\label{subsec:can}
		Using the results in Section~\ref{supp:sec_E1}, we now show that $E[\Delta^*_{\text{mr}}]=\Delta$ under $\mathcal{M}_{\text{union}}$. To this end, we consider $E[\Delta^*_{\text{confounded}}]$, $E[\text{D}^*_{\bm{\delta^W_{\adiff}}(Z,\cov)}]$, and $E[\text{D}^*_{\R(1\!-\!\trt,\cov)}]$ under $\Mgest$, $\Mipw$, and $\Mor$ respectively. 
		
		Under $\Mgest$ where working models $f(\trt,Z\mid \cov;\alpha^{\trt,Z})$ and $R(\trt,\cov;\beta^R)$ are correctly specified, we have $f^*(\trt,Z\mid \cov)=f(\trt,Z\mid \cov)$ and $R^*(\trt,\cov)=R(\trt,\cov)$. First we consider
		\begin{equation*} 
			E[\Delta^*_{\text{confounded}}]=E\Big[\frac{2\trt-1}{f^*(\trt\mid Z,\cov)}\Big(E[Y\mid \trt,Z,\cov]-E^*[Y\mid \trt,Z,\cov]\Big)+E^*[Y\mid \trt\!=\!1,Z,\cov]-E^*[Y\mid \trt\!=\!0,Z,\cov]\Big].
			\end{equation*} 
		Note that for any function of $\trt$, $Z$ and $\cov$, denoted as $h(\trt,Z,\cov)$, 
		we have 
		\vspace{-0.05in}\begin{equation}E[\frac{2\trt-1}{f(\trt\mid Z,\cov)}h(\trt,Z,\cov)]=E[h(1,Z,\cov)-h(0,Z,\cov)].\label{eq:tool}\vspace{-0.05in}\end{equation}
		Accordingly, when $f^*(\trt\mid Z,\cov)=f(\trt\mid Z,\cov)$, we have
		\begin{equation*}\begin{split} 
				E[\Delta^*_{\text{confounded}}]=&E\Big[\frac{2\trt-1}{f(\trt\mid Z,\cov)}\Big(E[Y\mid \trt,Z,\cov]-E^*[Y\mid \trt,Z,\cov]\Big)+E^*[Y\mid \trt\!=\!1,Z,\cov]-E^*[Y\mid \trt\!=\!0,Z,\cov]\Big]\\
				=&E[\dY_{\adiff}(Z,\cov)-\delta^{Y^*}_{\adiff}(Z,\cov)+\delta^{Y^*}_{\adiff}(Z,\cov)]=\Delta_{\text{confounded}}.
			\end{split}\end{equation*}
		
		Second, consider
		\begin{equation*}\begin{split} 
				E[\text{D}^*_{\dW_{\adiff}(Z,\cov)}]=&E\Big[\frac{2\trt-1}{f^*(\trt\mid Z,\cov)}\Big(W-E^*[W\mid \trt,Z,\cov]\Big) \sum_\trt R^*(1\!-\!\trt,\cov)f^*(\trt\mid Z,\cov)\Big]\\
				=&E\Big[\frac{2\trt-1}{f^*(\trt\mid Z,\cov)}\Big(E[W\mid \trt,Z,\cov]-E^*[W\mid \trt,Z,\cov]\Big) \sum_\trt R^*(1\!-\!\trt,\cov)f^*(\trt\mid Z,\cov)\Big].
			\end{split}\end{equation*}
		When $f^*(\trt\mid Z,\cov)=f(\trt\mid Z,\cov)$, by Eq. (\ref{eq:tool}) we have
		\begin{equation*}\begin{split} 
				E[\text{D}^*_{\dW_{\adiff}(Z,\cov)}]=&E\Big[\frac{2\trt-1}{f(\trt\mid Z,\cov)}\Big(E[W\mid \trt,Z,\cov]-E^*[W\mid \trt,Z,\cov]\Big) E[R^*(1\!-\!\trt,\cov) \mid Z,\cov]\Big]\\
				=&E\Big[\Big(\dW_{\adiff}(Z,\cov)-\delta^{W^*}_{\adiff}(Z,\cov)\Big) E[R^*(1\!-\!\trt,\cov) \mid Z,\cov]\Big].
			\end{split}\end{equation*}
		Because we also have $R^*(\trt,\cov)=R(\trt,\cov)$,
		\begin{equation*}\begin{split} 
				&E[\text{D}^*_{\dW_{\adiff}(Z,\cov)}+R^*(1\!-\!\trt,\cov) \delta^{W^*}_{\adiff}(Z,\cov)]\\
				=&E\Big[\Big(\dW_{\adiff}(Z,\cov)-\delta^{W^*}_{\adiff}(Z,\cov)\Big) E[R(1\!-\!\trt,\cov) \mid Z,\cov]+R(1\!-\!\trt,\cov) \delta^{W^*}_{\adiff}(Z,\cov)\Big]=\Delta_{\text{bias}}.
			\end{split}\end{equation*}
		
		Third, consider 
		\begin{equation*}\begin{split} 
				E[\text{D}^*_{R(1\!-\!\trt,\cov)}]=&E\Big\{\frac{2Z-1}{f^*(Z\mid \trt,\cov)} \frac{1}{\myeta^{W^*}_{\zdiff}(\trt,\cov)} \Big(\sum_Z \delta^{W^*}_{\adiff}(Z,\cov)f^*(Z\mid  1-\trt,\cov)\frac{f^*(1-\trt\mid \cov)}{f^*(\trt\mid \cov)}\Big)\\
				&\Big[Y-E^*[Y\mid Z=0,\trt,\cov] - R^*(\trt,\cov)\Big(W-E^*[W\mid Z=0,\trt,\cov]\Big)\Big]\Big\}\\
				=&E\Big\{\frac{2Z-1}{f^*(Z\mid \trt,\cov)} \frac{1}{\myeta^{W^*}_{\zdiff}(\trt,\cov)} \Big(\sum_Z \delta^{W^*}_{\adiff}(Z,\cov)f^*(Z\mid  1-\trt,\cov)\frac{f^*(1-\trt\mid \cov)}{f^*(\trt\mid \cov)}\Big)\\
				&\Big[\{R(\trt,\cov)-R^*(\trt,\cov)\}\myeta^{W}_{\zdiff}(\trt,\cov)Z + \{E[Y\mid Z=0,\trt,\cov]-E^*[Y\mid Z=0,\trt,\cov]\} \\
				&+ \{E[W\mid Z=0,\trt,\cov]-E^*[W\mid Z=0,\trt,\cov]\}R^*(\trt,\cov)\Big]\Big\}.
			\end{split}\end{equation*}
		When $f^*(Z\mid \trt,\cov)=f(Z\mid \trt,\cov)$, by similar argument as Eq. (\ref{eq:tool}) we have
		\begin{equation}
			E[\text{D}^*_{R(1\!-\!\trt,\cov)}]=E\Big\{\frac{\myeta^{W}_{\zdiff}(\trt,\cov)}{\myeta^{W^*}_{\zdiff}(\trt,\cov)} [R(\trt,\cov)-R^*(\trt,\cov)]\cdot[\sum_Z \delta^{W^*}_{\adiff}(Z,\cov)f^*(Z\mid  1-\trt,\cov)]\frac{f^*(1-\trt\mid \cov)}{f^*(\trt\mid \cov)}\Big\}\label{need_change_to_A0}.
			\end{equation}
		We can see that when $R^*(\trt,\cov)=R(\trt,\cov)$, $E[\text{D}^*_{R(1\!-\!\trt,\cov)}]=0$.
		
		In summary, under $\Mgest$, we have
		\begin{equation*}\begin{split} 
				E[\Delta^*_{\text{mr}}]=&E[\Delta^*_{\text{confounded}}]-\{E[\text{D}^*_{\dW_{\adiff}(Z,\cov)}+R^*(1\!-\!\trt,\cov) \delta^{W^*}_{\adiff}(Z,\cov)]+E[\text{D}^*_{R(1\!-\!\trt,\cov)}]\}\\
				=&\Delta_{\text{confounded}}-\{\Delta_{\text{bias}}+0\}=\Delta
			\end{split}\end{equation*}

		Under $\Mipw$ where $f(\trt,Z\mid \cov;\alpha^{\trt,Z})$, $\myeta^W_{\zdiff}(\trt,\cov;\beta^{\WZ})$ and $\dW_{\adiff}(Z,\cov;\beta^{\WA})$ are correctly specified, we have $f^*(\trt,Z\mid \cov)=f(\trt,Z\mid \cov)$, $\myeta^{W^*}_{\zdiff}(\trt,\cov)=\myeta^W_{\zdiff}(\trt,\cov)$, and $\myeta^{W^*}_{\adiff}(\trt,\cov)=\myeta^W_{\adiff}(\trt,\cov)$. Particularly, $f^*(A\mid X)=f(A\mid X)$. First we consider
		\begin{equation*} 
			E[\Delta^*_{\text{confounded}}]=E\Big[\frac{2\trt-1}{f^*(\trt\mid Z,\cov)}\Big(E[Y\mid \trt,Z,\cov]-E^*[Y\mid \trt,Z,\cov]\Big)+E^*[Y\mid \trt\!=\!1,Z,\cov]-E^*[Y\mid \trt\!=\!0,Z,\cov]\Big].
			\end{equation*} 
		When $f^*(\trt\mid Z,\cov)=f(\trt\mid Z,\cov)$, by Eq. (\ref{eq:tool}) we have
		\begin{equation*}\begin{split} 
				E[\Delta^*_{\text{confounded}}]=&E\Big[\frac{2\trt-1}{f(\trt\mid Z,\cov)}\Big(E[Y\mid \trt,Z,\cov]-E^*[Y\mid \trt,Z,\cov]\Big)+E^*[Y\mid \trt\!=\!1,Z,\cov]-E^*[Y\mid \trt\!=\!0,Z,\cov]\Big]\\
				=&E[\dY_{\adiff}(Z,\cov)-\delta^{Y^*}_{\adiff}(Z,\cov)+\delta^{Y^*}_{\adiff}(Z,\cov)]=\Delta_{\text{confounded}}.
			\end{split}\end{equation*}
		
		Second, consider
		\begin{equation*}\begin{split} 
				E[\text{D}^*_{\dW_{\adiff}(Z,\cov)}]=&E\Big[\frac{2\trt-1}{f^*(\trt\mid Z,\cov)}\Big(W-E^*[W\mid \trt,Z,\cov]\Big) \sum_\trt R^*(1\!-\!\trt,\cov)f^*(\trt\mid Z,\cov)\Big]\\
				=&E\Big[\frac{2\trt-1}{f^*(\trt\mid Z,\cov)}\Big(E[W\mid \trt,Z,\cov]-E^*[W\mid \trt,Z,\cov]\Big) \sum_\trt R^*(1\!-\!\trt,\cov)f^*(\trt\mid Z,\cov)\Big].
			\end{split}\end{equation*}
		When $f^*(\trt\mid Z,\cov)=f(\trt\mid Z,\cov)$, by Eq. (\ref{eq:tool}) we have
		\begin{equation*}\begin{split} 
				E[\text{D}^*_{\dW_{\adiff}(Z,\cov)}]=&E\Big[\frac{2\trt-1}{f(\trt\mid Z,\cov)}\Big(E[W\mid \trt,Z,\cov]-E^*[W\mid \trt,Z,\cov]\Big) E[R^*(1\!-\!\trt,\cov) \mid Z,\cov]\Big]\\
				=&E\Big[\Big(\dW_{\adiff}(Z,\cov)-\delta^{W^*}_{\adiff}(Z,\cov)\Big) E[R^*(1\!-\!\trt,\cov) \mid Z,\cov]\Big]\\
				=&0
			\end{split}\end{equation*}
		because $\myeta^{W^*}_{\adiff}(\trt,\cov)=\myeta^W_{\adiff}(\trt,\cov)$.
		
		Third, consider 
		\begin{equation*}\begin{split} 
				E[\text{D}^*_{R(1\!-\!\trt,\cov)}]=&E\Big\{\frac{2Z-1}{f^*(Z\mid \trt,\cov)} \frac{1}{\myeta^{W^*}_{\zdiff}(\trt,\cov)} \Big(\sum_Z \delta^{W^*}_{\adiff}(Z,\cov)f^*(Z\mid  1-\trt,\cov)\frac{f^*(1-\trt\mid \cov)}{f^*(\trt\mid \cov)}\Big)\\
				&\Big[Y-E^*[Y\mid Z=0,\trt,\cov] - R^*(\trt,\cov)\Big(W-E^*[W\mid Z=0,\trt,\cov]\Big)\Big]\Big\}\\
				=&E\Big\{\frac{2Z-1}{f^*(Z\mid \trt,\cov)} \frac{1}{\myeta^{W^*}_{\zdiff}(\trt,\cov)} \Big(\sum_Z \delta^{W^*}_{\adiff}(Z,\cov)f^*(Z\mid  1-\trt,\cov)\frac{f^*(1-\trt\mid \cov)}{f^*(\trt\mid \cov)}\Big)\\
				&\Big[\{R(\trt,\cov)-R^*(\trt,\cov)\}\myeta^{W}_{\zdiff}(\trt,\cov)Z + \{E[Y\mid Z=0,\trt,\cov]-E^*[Y\mid Z=0,\trt,\cov]\} \\
				&+ \{E[W\mid Z=0,\trt,\cov]-E^*[W\mid Z=0,\trt,\cov]\}R^*(\trt,\cov)\Big]\Big\}.
			\end{split}\end{equation*}
		When $f^*(Z\mid \trt,\cov)=f(Z\mid \trt,\cov)$, by similar argument as Eq. (\ref{eq:tool}) we have
		\begin{equation}
			E[\text{D}^*_{R(1\!-\!\trt,\cov)}]=E\Big\{\frac{\myeta^{W}_{\zdiff}(\trt,\cov)}{\myeta^{W^*}_{\zdiff}(\trt,\cov)} [R(\trt,\cov)-R^*(\trt,\cov)]\cdot[\sum_Z \delta^{W^*}_{\adiff}(Z,\cov)f^*(Z\mid  1-\trt,\cov)]\frac{f^*(1-\trt\mid \cov)}{f^*(\trt\mid \cov)}\Big\}\label{need_change_to_A}.
			\end{equation}
		Note that when the model for $f(\trt\mid \cov)$ is correctly specified, i.e., $f^*(\trt\mid \cov)=f(\trt\mid \cov)$, in Appendix \ref{appendix:change_measure} we show that for any function $h(Y,W,\trt,Z,\cov )$, we have
		\begin{equation*}
			E[h(Y,W,\trt,Z,\cov ) \frac{f(1-\trt\mid \cov)}{f(\trt\mid \cov)}]=E[h(Y,W,Z,1\!-\!\trt,\cov )].
			\end{equation*} 
		Let $h(Y,W,\trt,Z,\cov )=\frac{\myeta^{W}_{\zdiff}(\trt,\cov)}{\myeta^{W^*}_{\zdiff}(\trt,\cov)} [R(\trt,\cov)-R^*(\trt,\cov)]\cdot[\sum_Z \delta^{W^*}_{\adiff}(Z,\cov)f^*(Z\mid  1-\trt,\cov)]$, then Eq. (\ref{need_change_to_A}) is equivalent to
		\begin{equation*}
			E[\text{D}^*_{R(1\!-\!\trt,\cov)}]=E\Big\{\frac{\myeta^{W}_{\zdiff}(1\!-\!\trt,\cov)}{\myeta^{W^*}_{\zdiff}(1\!-\!\trt,\cov)} [R(1\!-\!\trt,\cov)-R^*(1\!-\!\trt,\cov)]\cdot[\sum_Z \delta^{W^*}_{\adiff}(Z,\cov)f^*(Z\mid  1-(1\!-\!\trt),\cov)]\Big\}.
			\end{equation*} 
		In this case, because we also have that $\myeta^{W^*}_{\zdiff}(\trt,\cov)=\myeta^W_{\zdiff}(\trt,\cov)$, and $\myeta^{W^*}_{\adiff}(\trt,\cov)=\myeta^W_{\adiff}(\trt,\cov)$, 
		\begin{equation*}\begin{split}
				&E[\text{D}^*_{R(1\!-\!\trt,\cov)}+R^*(1\!-\!\trt,\cov) \delta^{W^*}_{\adiff}(Z,\cov)]\\
				=&E\Big\{[R(1\!-\!\trt,\cov)-R^*(1\!-\!\trt,\cov)]E[\delta^{W}_{\adiff}(Z,\cov) \mid  \trt,\cov]+R^*(1\!-\!\trt,\cov) \delta^{W}_{\adiff}(Z,\cov)\Big\}=\Delta_{\text{bias}}.
			\end{split}\end{equation*}

		In summary, under $\Mipw$, we have
		\begin{equation*}\begin{split} 
				E[\Delta^*_{\text{mr}}]=&E[\Delta^*_{\text{confounded}}]-\{E[\text{D}^*_{\dW_{\adiff}(Z,\cov)}]+E[\text{D}^*_{R(1\!-\!\trt,\cov)}+R^*(1\!-\!\trt,\cov) \delta^{W^*}_{\adiff}(Z,\cov)]\}\\
				=&\Delta_{\text{confounded}}-\{0+\Delta_{\text{bias}}\}=\Delta
			\end{split}\end{equation*}

		Under $\Mor$ where $R(\trt,\cov;\beta^R)$, $E[Y\mid Z=0,\trt,\cov;\beta^Y]$, $\myeta^W_{\zdiff}(\trt,\cov;\beta^{\WZ})$, $\dW_{\adiff}(Z,\cov;\beta^{\WA})$, and $E[W\mid \trt=0,Z=0,\cov;\beta^W]$ are correctly specified, we have $R^*(\trt,\cov)=R(\trt,\cov)$, $E^*[Y\mid Z=0,\trt,\cov]=E[Y\mid Z=0,\trt,\cov]$, $\myeta^{W^*}_{\zdiff}(\trt,\cov)=\myeta^W_{\zdiff}(\trt,\cov)$, $\delta^{W^*}_{\adiff}(Z,\cov)=\delta^{W}_{\adiff}(Z,\cov)$, and $E^*[W\mid \trt=0,Z=0,\cov]=E[W\mid \trt=0,Z=0,\cov]$. First we consider
		\begin{equation*} 
			E[\Delta^*_{\text{confounded}}]=E\Big[\frac{2\trt-1}{f^*(\trt\mid Z,\cov)}\Big(E[Y\mid \trt,Z,\cov]-E^*[Y\mid \trt,Z,\cov]\Big)+E^*[Y\mid \trt\!=\!1,Z,\cov]-E^*[Y\mid \trt\!=\!0,Z,\cov]\Big].
			\end{equation*} 
		Note that
		\begin{equation*}\begin{split}
				E^*[Y\mid Z,\trt,\cov] = &E^*[Y\mid Z=0,\trt,\cov]+R^*(\trt,\cov)\myeta^{W^*}_{\zdiff}(\trt,\cov)Z\\
				=&E[Y\mid Z=0,\trt,\cov]+R(\trt,\cov)\myeta^{W}_{\zdiff}(\trt,\cov)Z=E[Y\mid Z,\trt,\cov],
			\end{split}\end{equation*}
		therefore we have 
		\begin{equation*}\begin{split} 
				E[\Delta^*_{\text{confounded}}]=&E\Big[\frac{2\trt-1}{f(\trt\mid Z,\cov)}\Big(E[Y\mid \trt,Z,\cov]-E[Y\mid \trt,Z,\cov]\Big)+E[Y\mid \trt\!=\!1,Z,\cov]-E[Y\mid \trt\!=\!0,Z,\cov]\Big]\\
				=&E\{E[Y\mid \trt\!=\!1,Z,\cov]-E[Y\mid \trt\!=\!0,Z,\cov]\}=\Delta_{\text{confounded}}.
			\end{split}\end{equation*}
		
		Second, consider
		\begin{equation*}\begin{split} 
				E[\text{D}^*_{\dW_{\adiff}(Z,\cov)}]=&E\Big[\frac{2\trt-1}{f^*(\trt\mid Z,\cov)}\Big(W-E^*[W\mid \trt,Z,\cov]\Big) \sum_\trt R^*(1\!-\!\trt,\cov)f^*(\trt\mid Z,\cov)\Big]\\
				=&E\Big[\frac{2\trt-1}{f^*(\trt\mid Z,\cov)}\Big(E[W\mid \trt,Z,\cov]-E^*[W\mid \trt,Z,\cov]\Big) \sum_\trt R^*(1\!-\!\trt,\cov)f^*(\trt\mid Z,\cov)\Big]\\
				=&0
			\end{split}\end{equation*}
		because $E^*[W\mid \trt,Z,\cov]=E[W\mid \trt,Z,\cov]$ by (\ref{underM3}).
		
		Third, consider 
		\begin{equation*}\begin{split} 
				E[\text{D}^*_{R(1\!-\!\trt,\cov)}]=&E\Big\{\frac{2Z-1}{f^*(Z\mid \trt,\cov)} \frac{1}{\myeta^{W^*}_{\zdiff}(\trt,\cov)} \Big(\sum_Z \delta^{W^*}_{\adiff}(Z,\cov)f^*(Z\mid  1-\trt,\cov)\frac{f^*(1-\trt\mid \cov)}{f^*(\trt\mid \cov)}\Big)\\
				&\Big[Y-E^*[Y\mid Z=0,\trt,\cov] - R^*(\trt,\cov)\Big(W-E^*[W\mid Z=0,\trt,\cov]\Big)\Big]\Big\}\\
				=&E\Big\{\frac{2Z-1}{f^*(Z\mid \trt,\cov)} \frac{1}{\myeta^{W^*}_{\zdiff}(\trt,\cov)} \Big(\sum_Z \delta^{W^*}_{\adiff}(Z,\cov)f^*(Z\mid  1-\trt,\cov)\frac{f^*(1-\trt\mid \cov)}{f^*(\trt\mid \cov)}\Big)\\
				&\Big[\{R(\trt,\cov)-R^*(\trt,\cov)\}\myeta^{W}_{\zdiff}(\trt,\cov)Z + \{E[Y\mid Z=0,\trt,\cov]-E^*[Y\mid Z=0,\trt,\cov]\} \\
				&+ \{E[W\mid Z=0,\trt,\cov]-E^*[W\mid Z=0,\trt,\cov]\}R^*(\trt,\cov)\Big]\Big\}\\
				=&0
			\end{split}\end{equation*}
		because $R^*(\trt,\cov)=R(\trt,\cov)$, $E^*[Y\mid Z=0,\trt,\cov]=E[Y\mid Z=0,\trt,\cov]$, $\delta^{W^*}_{\adiff}(Z=0,\cov)=\delta^{W^*}_{\adiff}(Z=0,\cov)$, and $E^*[W\mid \trt=0,Z=0,\cov]=E[W\mid \trt=0,Z=0,\cov]$
		
		Thus, under $\Mor$, we have
		\begin{equation*}\begin{split} 
				E[\Delta^*_{\text{mr}}]=&E[\Delta^*_{\text{confounded}}]-\{E[\text{D}^*_{\dW_{\adiff}(Z,\cov)}]+E[\text{D}^*_{R(1\!-\!\trt,\cov)}]+E[R^*(1\!-\!\trt,\cov) \delta^{W^*}_{\adiff}(Z,\cov)]\}\\
				=&E[\Delta^*_{\text{confounded}}]-\{E[\text{D}^*_{\dW_{\adiff}(Z,\cov)}]+E[\text{D}^*_{R(1\!-\!\trt,\cov)}]+E[R(1\!-\!\trt,\cov) \delta^{W}_{\adiff}(Z,\cov)]\}\\
				=&\Delta_{\text{confounded}}-\{0+0+\Delta_{\text{bias}}\}=\Delta
			\end{split}\end{equation*}
		
		In summary, $E[\Delta^*_{\text{mr}}]=\Delta$ under $\mathcal{M}_{\text{union}}=\Mgest\cup \Mipw\cup \Mor$.

		\subsection{Variance estimator}\label{appendix:var_delta}
		Here we provide a consistent estimator of the asymptotic variance $\sigma^2_{\Delta}(\Delta,\theta^*)$ by writing our problem in the form of standard M-estimation. Recall that $\hat{\theta}$ are the estimated nuisance parameters that solve $\mathbb{P}_n\left\{U_{\theta}(O;\hat{\theta})\right\}=0$, and $\hat{\Delta}_{\text{mr}}$ is the proposed multiply robust estimator that solves $\mathbb{P}_n\Big\{EIF_{\Delta}(O;\hat{\Delta}_{\text{mr}},\hat{\theta})\Big\}=0$. Let $\gamma=(\theta,\Delta)\transpose $ denote the vector of all parameters of dimension $k$, $\psi(\gamma)=\{U_{\theta}(O;\theta)\transpose ,EIF_{\Delta}(O;\Delta,\theta)\}\transpose $, and let $G_n(\gamma)=\mathbb{P}_n\Big\{\psi(\gamma)\Big\}=\frac{1}{n}\sum_{i=1}^n\psi(O_i;\gamma)$ 
		denote a $k\times 1$ vector of estimating functions where the $k$-th element is the estimating function for $\Delta$, then $\hat{\gamma}=(\hat{\theta},\hat{\Delta}_{\text{mr}})$ is the solution to the estimating equations $G_n(\gamma)=0$.
		Let $A_n(\hat{\gamma})=-\frac{\partial G_n(\gamma)}{\partial \gamma\transpose } \longmid_{\gamma=\hat{\gamma}}=-\frac{1}{n}\sum_{i=1}^n\left\{\frac{\partial}{\partial \gamma\transpose }\psi(O_i;\hat{\gamma})\right\}$ and $B_n(\hat{\gamma})=\frac{1}{n}\sum_{i=1}^n\psi(O_i;\hat{\gamma})\psi(O_i;\hat{\gamma})\transpose $. We define the empirical sandwich estimator as follows
		\begin{equation*}
			\widehat{Var}(\hat{\gamma})=A_n(\hat{\gamma})^{-1}B_n(\hat{\gamma})\left(A_n(\hat{\gamma})^{-1}\right)\transpose .
		\end{equation*}
		Then a consistent estimator for the asymptotic variance of $\hat{\Delta}_{\text{mr}}$ corresponds to $\widehat{Var}(\hat{\gamma})_{k,k}$, the $(k,k)$-th element of $\widehat{Var}(\hat{\gamma})$.
		In practice, one can also apply the nonparametric bootstrap to estimate the variance.
		
	\end{proof}

	\newpage
	\section{Proof of Theorem \ref{thm_cate} (efficient influence function in $\mathcal{M}_{\text{nonpar}}$ for polytomous case)}\label{appendix:if2}
	\begin{proof}
		Let $f(Y,W,\trt,Z,\cov;\pathwiseparam)$ denote a one-dimensional regular parametric submodel of $\mathcal{M}_{\text{nonpar}}$ indexed by $\pathwiseparam$, under which $\Delta_\pathwiseparam=E_\pathwiseparam[\dY_{\adiff,\pathwiseparam}(Z,\cov)]-E_\pathwiseparam[\R_{\pathwiseparam}(1-\trt,\cov)\bm{\delta^{W}}_{\bm{\adiff},\pathwiseparam}(Z,\cov)]$.
		The efficient influence function in $\mathcal{M}_{\text{nonpar}}$ is defined as the unique mean zero, finite variance random variable $D$ satisfying
		\begin{equation*}
			\dt \Delta_\pathwiseparam = E[D\cdot S(Y,W,\trt,Z,\cov)],
			\end{equation*} 
		where $S(\cdot)$ is the score function of the path $f(Y,W,\trt,Z,\cov;\pathwiseparam)$ at $\pathwiseparam=0$, and $\dt \Delta_\pathwiseparam$ is the pathwise derivative of $\Delta$.
		To find $D$, we derive the following pathwise derivatives. First of all, we have the same result as Eq. (\ref{eq:pathwisederivative1}) for the pathwise derivative of $\dY_{\adiff}(Z,\cov)=E[Y\mid Z,\trt\!=\!1,\cov]-E[Y\mid Z,\trt~=~0,\cov]$, which is
		{{\begin{equation*}\begin{split} 
						&\dt \dY_{\adiff,\pathwiseparam}(Z,\cov)\\
						=& E\Big[\frac{2\trt-1}{f(\trt\mid Z,\cov)}(Y-E[Y\mid \trt,Z,\cov])S(Y,\trt,Z,\cov)\longmid Z,\cov\Big]\\
						=& E\Big[\frac{2\trt-1}{f(\trt\mid Z,\cov)}(Y-\dY_{\adiff}(Z,\cov)\trt-E[Y\mid \trt\!=\!0,Z,\cov])S(Y,\trt,Z,\cov)\longmid Z,\cov\Big].
					\end{split}\end{equation*}}}
		Accordingly, the pathwise derivative of $E[\dY_{\adiff}(Z,\cov)]$ is given by
		{{\begin{equation*}\begin{split} 
						&\dt E_\pathwiseparam[\dY_{\adiff,\pathwiseparam}(Z,\cov)]\\
						=& E[\dt \dY_{\adiff}(Z,\cov) ] + E[\dY_{\adiff}(Z,\cov) S(Z,\cov)]\\
						=& E[\frac{2\trt-1}{f(\trt\mid Z,\cov)}(Y-E[Y\mid \trt,Z,\cov])S(Y,\trt,Z,\cov) ] + E[\dY_{\adiff}(Z,\cov) S(Z,\cov)]\\
						=&E\{[\frac{2\trt-1}{f(\trt\mid Z,\cov)}(Y-\dY_{\adiff}(Z,\cov)\trt-E[Y\mid \trt\!=\!0,Z,\cov])+\dY_{\adiff}(Z,\cov)]\cdot S(Y,\trt,Z,\cov) \}.
					\end{split}\end{equation*}}}
		Second, for $\delta^{w_i}_{\adiff}(Z,\cov)=E[\mathbbm{1}(W=w_i)\mid \trt\!=\!1,Z,\cov]-E[\mathbbm{1}(W=w_i)\mid \trt\!=\!0,Z,\cov]$ we have
		{{\begin{equation*}\begin{split} 
						&\dt \delta^{W=w_i}_\pathwiseparam(Z,\cov)\\
						=& E\Big[\frac{2\trt-1}{f(\trt\mid Z,\cov)}(\mathbbm{1}(W=w_i)-E[\mathbbm{1}(W=w_i)\mid \trt,Z,\cov])S(W,\trt,Z,\cov)\longmid Z,\cov\Big]\\
						=& E\Big[\frac{2\trt-1}{f(\trt\mid Z,\cov)}(\mathbbm{1}(W=w_i)-\delta^{w_i}_{\adiff}(Z,\cov)\trt-E[\mathbbm{1}(W=w_i)\mid \trt\!=\!0,Z,\cov])S(W,\trt,Z,\cov)\longmid Z,\cov\Big].
					\end{split}\end{equation*}}}
		Generalizing from $\mathbbm{1}(W=w_i)$ to vector $\IW$, we have
		{{\vspace{-0.05in}\begin{equation}\begin{split}
						&\dt \bm{\delta^{W}}_{\bm{\adiff},\pathwiseparam}(Z,\cov)_{k\times 1}\\
						=& E\Big[\frac{2\trt-1}{f(\trt\mid Z,\cov)}(\IW-E[\IW\mid \trt,Z,\cov])_{k\times 1}S(W,\trt,Z,\cov)\longmid Z,\cov\Big]\\
						=& E\Big[\frac{2\trt-1}{f(\trt\mid Z,\cov)}(\IW-\deltaWAvec(Z,\cov)\trt-E[\IW\mid \trt\!=\!0,Z,\cov])_{k\times 1}S(W,\trt,Z,\cov)\longmid Z,\cov\Big].\label{eq:pathwisederivative2_cate}
					\end{split}\vspace{-0.05in}\end{equation}}}
		Third, for $\myeta^{Y}_{\zjdiff}(1\!-\!\trt,\cov)=E[Y\mid Z\!=\!z_j,1\!-\!\trt,\cov]-E[Y\mid Z\!=\!z_0,1\!-\!\trt,\cov]$, we have
		{{\begin{equation*}\begin{split}
						&\dt \myeta^{Y}_{\zjdiff,\pathwiseparam}(1\!-\!\trt,\cov)\\
						=& E\Big[\Big(\frac{I(Z\!=\!z_j)}{f(Z\!=\!z_j\mid 1\!-\!\trt,\cov)}-\frac{\mathbbm{1}(Z=z_0)}{f(Z=z_0\mid 1\!-\!\trt,\cov)}\Big)\Big(Y-E[Y\mid Z,1\!-\!\trt,\cov]\Big)S(Y,Z,1\!-\!\trt,\cov)\longmid 1\!-\!\ttrt\Big].
					\end{split}\end{equation*}}}
		Generalizing from $\zjdiff$ to vector $\bigzdiff$, we have
		{{\vspace{-0.05in}\begin{equation}\begin{split}
						&\dt\bm{\myeta^{Y}}_{\bm{Z},\pathwiseparam}(1\!-\!\trt,\cov)\transpose _{1\times k}\\
						=&E\Big\{[\Big(\frac{\mathbbm{1}(Z=z_k)}{f(Z=z_k\mid 1\!-\!\trt,\cov)}-\frac{\mathbbm{1}(Z=z_0)}{f(Z=z_0\mid 1\!-\!\trt,\cov)}\Big),\Big(\frac{\mathbbm{1}(Z=z_k-1)}{f(Z=z_k-1\mid 1\!-\!\trt,\cov)}-\frac{\mathbbm{1}(Z=z_0)}{f(Z=z_0\mid 1\!-\!\trt,\cov)}\Big),\\
						&\dots,\Big(\frac{\mathbbm{1}(Z=z_1)}{f(Z=z_1\mid 1\!-\!\trt,\cov)}-\frac{\mathbbm{1}(Z=z_0)}{f(Z=z_0\mid 1\!-\!\trt,\cov)}\Big)]_{1\times k}\Big(Y-E[Y\mid Z,1\!-\!\trt,\cov]\Big)S(Y,Z,1\!-\!\trt,\cov)\longmid 1\!-\!\trt,\cov\Big\}\\
						=&E\Big\{\IfZ\transpose _{1\times k}\Big(Y-E[Y\mid Z,1\!-\!\trt,\cov]\Big)S(Y,Z,1\!-\!\trt,\cov)\longmid 1\!-\!\trt,\cov\Big\},
					\end{split}\vspace{-0.05in}\end{equation}
		}}
		where $\IfZ=\{\frac{\mathbbm{1}(Z=z_1)}{f(Z=z_1\mid  \trt,\cov)}-\frac{\mathbbm{1}(Z=z_0)}{f(Z=z_0\mid  \trt,\cov)},\frac{\mathbbm{1}(Z=z_2)}{f(Z=z_2\mid  \trt,\cov)}-\frac{\mathbbm{1}(Z=z_0)}{f(Z=z_0\mid  \trt,\cov)},\dots,\frac{\mathbbm{1}(Z=z_k)}{f(Z=z_k\mid  \trt,\cov)}-\frac{\mathbbm{1}(Z=z_0)}{f(Z=z_0\mid  \trt,\cov)}\}\transpose $ denote a $k\times 1$ vector generalizing $\frac{2Z-1}{f(Z\mid \trt,\cov)}$ in the binary case, with $\IfZj=\frac{\mathbbm{1}(Z=z_j)}{f(Z=z_j\mid  \trt,\cov)}-\frac{\mathbbm{1}(Z=z_0)}{f(Z=z_0\mid  \trt,\cov)}$.
		
		Forth, for $\myeta^{w_i}_{\zjdiff}(1\!-\!\trt,\cov)=E[\mathbbm{1}(W=w_i)\mid Z\!=\!z_j,1\!-\!\trt,\cov]-E[\mathbbm{1}(W=w_i)\mid Z\!=\!z_0,1\!-\!\trt,\cov]$, we have
		\vspace{-0.05in}\begin{equation}\begin{split}
				&\dt \myeta^{w_i}_{\zjdiff,\pathwiseparam}(1\!-\!\trt,\cov)\\
				=& E\Big[\Big(\frac{\mathbbm{1}(Z=z_j)}{f(Z=z_j\mid 1\!-\!\trt,\cov)}-\frac{\mathbbm{1}(Z=z_0)}{f(Z=z_0\mid 1\!-\!\trt,\cov)}\Big)\Big(\mathbbm{1}(W=w_i)-E[\mathbbm{1}(W=w_i)\mid Z,1\!-\!\trt,\cov]\Big)\\
				&S(W,Z,1\!-\!\trt,\cov)\longmid 1\!-\!\trt,\cov\Big] .
			\end{split}\vspace{-0.05in}\end{equation}
		Generalizing to column vector $\IW$, we have
		\vspace{-0.05in}\begin{equation}\begin{split}
				&\dt \bm{\myeta^{W}}_{\zjdiff,\pathwiseparam}(1\!-\!\trt,\cov)_{k \times 1}\\
				=& E\Big[\Big(\frac{\mathbbm{1}(Z=z_j)}{f(Z=z_j\mid 1\!-\!\trt,\cov)}-\frac{\mathbbm{1}(Z=z_0)}{f(Z=z_0\mid 1\!-\!\trt,\cov)}\Big)\Big(\IW-E[\IW\mid Z,1\!-\!\trt,\cov]\Big)_{k\times 1}\\
				&S(W,Z,1\!-\!\trt,\cov)\longmid 1\!-\!\trt,\cov\Big] .
			\end{split}\vspace{-0.05in}\end{equation}
		Generalizing to row vector $\bigzdiff$, we have
		\vspace{-0.05in}\begin{equation}\begin{split}
				&\dt \bm{\myeta}^{w_i}_{\bm{Z},\pathwiseparam}(1\!-\!\trt,\cov)_{1 \times k}\\
				=& E\Big[\IfZ\transpose _{1\times k}\Big(\mathbbm{1}(W=w_i)-E[\mathbbm{1}(W=w_i)\mid Z,1\!-\!\trt,\cov]\Big)S(W,Z,1\!-\!\trt,\cov)\longmid 1\!-\!\trt,\cov\Big].
			\end{split}\vspace{-0.05in}\end{equation}
		Generalizing to a matrix, we have
		\vspace{-0.05in}\begin{equation}\begin{split}
				&\dt \bm{\myeta^{W}}_{\bm{Z},\pathwiseparam}(1\!-\!\trt,\cov)_{k \times k}\\
				=& E\Big[\Big(\IW-E[\IW\mid Z,1\!-\!\trt,\cov]\Big)_{k\times 1}\IfZ\transpose _{1\times k}S(W,Z,1\!-\!\trt,\cov)\longmid 1\!-\!\trt,\cov\Big] .
			\end{split}\vspace{-0.05in}\end{equation}
		From the above, we finally have
		{\small{\vspace{-0.05in}\begin{equation}\begin{split}
						&\dt \R_\pathwiseparam(1\!-\!\trt,\cov)_{1\times k} = \dt\bm{\myeta^{Y}}_{\bm{Z},\pathwiseparam}(1\!-\!\trt,\cov)\transpose _{1 \times k}\bm{\myeta^{W}}_{\bm{Z},\pathwiseparam}(1\!-\!\trt,\cov)\inv_{k \times k}\\
						=& \dt\bm{\myeta^{Y}}_{\bm{Z},\pathwiseparam}(1\!-\!\trt,\cov)\transpose _{1 \times k}\bm{\myeta^{W}}_{\bm{Z}}(1\!-\!\trt,\cov)\inv_{k \times k}+ \bm{\myeta^{Y}}_{\bm{Z}}(1\!-\!\trt,\cov)\transpose _{1 \times k}\dt\bm{\myeta^{W}}_{\bm{Z},\pathwiseparam}(1\!-\!\trt,\cov)\inv_{k \times k}\\\\
						=& E\Big[\IfZ\transpose _{1\times k}\Big(Y-E[Y\mid Z,1\!-\!\trt,\cov]\Big)S(Y,Z,1\!-\!\trt,\cov)\longmid 1\!-\!\trt,\cov\Big] \myetaWZvec(1\!-\!\trt,\cov)\inv-\\
						&\R(1\!-\!\trt,\cov)_{1\times k}\cdot E\Big[\Big(\IW-E[\IW\mid Z,1\!-\!\trt,\cov]\Big)_{k\times 1}\IfZ\transpose _{1\times k}S(W,Z,1\!-\!\trt,\cov)\Big]\longmid 1\!-\!\trt,\cov\Big] 
						\myetaWZvec(1\!-\!\trt,\cov)\inv\\
						=& E\Big[\Big(Y-E[Y\mid Z,1\!-\!\trt,\cov]\Big)\IfZ\transpose _{1\times k}\myetaWZvec(1\!-\!\trt,\cov)\inv S(Y,Z,1\!-\!\trt,\cov) \\
						-& \R(1\!-\!\trt,\cov)_{1\times k}\cdot \Big(\IW-E[\IW\mid Z,1\!-\!\trt,\cov]\Big)_{k\times 1}\IfZ\transpose _{1\times k}\myetaWZvec(1\!-\!\trt,\cov)\inv S(W,Z,1\!-\!\trt,\cov)\longmid 1\!-\!\trt,\cov\Big]
						\\
						=& E\Big\{\Big[Y-E[Y\mid Z,1\!-\!\trt,\cov] -\R(1\!-\!\trt,\cov)_{1\times k}\cdot \Big(\IW-E[\IW\mid Z,1\!-\!\trt,\cov]\Big)_{k\times 1}\Big]\cdot\\
						&\IfZ\transpose _{1\times k}\myetaWZvec(1\!-\!\trt,\cov)\inv S(Y,W,Z,1\!-\!\trt,\cov)\longmid 1\!-\!\trt,\cov\Big\}
						\label{eq:pathwisederivative3_cate}
					\end{split}\vspace{-0.05in}\end{equation}}}
		Recall that $\IZ=\{\mathbbm{1}(Z=z_1),\mathbbm{1}(Z=z_2),\dots,\mathbbm{1}(Z=z_k)\}\transpose $ denote a $k\times 1$ vector generalizing the binary $Z$, with $\IZi=\mathbbm{1}(Z=z_i)$. We note that
		{{\begin{equation*} 
					E[Y\mid Z,1\!-\!\trt,\cov]\mathbbm{1}_{1\times k}=\myetaYZvec(1\!-\!\trt,\cov)\transpose _{1\times k}\text{diag}[\IZ]_{k\times k}+E[Y\mid Z=z_0,1\!-\!\trt,\cov]\mathbbm{1}_{1\times k},
					\end{equation*} }}
		and
		{\small{\begin{equation*}
					E[\IW\mid Z,1\!-\!\trt,\cov]_{k \times 1}\mathbbm{1}_{1\times k}=\myetaWZvec(1\!-\!\trt,\cov)_{k \times k}\text{diag}[\IZ]_{k\times k}+E[\IW\mid Z=z_0,1\!-\!\trt,\cov]_{k\times 1}\mathbbm{1}_{1\times k}.
					\end{equation*} }}
		Therefore, 
		{\small{\begin{equation*}\begin{split}
						&\Big(Y-E[Y\mid Z,1\!-\!\trt,\cov] -\R(1\!-\!\trt,\cov)_{1\times k}\cdot \Big(\IW-E[\IW\mid Z,1\!-\!\trt,\cov]\Big)_{k\times 1}\Big)\mathbbm{1}_{1\times k}\\
						=&\Big(Y -\R(1\!-\!\trt,\cov)\IW\Big)\mathbbm{1}_{1\times k}-\Big(\myetaYZvec(1\!-\!\trt,\cov)\transpose _{1\times k}\text{diag}[\IZ]_{k\times k}+E[Y\mid Z=z_0,1\!-\!\trt,\cov]\mathbbm{1}_{1\times k}\Big)\\
						&+\R(1\!-\!\trt,\cov)_{1\times k}\Big(\myetaWZvec(1\!-\!\trt,\cov)_{k \times k}\text{diag}[\IZ]_{k\times k}+E[\IW\mid Z=z_0,1\!-\!\trt,\cov]_{k\times 1}\mathbbm{1}_{1\times k}\Big)\\
						=&\Big[Y-E[Y\mid Z=z_0,1\!-\!\trt,\cov] -\R(1\!-\!\trt,\cov)_{1\times k}\Big(\IW-E[\IW\mid Z=z_0,1\!-\!\trt,\cov]\Big)_{k\times 1}\Big]\mathbbm{1}_{1\times k}.
					\end{split}\end{equation*}}}
		Thus, Eq. (\ref{eq:pathwisederivative3_cate}) can be simplified to
		{\small{\vspace{-0.05in}\begin{equation}\begin{split}
						&\dt \R_\pathwiseparam(1\!-\!\trt,\cov)_{1\times k}\\
						=& E\Big\{\Big[Y-E[Y\mid Z=z_0,1\!-\!\trt,\cov]-\R(1\!-\!\trt,\cov)_{1\times k}\cdot \Big(\IW-E[\IW\mid Z=z_0,1\!-\!\trt,\cov]\Big)_{k\times 1}\Big]\cdot\\
						&\IfZ\transpose _{1\times k}\Big(\myetaWZvec(1\!-\!\trt,\cov)\Big)_{k\times k}\inv S(Y,W,Z,1\!-\!\trt,\cov)\longmid 1\!-\!\trt,\cov\Big\}\label{eq:delta_R_IF}
					\end{split}\vspace{-0.05in}\end{equation}}}
		
		Now we consider the pathwise derivative of $E[ \R(1\!-\!\trt,\cov)_{1\times k} \deltaWAvec(Z,\cov)_{k\times k}]$. Note that
		{\small{\vspace{-0.05in}\begin{equation}\begin{split}
						&\dt E_\pathwiseparam[ \R_\pathwiseparam(1\!-\!\trt,\cov) \bm{\delta^{W}_{\adiff,\pathwiseparam}}(Z,\cov)]\\
						=& E[\dt \R(1\!-\!\trt,\cov) \bm{\delta^{W}_{\adiff,\pathwiseparam}}(Z,\cov)] + E[\dt \R(1\!-\!\trt,\cov)\deltaWAvec(Z,\cov) ] + E[\R(1\!-\!\trt,\cov)\deltaWAvec(Z,\cov) S(\trt,Z,\cov)],\label{eq:part0_cate}
					\end{split}\vspace{-0.05in}\end{equation}}}
		thus we consider $E[\R(1\!-\!\trt,\cov)\dt \dW_{\adiff,\pathwiseparam}(Z,\cov) ]$ and $E[\dt \R(1\!-\!\trt,\cov)\deltaWAvec(Z,\cov) ] $ seperately.
		
		First, by Eq. (\ref{eq:pathwisederivative2_cate}), and using similar argument as the derivation of Eq. (\ref{eq:part1}) in Appendix \ref{appendix:if}, we have
		{\small{\vspace{-0.05in}\begin{equation}\begin{split}
						&E[\R(1\!-\!\trt,\cov)\dt \bm{\delta^{W}}_{\bm{\adiff},\pathwiseparam}(Z,\cov) ]\\
						=&E\Big[e_{\R}(Z,\cov)_{1\times k}(\IW-\deltaWAvec(Z,\cov)\trt-E[\IW\mid \trt\!=\!0,Z,\cov])_{k\times 1}\frac{2\trt-1}{f(\trt\mid Z,\cov)}S(W,\trt,Z,\cov)\Big].\label{eq:part1_cate}
					\end{split}\vspace{-0.05in}\end{equation}}}
		where $e_{\R}(z,\ccov)=E[\R(1\!-\!\trt,\cov)\mid Z\!=\!z,\cov\!=\!\ccov] $.
		
		Second, we consider $ E[\dt \R_\pathwiseparam(1\!-\!\trt,\cov)\deltaWAvec(Z,\cov)]$. By Eq. (\ref{eq:delta_R_IF}) and using the argument as the derivation of Eq. (\ref{eq:part2}), we have
		{\small{\vspace{-0.05in}\begin{equation}\begin{split}
						&E[\dt \R_\pathwiseparam(1\!-\!\trt,\cov)\deltaWAvec(Z,\cov)]\\
						=&E\Big\{\Big[Y-E[Y\mid Z=z_0,\trt,\cov]-\R(\trt,\cov)\cdot \Big(\IW-E[\IW\mid Z=z_0,\trt,\cov]\Big)\Big]\\
						&\IfZ\transpose \Big(\myetaWZvec(\trt)\Big)\inv \Big(E[\deltaWAvec(Z,\cov)\mid  1-\trt,\cov]\frac{f(1-\trt\mid \cov)}{f(\trt\mid \cov)}\Big)S(Y,W,\trt,Z,\cov)\Big\}.\label{eq:part2_cate}
					\end{split}\vspace{-0.05in}\end{equation}
		}}

		Combining Eq. (\ref{eq:part0_cate}), (\ref{eq:part1_cate}), and (\ref{eq:part2_cate}) we have
		{\small{
				\begin{equation*}\begin{split} 
						&\dt E_\pathwiseparam[\bm{\delta^{W}_{\adiff,\pathwiseparam}}(Z,\cov)\cdot \R_\pathwiseparam(1\!-\!\trt,\cov) ]\\
						=& E\Big\{\R(1\!-\!\trt,\cov)\deltaWAvec(Z,\cov)\\
						+&E[\R(1\!-\!\trt,\cov)\mid Z,\cov](\IW-\deltaWAvec(Z,\cov)\trt-E[\IW\mid \trt\!=\!0,Z,\cov])\frac{2\trt-1}{f(\trt\mid Z,\cov)} \\
						+& \Big[Y-E[Y\mid Z=z_0,\trt,\cov]-\R(\trt,\cov)\cdot \Big(\IW-E[\IW\mid Z=z_0,\trt,\cov]\Big)\Big]\\
						&\IfZ\transpose \Big(\myetaWZvec(\trt)\Big)\inv \Big(E[\deltaWAvec(Z,\cov)\mid  1-\trt,\cov]\frac{f(1-\trt\mid \cov)}{f(\trt\mid \cov)}\Big) S(Y,W,\trt,Z,\cov)\Big\}.
					\end{split}\end{equation*}
		}}
		
		Therefore, the influence function for $\Delta= E[\dY_{\adiff}(Z,\cov)] - E[\R(1\!-\!\trt,\cov)\deltaWAvec(Z,\cov) ]$ is given by
		\begin{equation*}\begin{split} 
				&\text{IF}_{\Delta}(Y,W,A,Z,\cov)\\
				=& \frac{2\trt-1}{f(\trt\mid Z,\cov)}\Big(Y-\dY_{\adiff}(Z,\cov)\trt-E[Y\mid \trt\!=\!0,Z,\cov]\Big)\\
				-&E[\R(1\!-\!\trt,\cov)\mid Z,\cov]\cdot \Big(\IW-\deltaWAvec(Z,\cov)\trt-E[\IW\mid \trt\!=\!0,Z,\cov]\Big)\frac{2\trt-1}{f(\trt\mid Z,\cov)}\\
				-&\Big[Y-E[Y\mid Z\!=\!z_0,\trt,\cov] - \R(\trt,\cov)\Big(\IW-E[\IW\mid Z\!=\!z_0,\trt,\cov]\Big)\Big]\cdot\\
				&\IfZ\transpose \Big(\myetaWZvec(\trt)\Big)\inv \Big(E[\deltaWAvec(Z,\cov)\mid  1-\trt,\cov]\frac{f(1-\trt\mid \cov)}{f(\trt\mid \cov)}\Big)\\
				+&\dY_{\adiff}(Z,\cov)- \R(1\!-\!\trt,\cov) \deltaWAvec(Z,\cov)-\Delta.
			\end{split}\end{equation*}
	\end{proof}

	\newpage
	\section{Proof of Theorem~\ref{theorem_MR_2}}\label{appendix:proof_robustness_cate}
	\begin{proof}
		Under the regularity conditions given in Theorem 3.2 of \cite{newey1994large}, the estimated nuisance parameters
		\[
		\hat{\theta}=\{(\hat{\alpha}^{\trt,Z}_{\text{mle}})\transpose ,(\hat{\beta}^Y_{\text{mle}})\transpose ,(\hat{\beta}^{W0}_{\text{mle}})\transpose ,(\hat{\beta}_{\textdr}^{\WZ})\transpose ,(\hat{\beta}_{\textdr}^{\WA})\transpose ,(\hat{\beta}_{\textdr}^{R})\transpose \}\transpose \]
		from solving the moment function $\mathbb{P}_n\{U_\theta(O;\theta)\}=0$ are asymptotically normal and converge at $o(n^{-1/2})$ rate to some fixed values \[{\theta}^*=\{({\alpha}^{\trt,Z}_{*})\transpose ,({\beta}^Y_{*})\transpose ,({\beta}^{W0}_{*})\transpose ,({\beta}_{*}^{\WZ})\transpose ,({\beta}_{*}^{\WA})\transpose ,({\beta}_{*}^{R})\transpose \}\transpose \] satisfying $E[U_{\theta}(O;\theta^*)]=0$ regardless of whether the corresponding nuisance models are correctly specified.
		Accordingly, we let $\text{D}^*_{\bm{\delta^W_{\adiff}}(Z,\cov)}$, $\text{D}^*_{\R(1\!-\!\trt,\cov)}$, $\R^*(1\!-\!\trt,\cov)$, $E^*[Y\mid Z=z_0,\trt,\cov]$, $E^*[\IW\mid \trt\!=\!0,Z=z_0,\cov]$, $\bm{\delta^{W^*}_{\adiff}}(Z,\cov)$, $\bm{\myeta^{W^*}_{\zdiff}}(\trt,\cov)$, $E^*[\IW\mid Z=z_0,\trt,\cov]$, $f^*(\trt\mid Z,\cov)$, $f^*(Z\mid \trt,\cov)$, and $f^*(\trt\mid \cov)$ denote the probability limit of the estimated nuisance models. Similarly, we let $\Delta^*_{\text{mr}}$, $\Delta^*_{\text{confounded,dr}}$, and $\Delta^*_{\text{bias,mr}}$ denote the limit of the estimated parameters of interest. In addition, recall that $\IfZ=\{\mathbbm{1}(Z=z_1)/f(Z=z_1\mid  \trt,\cov)-\mathbbm{1}(Z=z_0)/f(Z=z_0\mid  \trt,\cov),\mathbbm{1}(Z=z_2)/f(Z=z_2\mid  \trt,\cov)-\mathbbm{1}(Z=z_0)/f(Z=z_0\mid  \trt,\cov),\dots,\mathbbm{1}(Z=z_k)/f(Z=z_k\mid  \trt,\cov)-\mathbbm{1}(Z=z_0)/f(Z=z_0\mid  \trt,\cov)\}\transpose $ denote a $k\times 1$ vector generalizing $(2Z-1)/f(Z\mid \trt,\cov)$ in the binary case, with $\IfZj=\mathbbm{1}(Z=z_j)/f(Z=z_j\mid  \trt,\cov)-\mathbbm{1}(Z=z_0)/f(Z=z_0\mid  \trt,\cov)$. With slight abuse of notation, we let $\IfZstar$ denote the limit of $\IfZ$.

		We start with showing that $\bm{\delta^{W^*}_{\adiff}}(Z,\cov)=\bm{\delta^{W}_{\adiff}}(Z,\cov)$ and $\bm{\myeta^{W^*}_{\zdiff}}(\trt,\cov)=\bm{\myeta^{W}_{\zdiff}}(\trt,\cov)$ under $\Mipw\cup\Mor$, and $\R^*(\trt,\cov)=\R(\trt,\cov)$ under $\Mgest\cup\Mor$. First, we show that $\bm{\delta^{W^*}_{\adiff}}(Z,\cov)=\bm{\delta^{W}_{\adiff}}(Z,\cov)$ and $\bm{\myeta^{W^*}_{\zdiff}}(\trt,\cov)=\bm{\myeta^{W}_{\zdiff}}(\trt,\cov)$ under $\Mipw\cup\Mor$. It suffice to show that $\delta^{w_i^*}_{\adiff}(Z,\cov)=\delta^{w_i}_{\adiff}(Z,\cov)$ and $\myeta^{w_i^*}_{\zdiff}(\trt,\cov)=\myeta^{w_i}_{\zdiff}(\trt,\cov)$ for all $i$. Or equivalently $(\beta^{\WiA}_*,\beta^{\WiZ}_*)=(\beta^{\WiA},\beta^{\WiZ})$
		where $(\beta^{\WiA},\beta^{\WiZ})$ is the subset of $(\beta^{\WA},\beta^{\WZ})$ corresponding to the $i$-th level of $W$. Recall that $\hat{\beta}_{\textdr}^{\WiA}$ and $\hat{\beta}_{\textdr}^{\WiZ}$ solves $\mathbb{P}_n\Big\{U_{\beta^{\WiA},\beta^{\WiZ}}(\hat{\beta}_{\textdr}^{\WiA},\hat{\beta}_{\textdr}^{\WiZ})\Big \}=0$
		with $\lim_{n\rightarrow\infty}\mathbb{P}_n\Big\{U_{\beta^{\WiA},\beta^{\WiZ}}(\hat{\beta}_{\textdr}^{\WiA},\hat{\beta}_{\textdr}^{\WiZ})\Big \}=E[U_{\beta^{\WiA},\beta^{\WiZ}}(\beta_{*}^{\WiA},\beta_{*}^{\WiZ})]$. Now consider $E[U_{\beta^{\WiA},\beta^{\WiZ}}(\beta_{*}^{\WiA},\beta_{*}^{\WiZ})]\longmid_{\beta_{*}^{\WiA}=\beta^{\WiA},\beta_{*}^{\WiZ}=\beta^{\WiZ}}$ under $\Mipw$ where $\myeta^{W_i}_{\zdiff}(\trt,\cov;\beta^{\WiZ})$ and $\delta^{W_i}_{\adiff}(Z,\cov;\beta^{\WiA})$ are correctly specified, i.e. $\myeta^{W_i}_{\zdiff}(\trt,\cov)=\myeta^{W_i}_{\zdiff}(\trt,\cov;\beta^{\WiZ})$ and $\delta^{W_i}_{\adiff}(Z,\cov)=\delta^{W_i}_{\adiff}(Z,\cov;\beta^{\WiA})$, we have
		\begin{equation*}\begin{split}
				&E[U_{\beta^{\WiA},\beta^{\WiZ}}(\beta^{\WiA},\beta^{\WiZ})]\\
				=&E\{\Big[g_0(\trt,Z,\cov)-E^*[g_0(\trt,Z,\cov)\mid \cov]\Big] \Big[\IWi-E[\IWi\mid \trt,Z,\cov; \beta^{W0}_{*},\beta^{\WiZ},\beta^{\WiA}]\Big]\}\\
				=&E\{\Big[g_0(\trt,Z,\cov)-E^*[g_0(\trt,Z,\cov)\mid \cov]\Big] \Big[E[\IWi\mid \trt=0,Z=z_0,\cov]-E[\IWi\mid \trt=0,Z=z_0,\cov; \beta_*^{W0}]+\\
				&[\myetaWiZvec(\trt=0,\cov)-\myetaWiZvec(\trt=0,\cov;\beta^{\WiZ})]\IZ+[\delta^{W_i}_{\adiff}(Z=z_0,\cov)-\delta^{W_i}_{\adiff}(Z=z_0,\cov;\beta^{\WiA})]\trt+\\
				&[\etaWiAZvec(\cov)-\etaWiAZvec(\cov;\beta^{\WAZ})]\trt \IZ\Big]\}\\
				=&E\{\Big[g_0(\trt,Z,\cov)-E[g_0(\trt,Z,\cov)\mid \cov]\Big] \Big[E[\IWi\mid \trt=0,Z=z_0,\cov]-E[\IWi\mid \trt=0,Z=z_0,\cov; \beta_*^{W0}]\Big]\}\\
				=&0
			\end{split}\end{equation*} 
		because $E\Big[\{g_0(\trt,Z,\cov)-E[g_0(\trt,Z,\cov)\mid \cov]\}h(X)\Big]=0$ for any function $h$. Thus, under $\Mipw$ where $\myeta^{W_i}_{\zdiff}(\trt,\cov;\beta^{\WiZ})$ and $\delta^{W_i}_{\adiff}(Z,\cov;\beta^{\WiA})$ are correctly specified, $\mathbb{P}_n\Big\{U_{\beta^{\WiA},\beta^{\WiZ}}(\hat{\beta}_{\textdr}^{\WiA},\hat{\beta}_{\textdr}^{\WiZ})\longmid_{\hat{\beta}_{\textdr}^{\WiA}=\beta^{\WiA},\hat{\beta}_{\textdr}^{\WiZ}=\beta^{\WiZ}}\Big\}$ converges to zero, i.e. $(\beta^{\WiA},\beta^{\WiZ})$ is a solution to the probability limit of $\mathbb{P}_n\Big\{U_{\beta^{\WiA},\beta^{\WiZ}}(\hat{\beta}_{\textdr}^{\WiA},\hat{\beta}_{\textdr}^{\WiZ})\Big\}=0$. Thus $\beta_{*}^{\WiA}=\beta^{\WiA}$ and $\beta_{*}^{\WiZ}=\beta^{\WiZ}$, and $\delta^{w_i^*}_{\adiff}(Z,\cov)=\delta^{w_i}_{\adiff}(Z,\cov)$ and $\myeta^{w_i^*}_{\zdiff}(\trt,\cov)=\myeta^{w_i}_{\zdiff}(\trt,\cov)$ for all $i$. Therefore $\bm{\delta^{W^*}_{\adiff}}(Z,\cov)=\bm{\delta^{W}_{\adiff}}(Z,\cov)$ and $\bm{\myeta^{W^*}_{\zdiff}}(\trt,\cov)=\bm{\myeta^{W}_{\zdiff}}(\trt,\cov)$ under $\Mipw$.
		
		Similar arguments apply to the scenario under $\Mor$. Under $\Mor$ where working models $\R(\trt,\cov;\beta^R)$, $E[Y\mid Z=z_0,\trt,\cov;\beta^Y]$, $\myeta^{W_i}_{\zdiff}(\trt,\cov;\beta^{\WiZ})$, $\delta^{W_i}_{\adiff}(Z,\cov;\beta^{\WiA})$, and $E[\IWi\mid \trt=0,Z=z_0,\cov;\beta^{W_i}]$ are correctly specified, we have ${\beta}^{W0}_{*}={\beta}^{W0}$ and thus $E^*[\IW\mid \trt=0,Z=z_0,\cov]=E[\IWi\mid \trt=0,Z=z_0,\cov]$. We again consider
		\begin{equation*}\begin{split}
				&E[U_{\beta^{\WiA},\beta^{\WiZ}}(\beta^{\WiA},\beta^{\WiZ})]\\
				=&E\{\Big[g_0(\trt,Z,\cov)-E^*[g_0(\trt,Z,\cov)\mid \cov]\Big] \Big[\IWi-E[\IWi\mid \trt,Z,\cov; \beta^{W0}_{*},\beta^{\WiZ},\beta^{\WiA}]\Big]\}\\
				=&E\{\Big[g_0(\trt,Z,\cov)-E^*[g_0(\trt,Z,\cov)\mid \cov]\Big] \Big[E[\IWi\mid \trt=0,Z=z_0,\cov]-E[\IWi\mid \trt=0,Z=z_0,\cov; \beta_*^{W0}]+\\
				&[\myetaWiZvec(\trt=0,\cov)-\myetaWiZvec(\trt=0,\cov;\beta^{\WiZ})]\IZ+[\delta^{W_i}_{\adiff}(Z=z_0,\cov)-\delta^{W_i}_{\adiff}(Z=z_0,\cov;\beta^{\WiA})]\trt+\\
				&[\etaWiAZvec(\cov)-\etaWiAZvec(\cov;\beta^{\WAZ})]\trt \IZ\Big]\}\\
				=&E\{\Big[g_0(\trt,Z,\cov)-E^*[g_0(\trt,Z,\cov)\mid \cov]\Big] \Big[E[\IWi\mid \trt=0,Z=z_0,\cov]-E[\IWi\mid \trt=0,Z=z_0,\cov; \beta_*^{W0}]\Big]\}\\
				=&0
			\end{split}\end{equation*} 
		because $E[\IWi\mid \trt=0,Z=z_0,\cov; \beta_*^{W0}]=E[\IWi\mid \trt=0,Z=z_0,\cov]$. Therefore $\bm{\delta^{W^*}_{\adiff}}(Z,\cov)=\bm{\delta^{W}_{\adiff}}(Z,\cov)$ and $\bm{\myeta^{W^*}_{\zdiff}}(\trt,\cov)=\bm{\myeta^{W}_{\zdiff}}(\trt,\cov)$ under $\Mor$. In addition, we have that 
		\begin{equation}\begin{split}\label{underM3_CAT}
				E^*[\IWi\mid \trt,Z,\cov]=&E^*[\IWi\mid \trt\!=\!0,Z\!=\!z_0,\cov]\!+\!\delta^{w_i^*}_{\adiff}(Z\!=\!z_0,\cov)\trt\!+\bm{\myeta^{w_i^*}_{\zdiff}}(\trt,\cov)\IZ\!+\!\bm{\eta^{w_i^*}_{AZ}}(\cov)\trt\; \IZ\\
				=&E[\IWi\mid \trt\!=\!0,Z\!=\!z_0,\cov]\!+\!\delta^{w_i}_{\adiff}(Z\!=\!z_0,\cov)\trt\!+\!\myetaWiZvec(\trt,\cov)\IZ\!+\!\etaWiAZvec(\cov)\trt\; \IZ\\
				=&E[\IWi\mid \trt,Z,\cov],
			\end{split}\end{equation} 
		i.e., $E^*[\IW\mid \trt,Z,\cov]=E[\IW\mid \trt,Z,\cov]$.
		
		Second, we show that $\R^*(\trt,\cov)=\R(\trt,\cov)$ under $\Mgest\cup\Mor$.
		Under $\Mgest$ where working models $f(\trt,Z\mid \cov;\alpha^{\trt,Z})$ and $\R(\trt,\cov;\beta^R)$ are correctly specified, we have ${\alpha}^{\trt,Z}_{*}={\alpha}^{\trt,Z}$, $f^*(\trt,Z\mid \cov)=f(\trt,Z\mid \cov)$, and thus $E^*[g_1(\trt,Z,\cov)\mid \trt,\cov]=E[g_1(\trt,Z,\cov)\mid \trt,\cov]$ for any function $g_1(\trt,Z,\cov)$. Recall that $\hat{\beta}_{\textdr}^R$ solves $\mathbb{P}_n\Big\{U_{\beta^{R}}(\hat{\beta}_{\textdr}^{R})\Big\}=0$ with $\lim_{n\rightarrow\infty}\mathbb{P}_n\Big\{U_{\beta^{R}}(\hat{\beta}_{\textdr}^{R})\Big\}=E[U_{\beta^{R}}(\beta_{*}^{R})]$. Now consider $E[U_{\beta^{R}}(\beta^{R}_*)]\longmid_{\beta^{R}_*=\beta^{R}}$ under $\Mgest$ where $\R(\trt,\cov;\beta^R)$ is correctly specified, i.e. $\R(\trt,\cov)=\R(\trt,\cov;\beta^R)$, we have
		\begin{equation*}\begin{split}
				&E[U_{\beta^{R}}(\beta^{R})]=E\{\Big[g_1(\trt,Z,\cov)-E^*[g_1(\trt,Z,\cov)\mid \trt,\cov]\Big]\Big[Y-E^*[Y\mid Z=z_0,\trt,\cov] -\\
				&\R(\trt,\cov;\beta^{R})(\IW-E^*[\IW\mid Z=z_0,\trt,\cov])\Big]\}\\
				=&E\{\Big[g_1(\trt,Z,\cov)-E[g_1(\trt,Z,\cov)\mid \trt,\cov]\Big]\Big[\{\R(\trt,\cov)-\R(\trt,\cov;\beta^{R})\}\bm{\myeta^{W}_{\zdiff}}(\trt,\cov)Z + \\
				&\{E[Y\mid Z=z_0,\trt,\cov]-E^*[Y\mid Z=z_0,\trt,\cov]\}+ \{E[\IW\mid Z=z_0,\trt,\cov]-E^*[\IW\mid Z=z_0,\trt,\cov]\}\R(\trt,\cov;\beta^{R})\Big]\}\\
				=&E\{\Big[g_1(\trt,Z,\cov)-E[g_1(\trt,Z,\cov)\mid \trt,\cov]\Big]\Big[\{E[Y\mid Z=z_0,\trt,\cov]-E^*[Y\mid Z=z_0,\trt,\cov]\}+ \\
				&\{E[\IW\mid Z=z_0,\trt,\cov]-E^*[\IW\mid Z=z_0,\trt,\cov]\}\R(\trt,\cov;\beta^{R})\Big]\}\\
				=&0
			\end{split}\end{equation*} 
		because
		$E\Big[\{g_1(\trt,Z,\cov)-E[g_1(\trt,Z,\cov)\mid \trt,\cov]\}h(A,X)\Big]=0$ for any function $h$. Thus, under $\Mgest$ where $\R(\trt,\cov;\beta^R)$ is correctly specified, $\mathbb{P}_n\Big\{U_{\beta^{R}}(\hat{\beta}_{\textdr}^{R})\longmid_{\hat{\beta}_{\textdr}^{R}=\beta^R}\Big\}$ converges to zero, i.e. $\beta^R$ is a solution to the probability limit of $\mathbb{P}_n\Big\{U_{\beta^{R}}(\hat{\beta}_{\textdr}^{R})\Big\}=0$. Thus $\beta^R_*=\beta^R$ and $\R^*(\trt,\cov)=\R(\trt,\cov)$ under $\Mgest$.
		
		Similar arguments apply to the scenario under $\Mor$. Under $\Mor$ where working models $\R(\trt,\cov;\beta^R)$, $E[Y\mid Z=z_0,\trt,\cov;\beta^Y]$ and $E[\IW\mid \trt,Z,\cov;\beta^{W_i}]$ are correctly specified, we have $E^*[Y\mid Z=z_0,\trt,\cov]=E[Y\mid Z=z_0,\trt,\cov]$ and by (\ref{underM3_CAT}) we have $E^*[\IW\mid \trt,Z,\cov]=E[\IW\mid \trt,Z,\cov]$. We again consider 
		\begin{equation*}\begin{split}
				&E[U_{\beta^{R}}(\beta^{R})]=E\{\Big[g_1(\trt,Z,\cov)-E^*[g_1(\trt,Z,\cov)\mid \trt,\cov]\Big]\Big[Y-E^*[Y\mid Z=z_0,\trt,\cov] -\\
				&\R(\trt,\cov;\beta^{R})(\IW-E^*[\IW\mid Z=z_0,\trt,\cov])\Big]\}\\
				=&E\{\Big[g_1(\trt,Z,\cov)-E^*[g_1(\trt,Z,\cov)\mid \trt,\cov]\Big]\Big[\{\R(\trt,\cov)-\R(\trt,\cov;\beta^{R})\}\bm{\myeta^{W}_{\zdiff}}(\trt,\cov)Z + \\
				&\{E[Y\mid Z=z_0,\trt,\cov]-E^*[Y\mid Z=z_0,\trt,\cov]\}+ \{E[\IW\mid Z=z_0,\trt,\cov]-E^*[\IW\mid Z=z_0,\trt,\cov]\}\R(\trt,\cov;\beta^{R})\Big]\}\\
				=&E\{\Big[g_1(\trt,Z,\cov)-E^*[g_1(\trt,Z,\cov)\mid \trt,\cov]\Big]\Big[\{E[Y\mid Z=z_0,\trt,\cov]-E^*[Y\mid Z=z_0,\trt,\cov]\}+ \\
				&\{E[\IW\mid Z=z_0,\trt,\cov]-E^*[\IW\mid Z=z_0,\trt,\cov]\}\R(\trt,\cov;\beta^{R})\Big]\}\\
				=&0
			\end{split}\end{equation*} 
		because $E^*[Y\mid Z=z_0,\trt,\cov]=E[Y\mid Z=z_0,\trt,\cov]$ and $E^*[\IW\mid Z=z_0,\trt,\cov]=E[\IW\mid Z=z_0,\trt,\cov]$. Therefore $\beta^R_*=\beta^R$ and $\R^*(\trt,\cov)=\R(\trt,\cov)$ under $\Mor$.
		
		We now show that $E[\Delta^*_{\text{mr}}]=\Delta$ under $\mathcal{M}_{\text{union}}$. To this end, we consider $E[\Delta^*_{\text{confounded}}]$, $E[\text{D}^*_{\bm{\delta^W_{\adiff}}(Z,\cov)}]$, and $E[\text{D}^*_{\R(1\!-\!\trt,\cov)}]$ respectively. 
		Under $\Mgest$ where working models $f(\trt,Z\mid \cov;\alpha^{\trt,Z})$ and $R(\trt,\cov;\beta^R)$ are correctly specified, we have $f^*(\trt,Z\mid \cov)=f(\trt,Z\mid \cov)$ and $R^*(\trt,\cov)=R(\trt,\cov)$. First we consider
		\begin{equation*} 
			E[\Delta^*_{\text{confounded}}]=E\Big[\frac{2\trt-1}{f^*(\trt\mid Z,\cov)}\Big(E[Y\mid \trt,Z,\cov]-E^*[Y\mid \trt,Z,\cov]\Big)+E^*[Y\mid \trt\!=\!1,Z,\cov]-E^*[Y\mid \trt\!=\!0,Z,\cov]\Big].
			\end{equation*} 
		When $f^*(\trt\mid Z,\cov)=f(\trt\mid Z,\cov)$, by Eq. (\ref{eq:tool}) we have
		\begin{equation*}\begin{split} 
				E[\Delta^*_{\text{confounded}}]=&E\Big[\frac{2\trt-1}{f(\trt\mid Z,\cov)}\Big(E[Y\mid \trt,Z,\cov]-E^*[Y\mid \trt,Z,\cov]\Big)+E^*[Y\mid \trt\!=\!1,Z,\cov]-E^*[Y\mid \trt\!=\!0,Z,\cov]\Big]\\
				=&E[\dY_{\adiff}(Z,\cov)-\delta^{Y^*}_{\adiff}(Z,\cov)+\delta^{Y^*}_{\adiff}(Z,\cov)]=\Delta_{\text{confounded}}.
			\end{split}\end{equation*}
		
		Second, consider 
		\begin{equation*}\begin{split}
				E[\text{D}^*&_{\deltaWAvec(Z,\cov)}]=
				E\Big[\Big(\sum_\trt \R^*(1\!-\!\trt,\cov)f^*(\trt\mid Z,\cov)\Big)\Big(\IW-E^*[\IW\mid \trt,Z,\cov]\Big)\frac{2\trt-1}{f^*(\trt\mid Z,\cov)}\Big]\\
				=&E\Big[\Big(\sum_\trt \R^*(1\!-\!\trt,\cov)f^*(\trt\mid Z,\cov)\Big)\Big(E[\IW\mid \trt,Z,\cov]-E^*[\IW\mid \trt,Z,\cov]\Big)\frac{2\trt-1}{f^*(\trt\mid Z,\cov)}\Big]
			\end{split}\end{equation*}
		When $f^*(\trt\mid Z,\cov)=f(\trt\mid Z,\cov)$, $\sum_\trt \R^*(1\!-\!\trt,\cov)f^*(\trt\mid Z,\cov)= E[\R^*(1\!-\!\trt,\cov) \mid Z,\cov]$. By Eq. (\ref{eq:tool}) we have
		\begin{equation*}\begin{split} 
				E[\text{D}^*_{\deltaWAvec(Z,\cov)}]=E\Big[E[\R^*(1\!-\!\trt,\cov) \mid Z,\cov]\Big(\deltaWAvec(Z,\cov)-\bm{\delta^{W^*}_{\adiff}}(Z,\cov)\Big) \Big].
			\end{split}\end{equation*}
		Because we also have $\R^*(\trt,\cov)=\R(\trt,\cov)$,
		\begin{equation*}\begin{split} 
				&E[\text{D}^*_{\deltaWAvec(Z,\cov)}+\R^*(1\!-\!\trt,\cov) \bm{\delta^{W^*}_{\adiff}}(Z,\cov)]\\
				=&E\Big[E[\R(1\!-\!\trt,\cov) \mid Z,\cov]\Big(\deltaWAvec(Z,\cov)-\bm{\delta^{W^*}_{\adiff}}(Z,\cov)\Big) +\R(1\!-\!\trt,\cov) \bm{\delta^{W^*}_{\adiff}}(Z,\cov)\Big]=\Delta_{\text{bias}}.
			\end{split}\end{equation*}

		Third, we consider $E[\text{D}^*_{\R(1\!-\!\trt,\cov)}]$. Because
		\begin{equation*} 
			E[Y\mid \trt,Z,\cov]\mathbbm{1}_{1\times k}=\myetaYZvec(\trt,\cov)\transpose _{1\times k}\text{diag}[\IZ]_{k\times k}+E[Y\mid Z=z_0,\trt,\cov]\mathbbm{1}_{1\times k},
			\end{equation*} 
		and
		\begin{equation*}
			E[\IW\mid \trt,Z,\cov]_{k \times 1}\mathbbm{1}_{1\times k}=\myetaWZvec(\trt,\cov)_{k \times k}\text{diag}[\IZ]_{k\times k}+E[\IW\mid Z=z_0,\trt,\cov]_{k\times 1}\mathbbm{1}_{1\times k},
			\end{equation*} 
		we have
		{\small{\vspace{-0.05in}\begin{equation}\begin{split}
						&E[\text{D}^*_{\R(1\!-\!\trt,\cov)}]\\
						=&E\Big\{\Big[Y-E^*[Y\mid Z=z_0,\trt,\cov] - \R^*(\trt,\cov)\Big(\IW-E^*[\IW\mid Z=z_0,\trt,\cov]\Big)\Big]\\
						&\Big(\IfZstar\Big) \bm{\myeta^{W^*}_{\zdiff}}(\trt,\cov)\inv \Big(\sum_Z \bm{\delta^{W^*}_{\adiff}}(Z,\cov)f^*(Z\mid  1-\trt,\cov)\cdot\frac{f^*(1-\trt\mid \cov)}{f^*(\trt\mid \cov)}\Big)\Big\}\\
						=&E\Big\{\Big[\{\R(\trt,\cov)-\R^*(\trt,\cov)\}_{1\times k}\myetaWZvec(\trt,\cov)_{k\times k}\text{diag}[\IZ]_{k\times k} + \{E[Y\mid Z=z_0,\trt,\cov]\!-\!E^*[Y\mid Z=z_0,\trt,\cov]\}\mathbbm{1}_{1\times k} \\
						&+ \R^*(\trt,\cov)_{1\times k}\{E[\IW\mid Z=z_0,\trt,\cov]-E^*[\IW\mid Z=z_0,\trt,\cov]\}_{k\times 1}\mathbbm{1}_{1\times k}\Big]\\
						&\text{diag}[\IfZstar]_{k\times k}\bm{\myeta^{W^*}_{\zdiff}}(\trt,\cov)\inv \Big(\sum_Z \bm{\delta^{W^*}_{\adiff}}(Z,\cov)f^*(Z\mid  1-\trt,\cov)\Big)\frac{f^*(1-\trt\mid \cov)}{f^*(\trt\mid \cov)}\Big\}.\label{eq:dr_simple}
					\end{split}\vspace{-0.05in}\end{equation}}}

		When $\IfZstar=\IfZ$, by similar argument as Eq. (\ref{eq:tool}), we have for any function $h(\trt,Z,\cov)$
		\vspace{-0.05in}\begin{equation}
			E[h(\trt,Z,\cov)\mathbbm{1}_{1\times k}\text{diag}[\IfZ]_{k\times k}]=E[\{h(\trt,z_1,\cov)-h(\trt,z_0,\cov),\dots,h(\trt,z_k,\cov)-h(\trt,z_0,\cov)\}]_{1\times k}.\label{eq:tool_cate}\vspace{-0.05in}\end{equation}
		Thus, Eq. (\ref{eq:dr_simple}) can be simplified to 
		\vspace{-0.05in}\begin{equation}\begin{split}
				&E[\text{D}^*_{\R(1\!-\!\trt,\cov)}]\\
				=&E\Big\{
				[\R(\trt,\cov)-\R^*(\trt,\cov)]_{1\times k}\myetaWZvec(\trt,\cov)_{k\times k}
				\text{diag}\left[\IfZ+\frac{\mathbbm{1}(Z=z_0)}{f(Z\!=\!z_0\mid \trt,\cov)}\right]_{k\times k}\\
				&\bm{\myeta^{W^*}_{\zdiff}}(\trt,\cov)\inv \Big(\sum_Z \bm{\delta^{W^*}_{\adiff}}(Z,\cov)f(Z\mid  1-\trt,\cov)\Big)\frac{f^*(1-\trt\mid \cov)}{f^*(\trt\mid \cov)}
				\Big\}
				\label{need_change_to_A_cate},
			\end{split}\vspace{-0.05in}\end{equation}
		where
		\begin{equation*}\begin{split}
				\text{diag}&\left[\IfZ+\frac{\mathbbm{1}(Z=z_0)}{f(Z\!=\!z_0\mid \trt,\cov)}\right]=\text{diag}[\IZ]\text{diag}[\IfZ]\\
				=&\text{diag}\{\frac{\mathbbm{1}(Z=z_1)}{f(Z=z_1\mid \trt,\cov)},\frac{\mathbbm{1}(Z=z_2)}{f(Z=z_1\mid \trt,\cov)},\dots,\frac{\mathbbm{1}(Z=z_k)}{f(Z=z_k\mid \trt,\cov)}\}.
			\end{split}\end{equation*}
		Because $E[\text{diag}\left[\IfZ+\frac{\mathbbm{1}(Z=z_0)}{f(Z\!=\!z_0\mid \trt,\cov)}\right]_{k\times k}\mid \trt,\cov]=\text{I}_{k\times k}$, we can simplify Eq.~(\ref{need_change_to_A_cate}) as follows
		\vspace{-0.05in}\begin{equation}\begin{split}
				&E[\text{D}^*_{\R(1\!-\!\trt,\cov)}]\\
				=&E\Big\{
				[\R(\trt,\cov)-\R^*(\trt,\cov)]_{1\times k}\myetaWZvec(\trt,\cov)_{k\times k}
				\text{diag}\left[\IfZ+\frac{\mathbbm{1}(Z=z_0)}{f(Z\!=\!z_0\mid \trt,\cov)}\right]_{k\times k}\\
				&\bm{\myeta^{W^*}_{\zdiff}}(\trt,\cov)\inv \Big(\sum_Z \bm{\delta^{W^*}_{\adiff}}(Z,\cov)f(Z\mid  1-\trt,\cov)\Big)\frac{f^*(1-\trt\mid \cov)}{f^*(\trt\mid \cov)}
				\Big\}\\
				=&E\Big\{
				[\R(\trt,\cov)-\R^*(\trt,\cov)]\myetaWZvec(\trt,\cov)
				\bm{\myeta^{W^*}_{\zdiff}}(\trt,\cov)\inv \Big(\sum_Z \bm{\delta^{W^*}_{\adiff}}(Z,\cov)f(Z\mid  1-\trt,\cov)\Big)\frac{f^*(1-\trt\mid \cov)}{f^*(\trt\mid \cov)}
				\Big\}\label{need_change_to_A_cate2},
			\end{split}\vspace{-0.05in}\end{equation}
		We can see that when $\R^*(\trt,\cov)=\R(\trt,\cov)$ we have $E[\text{D}^*_{\R(1\!-\!\trt,\cov)}]=0$.
		
		In summary, under $\Mgest$, we have
		\begin{equation*}\begin{split} 
				E[\Delta^*_{\text{mr}}]=&E[\Delta^*_{\text{confounded}}]-\{E[\text{D}^*_{\deltaWAvec(Z,\cov)}+\R^*(1\!-\!\trt,\cov) \bm{\delta^{W^*}_{\adiff}}(Z,\cov)]+E[\text{D}^*_{\R(1\!-\!\trt,\cov)}]\}\\
				=&\Delta_{\text{confounded}}-\{\Delta_{\text{bias}}+0\}=\Delta
			\end{split}\end{equation*}

		Under $\Mipw$ where $f(\trt,Z\mid \cov;\alpha^{\trt,Z})$, $\myetaWZvec(\trt,\cov;\beta^{\WZ})$ and $\bm{\delta^{W}_{\adiff}}(Z,\cov;\beta^{\WA})$ are correctly specified, we have $f^*(\trt,Z\mid \cov)=f(\trt,Z\mid \cov)$, $\bm{\delta^{W^*}_{\adiff}}(Z,\cov)=\bm{\delta^{W}_{\adiff}}(Z,\cov)$ and $\bm{\myeta^{W^*}_{\zdiff}}(\trt,\cov)=\bm{\myeta^{W}_{\zdiff}}(\trt,\cov)$. First we consider
		\begin{equation*} 
			E[\Delta^*_{\text{confounded}}]=E\Big[\frac{2\trt-1}{f^*(\trt\mid Z,\cov)}\Big(E[Y\mid \trt,Z,\cov]-E^*[Y\mid \trt,Z,\cov]\Big)+E^*[Y\mid \trt\!=\!1,Z,\cov]-E^*[Y\mid \trt\!=\!0,Z,\cov]\Big].
			\end{equation*} 
		When $f^*(\trt\mid Z,\cov)=f(\trt\mid Z,\cov)$, by Eq. (\ref{eq:tool}) we have
		\begin{equation*}\begin{split} 
				E[\Delta^*_{\text{confounded}}]=&E\Big[\frac{2\trt-1}{f(\trt\mid Z,\cov)}\Big(E[Y\mid \trt,Z,\cov]-E^*[Y\mid \trt,Z,\cov]\Big)+E^*[Y\mid \trt\!=\!1,Z,\cov]-E^*[Y\mid \trt\!=\!0,Z,\cov]\Big]\\
				=&E[\dY_{\adiff}(Z,\cov)-\delta^{Y^*}_{\adiff}(Z,\cov)+\delta^{Y^*}_{\adiff}(Z,\cov)]=\Delta_{\text{confounded}}.
			\end{split}\end{equation*}
		
		Second, consider 
		\begin{equation*}\begin{split}
				E[\text{D}^*&_{\deltaWAvec(Z,\cov)}]=
				E\Big[\Big(\sum_\trt \R^*(1\!-\!\trt,\cov)f^*(\trt\mid Z,\cov)\Big)\Big(\IW-E^*[\IW\mid \trt,Z,\cov]\Big)\frac{2\trt-1}{f^*(\trt\mid Z,\cov)}\Big]\\
				=&E\Big[\Big(\sum_\trt \R^*(1\!-\!\trt,\cov)f^*(\trt\mid Z,\cov)\Big)\Big(E[\IW\mid \trt,Z,\cov]-E^*[\IW\mid \trt,Z,\cov]\Big)\frac{2\trt-1}{f^*(\trt\mid Z,\cov)}\Big]
			\end{split}\end{equation*}
		When $f^*(\trt\mid Z,\cov)=f(\trt\mid Z,\cov)$, by Eq. (\ref{eq:tool}) we have
		\begin{equation*}\begin{split} 
				E[\text{D}^*_{\deltaWAvec(Z,\cov)}]=E\Big[ E[\R^*(1\!-\!\trt,\cov) \mid Z,\cov]\Big(\deltaWAvec(Z,\cov)-\bm{\delta^{W^*}_{\adiff}}(Z,\cov)\Big)\Big]=0
			\end{split}\end{equation*}
		because $\deltaWAvec(Z,\cov)=\bm{\delta^{W^*}_{\adiff}}(Z,\cov)$.

		Third, we consider $E[\text{D}^*_{\R(1\!-\!\trt,\cov)}]$. As discussed above, when $\IfZstar=\IfZ$, we have 
		Eq. (\ref{need_change_to_A_cate2}) hold. Note that when the model for $f(\trt\mid \cov)$ is correctly specified, i.e., $f^*(\trt\mid \cov)=f(\trt\mid \cov)$, in Appendix \ref{appendix:change_measure} we show that for any function $h(Y,W,\trt,Z,\cov )$,
		\begin{equation*}
			E[h(Y,W,\trt,Z,\cov ) \frac{f(1-\trt\mid \cov)}{f(\trt\mid \cov)}]=E[h(Y,W,Z,1\!-\!\trt,\cov )].
			\end{equation*} 
		Let 
		{\small{\begin{equation*}
					h(Y,W,\trt,Z,\cov)=[\R(\trt,\cov)-\R^*(\trt,\cov)]\myetaWZvec(\trt,\cov)
					\bm{\myeta^{W^*}_{\zdiff}}(\trt,\cov)\inv \Big(\sum_Z \bm{\delta^{W^*}_{\adiff}}(Z,\cov)f(Z\mid  1-\trt,\cov)\Big),
					\end{equation*} }}
		then Eq. (\ref{need_change_to_A_cate2}) is equivalent to
		{\small{\begin{equation*}\begin{split}
						&E[\text{D}^*_{\R(1\!-\!\trt,\cov)}]\\
						=&E\Big\{
						[\R(1\!-\!\trt,\cov)-\R^*(1\!-\!\trt,\cov)]\myetaWZvec(1\!-\!\trt,\cov)
						\Big(\bm{\myeta^{W^*}_{\zdiff}}(1\!-\!\trt,\cov)\Big)\inv \Big(\sum_Z \bm{\delta^{W^*}_{\adiff}}(Z,\cov)f(Z\mid  1-(1\!-\!\trt),\cov)\Big)
						\Big\}.
					\end{split}\end{equation*}}}
		In this case, because we also have that $\bm{\delta^{W^*}_{\adiff}}(Z,\cov)\deltaWAvec(Z,\cov)$ and $\bm{\myeta^{W^*}_{\zdiff}}(\trt,\cov)=\myetaWZvec(\trt,\cov)$ 
		\begin{equation*}\begin{split}
				&E[\text{D}^*_{R(1\!-\!\trt,\cov)}+\R^*(1\!-\!\trt,\cov) \bm{\delta^{W^*}_{\adiff}}(Z,\cov)]\\
				=&E\Big\{[\R(1\!-\!\trt,\cov)-\R^*(1\!-\!\trt,\cov)]E[\deltaWAvec(Z,\cov) \mid  \trt,\cov]+\R^*(1\!-\!\trt,\cov) \deltaWAvec(Z,\cov)\Big\}=\Delta_{\text{bias}}.
			\end{split}\end{equation*}

		In summary, under $\Mipw$, we have
		\begin{equation*}\begin{split} 
				E[\Delta^*_{\text{mr}}]=&E[\Delta^*_{\text{confounded}}]-\{E[\text{D}^*_{\deltaWAvec(Z,\cov)}]+E[\text{D}^*_{\R(1\!-\!\trt,\cov)}+\R^*(1\!-\!\trt,\cov) \bm{\delta^{W^*}_{\adiff}}(Z,\cov)]\}\\
				=&\Delta_{\text{confounded}}-\{0+\Delta_{\text{bias}}\}=\Delta
			\end{split}\end{equation*}

		Under $\Mor$ where $\R(\trt,\cov;\beta^R)$, $E[Y\mid Z=z_0,\trt,\cov;\beta^Y]$, $\bm{\myeta^W_{\zdiff}}(\trt,\cov;\beta^{\WZ})$, $\bm{\dW_{\adiff}}(Z,\cov;\beta^{\WA})$, and $E[\IW\mid \trt=0,Z=z_0,\cov;\beta^W]$ are correctly specified, we have $\R^*(\trt,\cov)=\R(\trt,\cov)$, $E^*[Y\mid Z=z_0,\trt,\cov]=E[Y\mid Z=z_0,\trt,\cov]$, $\bm{\myeta^{W^*}_{\zdiff}}(\trt,\cov)=\bm{\myeta^W_{\zdiff}}(\trt,\cov)$, $\bm{\delta^{W^*}_{\adiff}}(Z,\cov)=\bm{\delta^{W}_{\adiff}}(Z,\cov)$, and $E^*[\IW\mid \trt=0,Z=z_0,\cov]=E[\IW\mid \trt=0,Z=z_0,\cov]$. First we consider
		\begin{equation*} 
			E[\Delta^*_{\text{confounded}}]=E\Big[\frac{2\trt-1}{f^*(\trt\mid Z,\cov)}\Big(E[Y\mid \trt,Z,\cov]-E^*[Y\mid \trt,Z,\cov]\Big)+E^*[Y\mid \trt\!=\!1,Z,\cov]-E^*[Y\mid \trt\!=\!0,Z,\cov]\Big].
			\end{equation*} 
		Note that
		\begin{equation*}\begin{split}
				E^*[Y\mid Z,\trt,\cov] = &E^*[Y\mid Z=z_0,\trt,\cov]+\R^*(\trt,\cov)\bm{\myeta^{W^*}_{\zdiff}}(\trt=0,\cov)\IZ\\
				=&E[Y\mid Z=z_0,\trt,\cov]+\R(\trt,\cov)\myetaWZvec(\trt,\cov)\IZ=E[Y\mid Z,\trt,\cov],
			\end{split}\end{equation*}
		therefore we have 
		\begin{equation*}\begin{split} 
				E[\Delta^*_{\text{confounded}}]=&E\Big[\frac{2\trt-1}{f(\trt\mid Z,\cov)}\Big(E[Y\mid \trt,Z,\cov]-E[Y\mid \trt,Z,\cov]\Big)+E[Y\mid \trt\!=\!1,Z,\cov]-E[Y\mid \trt\!=\!0,Z,\cov]\Big]\\
				=&E\{E[Y\mid \trt\!=\!1,Z,\cov]-E[Y\mid \trt\!=\!0,Z,\cov]\}=\Delta_{\text{confounded}}.
			\end{split}\end{equation*}
		
		Second, consider
		\begin{equation*}\begin{split}
				E[\text{D}^*&_{\deltaWAvec(Z,\cov)}]=
				E\Big[\Big(\sum_\trt \R^*(1\!-\!\trt,\cov)f^*(\trt\mid Z,\cov)\Big)\Big(\IW-E^*[\IW\mid \trt,Z,\cov]\Big)\frac{2\trt-1}{f^*(\trt\mid Z,\cov)}\Big]\\
				=&E\Big[\Big(\sum_\trt \R^*(1\!-\!\trt,\cov)f^*(\trt\mid Z,\cov)\Big)\Big(E[\IW\mid \trt,Z,\cov]-E^*[\IW\mid \trt,Z,\cov]\Big)\frac{2\trt-1}{f^*(\trt\mid Z,\cov)}\Big]\\
				=&0
			\end{split}\end{equation*}
		because $E^*[\IW\mid \trt,Z,\cov]=E[\IW\mid \trt,Z,\cov]$ by (\ref{underM3_CAT}).
		
		Third, consider we consider $E[\text{D}^*_{\R(1\!-\!\trt,\cov)}]$.
		Because $\R^*(\trt,\cov)=\R(\trt,\cov)$, $E^*[Y\mid Z=z_0,\trt,\cov]=E[Y\mid Z=z_0,\trt,\cov]$, $\bm{\delta^{W^*}_{\adiff}}(Z=z_0,\cov)=\bm{\delta^{W}_{\adiff}}(Z=z_0,\cov)$, and $E^*[\IW\mid \trt=0,Z=z_0,\cov]=E[\IW\mid \trt=0,Z=z_0,\cov]$, it is straightforward to see from Eq. (\ref{eq:dr_simple}) that $E[\text{D}^*_{\R(1\!-\!\trt,\cov)}]=0$.

		In summary, under $\Mor$, we have
		\begin{equation*}\begin{split} 
				E[\Delta^*_{\text{mr}}]=&E[\Delta^*_{\text{confounded}}]-\{E[\text{D}^*_{\deltaWAvec(Z,\cov)}]+E[\text{D}^*_{\R(1\!-\!\trt,\cov)}]+E[\R^*(1\!-\!\trt,\cov) \bm{\delta^{W^*}_{\adiff}}(Z,\cov)]\}\\
				=&E[\Delta^*_{\text{confounded}}]-\{E[\text{D}^*_{\deltaWAvec(Z,\cov)}]+E[\text{D}^*_{\R(1\!-\!\trt,\cov)}]+E[\R(1\!-\!\trt,\cov) \bm{\delta^{W}_{\adiff}}(Z,\cov)]\}\\
				=&\Delta_{\text{confounded}}-\{0+0+\Delta_{\text{bias}}\}=\Delta
			\end{split}\end{equation*}
		
		In summary, $E[\Delta^*_{\text{mr}}]=\Delta$ under $\mathcal{M}_{\text{union}}=\Mgest\cup \Mipw\cup \Mor$.
		The rest of the arguments are the same as Appendix~\ref{appendix:proof_robustness_binary}.
	\end{proof}

	\newpage
	\section{Change from $\trt$ to $1\!-\!\trt$}\label{appendix:change_measure}
	In this section, we show that 
	\begin{equation*}
		E[h(Y,W,\trt,Z,\cov ) \frac{f(1-\trt\mid \cov)}{f(\trt\mid \cov)}]=E[h(Y,W,Z,1\!-\!\trt,\cov )].
		\end{equation*} 
	
	\begin{proof}
		Consider
		\begin{equation*}\begin{split}
				E[h(Y,W,\trt,Z,\cov ) \frac{f(1-\trt\mid \cov)}{f(\trt\mid \cov)}]=&\int E[h(Y,W,\trt,Z,\cov) \mid  \trt=\ttrt,\cov=\ccov] \frac{f(1\!-\!\ttrt\mid \ccov)}{f(\ttrt\mid \ccov)} f(\ttrt,\ccov)d\ttrt d\ccov \\
				=&\int E[h(Y,W,\trt,Z,\cov) \mid  \trt=\ttrt,\cov=\ccov] f(1\!-\!\trt,\cov)d\ttrt d\ccov .
			\end{split}\end{equation*}
		Because $\trt$ is binary, we have
		\begin{equation*}\begin{split}
				&\int E[h(Y,W,\trt,Z,\cov) \mid \trt\!=\!\ttrt,\cov\!=\!\ccov] f(1\!-\!\trt,\cov)d\ttrt d\ccov\\
				&\!=\!\int E[h(Y,W,\trt,Z,\cov) \mid \trt\!=\!1,\cov\!=\!\ccov] P(\trt\!=\!0,\ccov)+E[h(Y,W,\trt,Z,\cov) \mid \trt\!=\!0,\cov\!=\!\ccov] P(\trt\!=\!1,\ccov)d\ccov\\
				&\!=\!\int E[h(Y,W,Z,1\!-\!\trt) \mid 1\!-\!\trt\!=\!0,\cov\!=\!\ccov] P(\trt\!=\!0,\ccov)+E[h(Y,W,Z,1\!-\!\trt) \mid 1\!-\!\trt\!=\!1,\cov\!=\!\ccov] P(\trt\!=\!1,\ccov)d\ccov\\
				&\!=\!\int E[h(Y,W,Z,1\!-\!\trt) \mid 1\!-\!\trt\!=\!\ttrt,\cov\!=\!\ccov] f(\ttrt,\ccov)d\ttrt d\ccov \\
				&\!=\!E[h(Y,W,Z,1\!-\!\trt,\cov) ]
			\end{split}\end{equation*}
		Therefore
		\begin{equation*}
			E[h(Y,W,\trt,Z,\cov) \frac{f(1-\trt\mid \cov)}{f(\trt\mid \cov)}]=E[h(Y,W,Z,1-\trt,\cov)].
			\end{equation*} 
	\end{proof}		
	
	\clearpage
	\thispagestyle{empty}
	\bibliographystyle{apalike}
	\bibliography{ref_Neg_Ctrl}

\end{spacing}

\end{document}